\documentclass[10pt,a4paper,final]{iopart}

\usepackage{iopams}
\usepackage{graphicx}
\usepackage{widetext}
\usepackage[breaklinks=true,colorlinks=true,linkcolor=blue,urlcolor=blue,citecolor=blue]{hyperref}

\usepackage{helvet}
\def\Put(#1,#2)#3{\leavevmode\makebox(0,0){\put(#1,#2){#3}}}

\begin{document}

\title[Physics with Reactor Neutrinos]{Physics with Reactor Neutrinos}

\author{Xin Qian}
\address{Physics Department, Brookhaven National Laboratory, Upton, NY, 11973, USA}
\ead{xqian@bnl.gov}

\author{Jen-Chieh Peng}
\address{Department of Physics, University of Illinois at Urbana-Champaign, Urbana, IL, 61801, USA}
\ead{jcpeng@illinois.edu}

\begin{abstract}
 Neutrinos produced by nuclear reactors have played a major role in advancing
our knowledge of the properties of neutrinos. The first direct
detection of the neutrino, confirming its existence, was performed
using reactor neutrinos.  More recent experiments utilizing reactor
neutrinos have also found clear evidence for neutrino oscillation,
providing unique input for the determination of neutrino mass and
mixing. Ongoing and future reactor neutrino experiments will explore
other important issues, including the neutrino mass hierarchy and the
search for sterile neutrinos and other new physics beyond the
standard model. In this article, we review the recent progress in
physics using reactor neutrinos and the opportunities they offer for future
discoveries.
\end{abstract}

\vspace{2pc}

\submitto{\RPP}
\maketitle
\ioptwocol
\tableofcontents

\section{Introduction}\label{sec:intro}

Neutrinos are among the most fascinating and enigmatic particles
in nature. The standard model in particle physics includes
neutrinos as one of the fundamental point-like building blocks. 
Processes involving the production and interaction
of neutrinos provided crucial inputs for formulating the
electroweak theory, unifying the electromagnetic and weak
interactions. Neutrinos also play a prominent role in cosmology.
The abundant neutrinos produced soon after the big bang offer 
the potential to view the Universe at an epoch much earlier
than that accessible from the cosmic microwave background. The direct
detection of these \lq{relic}\rq  ~neutrinos from the big bang remains
a major experimental challenge. For a long time, these neutrinos were 
also considered a prime candidate for dark matter. 
While this is no longer viable given the current upper limit on
the neutrino mass, neutrinos nevertheless constitute a 
non-negligible fraction of the invisible mass in the Universe.    

Neutrinos also play an important role in astrophysics. 
Detection of neutrinos emitted in a supernova explosion reveals
not only the mechanisms of supernova evolution but also the properties
and interactions of neutrinos in a super dense environment. Extensive
efforts are also dedicated to the search for ultra-high-energy
extra-galactic neutrinos. The charge-neutral 
neutrinos can potentially be traced back to locate the sources of 
ultra-high-energy cosmic rays.  

Neutrino beams from accelerators have also been employed to probe the
quark structures of nucleons and nuclei via deep inelastic
scattering (DIS). Experiments using neutrino beams,
together with those with charged lepton beams, have provided
crucial tests to validate QCD as the theory for strong interactions. 

Observations of neutrino mixings and the existence of three non-degenerate 
neutrino mass eigenstates have provided the only unambiguous evidence so far
for physics beyond the standard model. The origin of such tiny neutrino
mass remains a mystery and could reveal new mechanisms other than the
Higgs mechanism for mass generation. Neutrinos may 
also be a portal for approaching the dark sector. Mixing 
between the standard model neutrinos with \lq{sterile}\rq ~neutrinos in the
dark sector could lead to observable effects.  

The purpose of this article is to review recent progress in neutrino
physics obtained from experiments performed near nuclear reactors.
As a prolific and steady source of electron antineutrinos, nuclear reactors have 
been a crucial tool for understanding some fundamental properties
of neutrinos. In fact, the first detection of neutrinos was from a
reactor neutrino experiment~\footnote{For convenience, we use \lq{reactor
  neutrino}\rq ~instead of \lq{reactor antineutrino}\rq ~throughout this review.}.
To illustrate the important roles of reactors
for neutrino physics, we first briefly review the history of the 
discovery of neutrino.

In his famous letter to ``radioactive ladies and gentlemen", Pauli
postulated~\cite{Pauli:1930pc} in 1930 the existence of a new 
charge-neutral weakly 
interacting particle emitted undetected in nuclear beta decay.
This spin-1/2 particle would not only resolve the outstanding
puzzle of energy non-conservation, but also explain the apparent violation
of angular momentum conservation in nuclear beta decay. Soon after Pauli's
neutrino postulate, Fermi formulated~\cite{Fermi:1934sk,Fermi:1934hr} in 1933 
his celebrated theory    
of nuclear beta decay, taking into account Pauli's neutrino, and successfully 
explained the experimental data. While Fermi's theory provided convincing evidence for
the existence of the neutrino, a direct detection of the neutrino had to wait for
many years. The prospect for directly detecting the neutrino was considered
by Bethe and Peierls~\cite{Bethe:1934qn}, who suggested the 
so-called \lq{inverse beta decay}\rq~(IBD),
$\bar \nu_e + p \to e^+ + n$, as a possible reaction to detect the neutrino.
However, they estimated a tiny IBD cross section ($\sim$10$^{-42}$~cm$^2$),
prompting them to conclude that ``...there is no practically possible 
way of observing the neutrino." Responding to this conclusion, Pauli 
commented that ``I have done something very bad by proposing a particle 
that cannot be detected; it is something no theorist 
should ever do~\cite{Reines:1996ia}."   
 
The advent of nuclear reactors as a steady and intense source of 
electron antineutrinos ($\bar \nu_e$) and the development of large
volume liquid scintillator detectors opened the door for Fred Reines and Clyde
Cowan to perform the pioneering experiments at the Hanford~\cite{Reines:1953pu} 
and Savanah River~\cite{nu_dis,Reines:1956rs}
nuclear reactors to detect neutrinos directly via the IBD reaction 
suggested by
Bethe and Peierls. A crucial feature of the IBD reaction is the time
correlation between the prompt signal from the ionization and
annihilation of $e^+$ and the delayed signal from the $\gamma$ rays
produced in the neutron capture. This distinctive pattern in time
correlation allows a powerful rejection of many experimental 
backgrounds~\cite{Reines:1953kf}.

Upon the definitive observation of neutrinos via the IBD reaction, Reines
and Cowan sent a telegram on June 14, 1956, to Pauli informing him that 
``..we have definitely detected neutrinos from fission fragments by
observing inverse beta decay". Pauli replied that ``Everything comes to
him who knows how to wait"~\cite{Reines:1996ia}. Indeed, it took 26 
years for Pauli's neutrino
to be detected experimentally. It would take another 30 years before
Reines received the Nobel Prize for his pioneering experiment.

In addition to discovering the neutrino via the IBD reaction, Reines,
Cowan, and collaborators also reported several pioneering
measurements using their large liquid scintillator detectors.
They performed the first search for the neutrino magnetic moment
via $\nu - e$ elastic scattering, setting an upper limit
at $\sim$10$^{-7}$~Bohr magnetons initially~\cite{Cowan:1954pq}, which was
later improved to $\sim$10$^{-9}$~Bohr magnetons using a larger
detector~\cite{Cowan:1957pp}. A search for proton stability was also
carried out, resulting in a lifetime of free protons (bound nucleons) 
greater than $10^{21}$ ($10^{22}$)~yr. By inserting a sample of
Nd$_2$O$_3$ enriched in $^{150}$Nd inside the liquid scintillator,
they searched for neutrinoless double beta decay from $^{150}$Nd
and set a lower limit on the half-life at $2.2\times10^{18}$~yr~\cite{Reines:1954pg}. 
It is truly
remarkable that searches for the neutrino magnetic moment, proton
decay, and neutrinoless double beta decay are still among the most
important topics being actively pursued, using techniques
similar to those developed by Reines and Cowan. The
favored reaction to detect reactor electron antineutrinos to date remains IBD, and 
large liquid scintillators are currently utilized or being constructed
for a variety of fundamental experiments.  

As recognized by Pauli when he first put forward his neutrino 
hypothesis, the neutrino must have a tiny mass, comparable or lighter than that 
of the electron~\cite{Pauli:1930pc}. Later, Fermi's theory for beta decay
was found to be in excellent agreement
with experimental data when a massless neutrino was assumed. Indeed,
Fermi was in favor of a massless neutrino as a simple and elegant 
scenario, putting the neutrino in the same class of particles
as the photon and the graviton~\cite{Fermi:1933jpa}. A finite neutrino 
mass could be revealed
from a precise measurement of the endpoint
energy of nuclear beta decay, notably tritium beta decay. While the precision 
of tritium beta decay experiments continued
to improve, yet no definitive evidence for a finite neutrino mass was 
found~\cite{Patrignani:2016xqp}. As one of the most abundant particles in the Universe, the exact value
of the neutrino mass has implications not only on particle physics, but also
on cosmology and astrophysics. The quest for determining the neutrino mass
remains an active and exciting endeavor today.

Inspired by the mixing phenomenon observed in the neutral kaon system,
Pontecorvo suggested the possibility of neutrino-antineutrino mixing
and oscillation~\cite{ponte1,Pontecorvo:1957qd}. 
After the muon neutrino was 
discovered, this idea was
extended to the possible mixing and oscillation between neutrinos 
of different flavors
(i.e., mixing between the electron neutrino and muon neutrino)~\cite{Maki,
ponte2,Gribov:1968kq}.   
Neutrino oscillation is a quantum mechanical phenomenon when neutrinos
are produced in a state that is a superposition of eigenstates of
different mass. As such, this oscillation is possible only when
at least one neutrino mass eigenstate possesses a non-zero mass. 
The pattern of the oscillation, if found, will directly reveal the
amount of mixing (in terms of mixing angle), as well as mass-squared
difference (i.e., $\Delta m^2_{21} \equiv m_2^2 - m_1^2$). Thus, neutrino oscillation provided an exciting new
venue to search for a tiny neutrino mass, beyond the
reach of any foreseeable nuclear beta decay experiments. 

Searches for the phenomenon of neutrino oscillation were pursued
in earnest using a variety of man-made and natural sources of neutrinos.
In the early 1980s, two reactor neutrino experiments reported possible
evidence for neutrino oscillation. The experiment performed by 
Reines and collaborators~\cite{Reines:1980pc} at the Savannah River 
reactor found an
intriguing difference between the detected number of electron antineutrinos
and the sum of electron and other types of antineutrinos using a deuteron
(heavy water) target. The distinctions among different types of neutrino
flavors were made possible through the observation of neutral-current as well
as charged-current disintegration of the deuteron, a method adopted later
by the SNO solar neutrino experiment. The larger number
of neutrinos observed for the neutral-current events than that for the
charged-current ones suggested that some electron neutrinos had oscillated into other types
of neutrinos as they traveled from the reactor to the detector.

The 
other tantalizing evidence~\cite{Cavaignac:1984sp} for neutrino 
oscillation was obtained by detecting
IBD events at two distances, 13.6 and 18.3~meters, from the core of the Bugey
reactor in France. From a comparison of detected IBD events at the two
distances, for which the uncertainties of the flux and energy spectrum 
of the neutrino source largely canceled, a smaller than expected number of detected
IBD events at the larger distance was interpreted
as evidence for oscillation.

Although later reactor experiments~\cite{Zacek:1986cu,Afonin:1988gx,
Declais:1994su,Greenwood:1996pb} performed in the 1980s and 1990s 
did not confirm the earlier results on neutrino oscillation, interest
continued to grow
in finding neutrino oscillation with larger and better detectors
using intense reactor neutrino sources. The first
observation of reactor neutrino oscillation was reported in 2002 by the
KamLAND experiment~\cite{Eguchi:2002dm}. Amusingly, while earlier 
experiments were located at relatively short distances from the reactors
in order to have reasonable event rates, KamLAND was situated at an
average distance of $\sim$180~km from the neutrino sources. At such a large
distance, corresponding to a long oscillation period, the relevant
neutrino mass scale is tiny, of the order of $\Delta m^2 \sim$10$^{-4}$~eV$^2$.
This long distance allows one to probe the large mixing angle (LMA) solution,
one of the few possible explanations to the solar neutrino problem
(see Sec.~\ref{sec:KamLAND} for more details). 
The KamLAND result, together with the analysis~\cite{Fogli:2002pt} of
experiments reporting the observation of solar neutrino oscillation,
allowed an accurate determination of the mixing angle ($\theta_{12}$)
governing these oscillations. The KamLAND result remains the best
measurement of $\Delta m^2_{21}$.

Starting from the late 1980s, evidence for neutrino oscillation was reported by
the large underground detectors including Kamiokande~\cite{Hirata:1988uy,Hatakeyama:1998ea} 
and Super-Kamiokande~\cite{Fukuda:1998mi}, which
detected energetic electron and muon neutrinos ($\sim$GeV) originating 
from the decay of mesons produced in the interaction of 
cosmic rays in Earth's atmosphere. These results suggested the possibility
of observing oscillation for reactor neutrinos at a distance of $\sim$1~km.
Two reactor neutrino experiments, CHOOZ~\cite{Chooz1,Chooz2} and 
Palo Verde~\cite{PaloVerde}, were constructed
specifically to look for such oscillations. However, no evidence for
oscillation was found within the sensitivities of both experiments.
The CHOOZ experiment set an upper limit at 0.12 (90\% C.L.)
for $\sin^2 2\theta_{13}$~\cite{Chooz2}. Together with other oscillation
  experiments, in particular Super-Kamiokande, these results indicated a very
  small value, possibly zero, for the mixing angle
  $\theta_{13}$, which dictates the amplitude of the reactor
  neutrino oscillation at this distance scale. 

As one of the fundamental parameters describing the properties of 
neutrinos, $\theta_{13}$ is also highly relevant for the phenomenon
of CP-violation in the neutrino sector. The importance of the as
yet unknown mixing angle $\theta_{13}$ led to a worldwide effort
to measure it in high-precision experiments. Around 2006,
three reactor neutrino experiments, Daya Bay, Double Chooz, and RENO, were
proposed to probe $\theta_{13}$. All three experiments have already 
collected unprecedentedly large numbers of 
neutrino events. Evidence for non-zero values of $\theta_{13}$,
deduced from the observation of neutrino oscillation at a 1$\sim$2~kilometer
distance, has emerged from all three 
experiments~\cite{An:2012eh,Ahn:2012nd,dc}.
Despite being the smallest among the  
three neutrino mixing angles in the standard three-neutrino
paradigm, $\theta_{13}$ is nevertheless the most precisely determined to date.

Discovery of a non-zero $\theta_{13}$ mixing angle is an important
milestone in neutrino physics. The precise measurement of $\theta_{13}$
not only provides a crucial input for model-building in neutrino
physics, but also inspires new reactor neutrino experiments to explore
other important issues in neutrino physics, such as determining
the neutrino mass hierarchy~\cite{An:2015jdp} and searching for sterile
neutrinos~\cite{Ashenfelter:2015uxt}.
It is remarkable that all ongoing and planned reactor neutrino
experiments adopt essentially the same techniques pioneered by
Reines and Cowan and their coworkers over 60 years ago.

The focus of this review is on the three ongoing reactor neutrino
experiments, Daya Bay, Double Chooz, and RENO. These experiments
share many common features, and we will in some cases discuss one of
these experiments as a specific example. Previous review articles
on reactor neutrino physics are also 
available~\cite{Bemporad:2001qy,Qian:2014xha,Vogel:2015wua,Lachenmaier:2015jja}.
The organization of this review article is as follows. Section~\ref{sec:production_detection}
describes the salient characteristics of the antineutrinos produced
in nuclear reactors as well as the experimental techniques
for detecting them. The subject of reactor neutrino oscillation is
discussed in Sec.~\ref{sec:nu_osc}. The discussion regarding the
reactor antineutrino anomaly and the search for a light sterile neutrino
is presented in Sec.~\ref{sec:sterile}. Some additional physics topics
accessible in reactor neutrino experiments are described in
Sec.~\ref{sec:additional}, followed by conclusions in Sec.~\ref{sec:conclusion}.


\section{Production and Detection of Reactor Neutrinos}\label{sec:production_detection}

To date, five main natural and man-made neutrino sources have played
crucial roles in advancing our knowledge of neutrino properties.
They are: i) reactor electron antineutrinos ($\bar{\nu}_e$) produced through fission
processes; ii) accelerator neutrinos
($\nu_\mu$, $\nu_e$, $\bar{\nu}_\mu$, and $\bar{\nu}_e$)
resulting from decays of mesons created by proton beams bombarding a production
target; iii) solar neutrinos ($\nu_e$) generated via
fusion processes in the sun; iv) supernova neutrinos (all flavors) produced during
supernova explosions;
and v) atmospheric neutrinos ($\nu_\mu$, $\nu_e$, $\bar{\nu}_\mu$, and $\bar{\nu}_e$)
created through decays of mesons produced by the interaction of high-energy
cosmic rays with Earth's atmosphere.
Beside these, geoneutrinos produced from radionuclide inside the Earth
  and extra-galactic ultra-high energy neutrinos have also been detected. 

Compared to atmospheric and accelerator neutrinos, reactor
neutrinos have the advantage of being a source of pure flavor
($\bar{\nu}_e$ with energy up to $\sim$10~MeV)\footnote{
  At very low energy ($\sim$0.1~MeV), a small component of $\nu_{e}$ is
  generated from neutron activation of shielding materials~\cite{nue_reactor}.}.
In addition, the primary reactor neutrino detection channel, IBD,
is well understood theoretically and allows an accurate measurement of the
neutrino energy, unlike high-energy neutrino--nucleus interactions.
Compared to rates for solar and supernova neutrinos, the event detection rate of reactor
neutrinos can be much larger, as detectors can be placed at distances
close to the source. In this Section, we review the production and
detection of reactor neutrinos. 

\subsection{Production of Reactor Neutrinos}~\label{sec:production}

Energy is generated in a reactor core through neutron-induced nuclear
fission. This process is maintained by neutrons emitted in
fission. For example, the average number of emitted neutrons is about 2.44 per
$^{235}$U fission~\cite{IAEA-CRP-STD},
among which, on average, only one neutron will induce a new fission reaction
for a controlled reactor operation.

While the fission of $^{235}$U is the
dominating process in a research reactor using highly enriched uranium (HEU)
fuel ($>$20$\%$ $^{235}$U concentration), more fissile isotopes are involved in
a commercial power reactor using low enriched uranium (LEU) fuel
(3--5\% $^{235}$U concentration). Inside the core of a commercial power reactor,
a portion of the neutrons are captured by $^{238}$U because of its much higher
concentration, producing new fissile isotopes: $^{239}$Pu and $^{241}$Pu.
Fissions of $^{235}$U, $^{239}$Pu, and $^{241}$Pu are induced
by thermal neutrons ($\sim$0.025-eV kinetic energy).
In contrast, fission of
$^{238}$U can be induced only by fast neutrons ($\sim$1-MeV kinetic energy).
The average number
of emitted neutrons are 2.88~\cite{IAEA-CRP-STD}, 2.95~\cite{IAEA-CRP-STD}, and
2.82~\cite{ENDF} per $^{239}$Pu, $^{241}$Pu, and $^{238}$U fission, respectively.

The reactor neutrinos are mainly produced through the
beta-decays of the neutron-rich fission daughters of these four isotopes, in
which a bound neutron is converted into a proton while producing an electron and an
electron antineutrino. Besides the
fission processes, another important source of $\bar{\nu}_e$ originates from
neutron capture on $^{238}$U: $^{238}$U$(n,\gamma)^{239}$U. The beta decay of
$^{239}$U (Q-value of 1.26~MeV and half-life of 23.5~mins) and the
subsequent beta decay of $^{239}$Np (Q-value of 0.72~MeV and half-life of
2.3~days) produce a sizable amount of $\bar{\nu}_e$ at low energies.  
An average of $\sim$6~$\bar{\nu}_e$ were produced per fission, leading to
$\sim$2$\times10^{20}$~$\bar{\nu}_e$ emitted every second isotropically for
each GW of thermal power.


\begin{figure}[htp]
\begin{centering}
\includegraphics[width=0.45\textwidth]{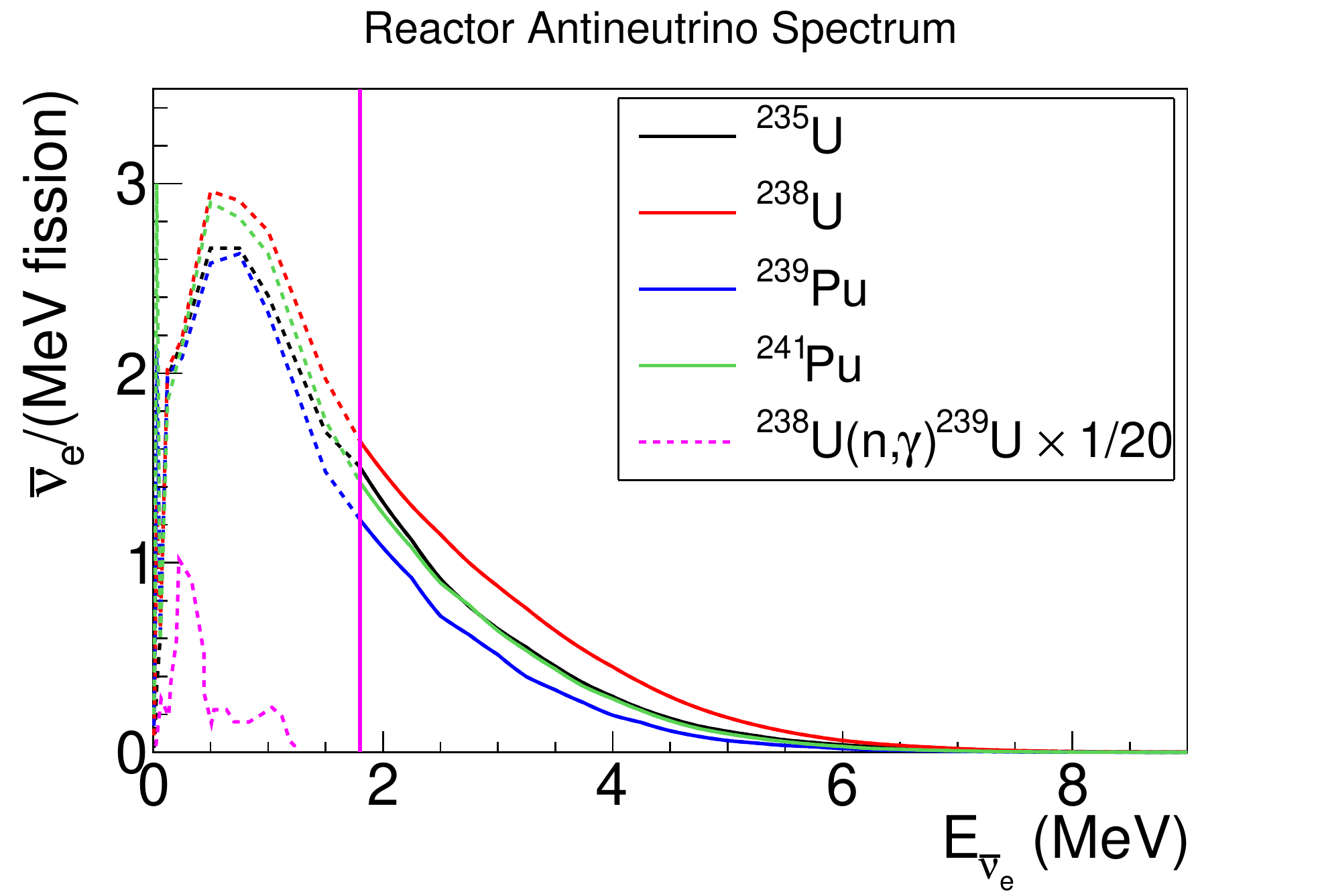}
\par\end{centering}
\caption{\label{fig:spectrum}
  The $\bar{\nu}_e$ energy spectra for $^{235}$U, $^{238}$U, $^{239}$Pu, and $^{241}$Pu
  fissions. Above the inverse beta decay (IBD) threshold (marked by the vertical
  line), spectra from
  Ref.~\cite{Huber:2011wv,Mueller:2011nm}
  are shown. Below the IBD threshold, spectra are plotted based on Table II 
  of Ref.~\cite{Vogel:1989iv}. Fine structures at the end points of various
  decay branches cannot be seen, given the coarse binning. 
  In addition, we show the antineutrino spectrum produced by neutron capture
  on $^{238}$U (taken from Ref.~\cite{Deniz:2009mu}), which is normalized properly
  relative to the $^{238}$U fission and scaled down by a factor of 20 for the
  display.}
\end{figure}

The expected $\bar{\nu}_e$ energy spectra are shown in Fig.~\ref{fig:spectrum}.
The magnitude of $\bar{\nu}_e$ spectra for $^{238}$U ($^{241}$Pu)
 are larger than that of $^{235}$U ($^{239}$Pu), because more
 neutron-rich fissile isotopes lead to more beta-unstable neutron-rich
 fission daughters. In addition, the $\bar{\nu}_e$
energy spectrum is considerably harder for the fast-neutron-induced $^{238}$U
fission chain than the other three thermal-neutron induced fission chains.

For commercial power reactors burning LEU, typical average values of fission
fractions during operation are around 58\%, 29\%, 8\%, and 5\% for $^{235}$U,
$^{239}$Pu, $^{238}$U, and $^{241}$Pu,
respectively. Roughly 30\% of the antineutrinos (two out of the
average six antineutrinos produced per fission) have energies above
1.8~MeV, which is the energy threshold of the IBD process.
In particular, the
low-energy $\bar{\nu}_e$ produced by neutron capture on $^{238}$U is irrelevant
for detection through IBD.  In the following, we describe two principal approaches
for calculating the antineutrino flux and energy spectrum. More details can
be found in a recent review~\cite{Hayes:2016qnu}.

In the first approach, the flux and spectrum can be predicted by the cumulative
fission yields $Y_n(t)$ at time $t$ for fission product of nucleus $n$ having a mass number
$A$ and an atomic number $Z$,  branching ratios $b_{n,i}$ of
$\beta$-decay branch $i$ with endpoints $E_0^{n,i}$, and the
energy spectrum of each of $\beta$ decays
$P(E_{\bar{\nu}},E_0^{n,i})$:
\begin{equation}
  \frac{dN}{dE_{\bar{\nu}}} = \sum_n Y_n(t) \cdot \left(
  \sum_i b_{n,i} \cdot P(E_{\bar{\nu}},E_0^{n,i})\right).
\end{equation}
This method was recently used in Ref.~\cite{Mueller:2011nm} and included about
10k beta decay branches, following the early work in
Refs.~\cite{Davis:1979gg,Vogel:1980bk,Klapdor:1982zz,Klapdor:1982sf,Kopeikin_1980}.
Despite being straightforward, several challenges in this method
lead to large uncertainties in predicting the flux and spectrum. First,
the fission yields, $\beta$-decay branching ratios, and the endpoint energies
are sometimes not well known,
especially for short-lived fragments having large beta-decay Q values. Second,
the precise calculation of the individual spectrum shape
$P(E_{\bar{\nu}},E_0^{n,i})$ requires a good model of the Coulomb distortions
(including radiative corrections, the nuclear finite-size effects, and weak
magnetism) in the case of an allowed decay type having zero orbital angular momentum
transfer. Finally, many of the decay channels are of the forbidden types having
non-zero orbital angular momentum transfer. For example, about 25\% of decays are
the first forbidden type involving parity change, in which the individual
spectrum shape $P(E_{\bar{\nu}},E_0^{n,i})$ is poorly known.
Generally, a 10--20\% relative uncertainty on the antineutrino spectra is
obtained using this method. 

Another method uses experimentally measured electron spectra associated
with the fission of the four isotopes to deduce the antineutrino spectra.
The electron energy spectra for the thermal neutron fission
of $^{235}$U, $^{239}$Pu, and $^{241}$Pu have been measured at
Institut Laue--Langevin (ILL)~\cite{VonFeilitzsch:1982jw,Schreckenbach:1985ep,Hahn:1989zr}.
The electron spectrum associated with the fast neutron fission of $^{238}$U has
been measured in M\"{u}nich~\cite{Haag:2013raa}. Since the electron and the
$\bar{\nu}_e$ share the total energy of each $\beta$-decay branch, ignoring the
negligible nuclear recoil energy, the
$\bar{\nu}_e$ spectrum can be deduced from the measured electron spectrum.

The procedure involved fitting the electron spectrum to a
set of $\sim$30 virtual branches having equally spaced endpoint energies, assuming
all decays are of the allowed type. For each virtual branch, the charge of parent
nucleus $Z$ is taken from a fit to the average $Z$ of real branches
as a function of the endpoint energy. The conversion to the $\bar{\nu}_e$ spectrum
is then performed in each of these virtual branches using their fitted branching
ratios. This conversion method was used in
Refs.~\cite{Mueller:2011nm,VonFeilitzsch:1982jw,Schreckenbach:1985ep,Hahn:1989zr,Vogel:2007du}.

In addition to the experimental uncertainties associated with the electron spectrum,
corrections to the individual $\beta$-decay branch resulting from radiative correction,
weak magnetism, and finite nuclear size also introduce uncertainties.
With these contributions, the model uncertainty in the flux is
estimated to be $\sim$2\%~\cite{Huber:2011wv,Mueller:2011nm}. 
However, the uncertainties resulting from spectrum shape and magnitude of the numerous
first forbidden $\beta$ decays can be substantial~\cite{anom2}.
When the first forbidden decays are included, the estimated uncertainty increases
to $\sim$5\%~\cite{anom2}. Besides these model uncertainties, the total experimental
uncertainty  of the $\bar{\nu}_e$ spectrum further includes the
contribution from the thermal power of the reactor, its time-dependent fuel
composition (i.e., fission fractions), and fission energies associated with
$^{235}$U, $^{238}$U, $^{239}$Pu, and $^{241}$Pu.

\begin{figure}[htp]
\begin{centering}
\includegraphics[width=0.45\textwidth]{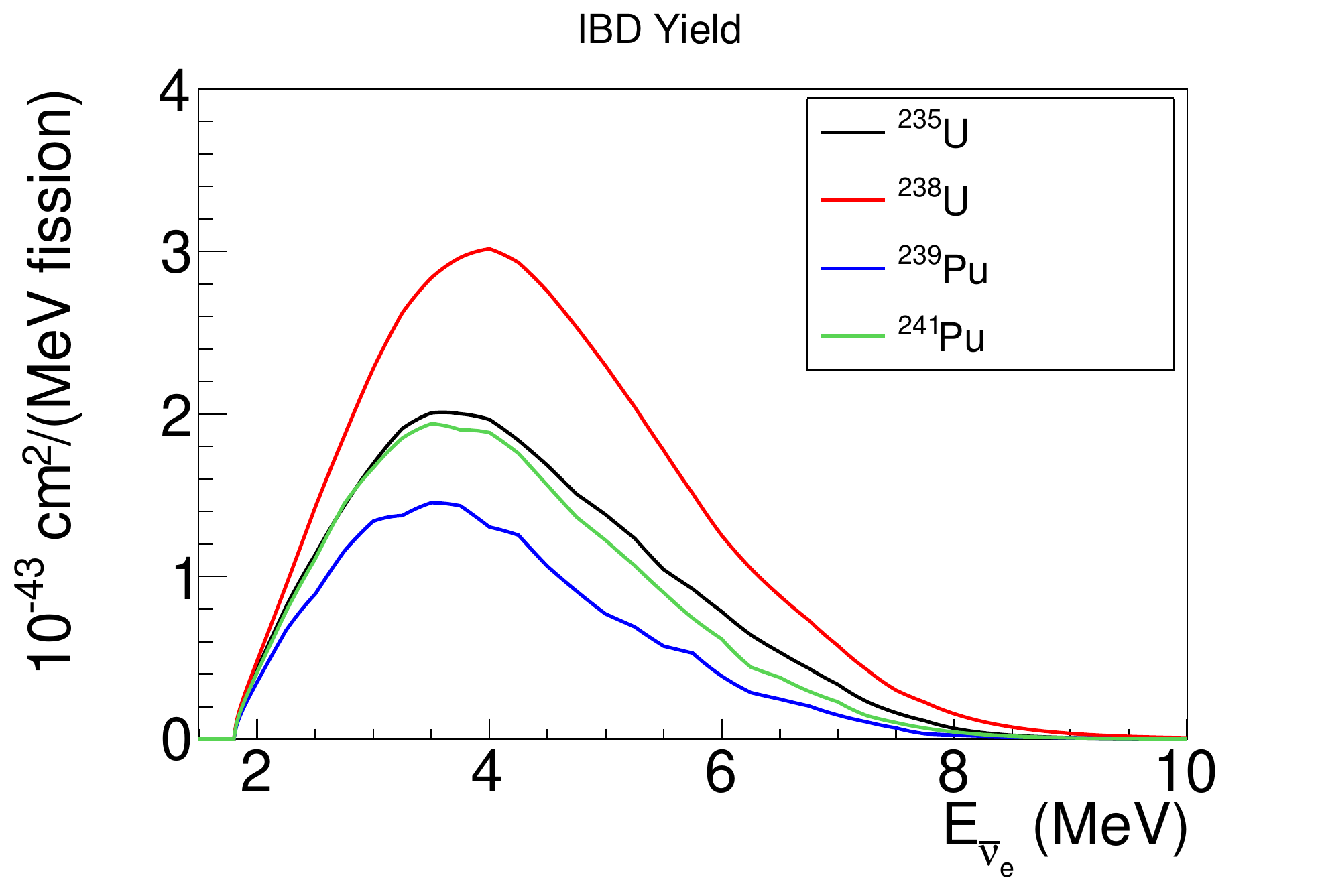}
\par\end{centering}
\caption{\label{fig:IBD}
  Inverse beta decay yields from the convolution of the IBD cross section
  and the antineutrino spectra for $^{235}$U, $^{238}$U, $^{239}$Pu, and $^{241}$Pu. }
\end{figure}

\subsection{Detection of Reactor Neutrinos}

\begin{table*}[!htp]
  \caption{Summary of various $\bar{\nu}_e$ detection methods. CC (NC) stands for
    the charged-current (neutral-current) interaction. The cross section is integrated
    over the entire reactor neutrino energy spectrum. $N$ stands for the number of neutrons
    in the target nucleus. For these estimations, fission fractions are assumed to be
    58\%, 29\%, 8\%, and 5\% for $^{235}$U, $^{239}$Pu, $^{238}$U, and $^{241}$Pu,
    respectively. }  
\medskip
\renewcommand{\arraystretch}{1.1} \centering 
\begin{tabular}{|c|c|c|c|}
  \hline
  Channel & Interaction & Cross Section & Threshold \\
  & Type      & ($10^{-44}$ cm$^2$/fission) & (MeV) \\\hline
  $\bar{\nu}_e + p \rightarrow e^+ + n$ & CC & $\sim$63 & 1.8 \\\hline
  $\bar{\nu}_e + d \rightarrow n + n + e^+$ & CC & $\sim$1.1 & 4.0\\\hline
  $\bar{\nu}_e + d \rightarrow n + p + \bar{\nu}_e$ & NC & $\sim$3.1 & 2.2\\\hline
  $\bar{\nu}_e + e^- \rightarrow \bar{\nu}_e + e^-$ & CC/NC & $\sim$0.4 & 0\\\hline
  $\bar{\nu}_e + A \rightarrow \bar{\nu}_e + A $   & NC & $\sim$9.2$\times N^2$ & 0\\\hline
\end{tabular}\label{tab:anue_reaction}
\end{table*}

In addition to the aforementioned IBD process, several methods can
potentially be used to detect reactor neutrinos. The first method is  the
charged-current (CC) ($\bar{\nu}_e + d \rightarrow n + n + e^+$)
and neutral-current (NC) deuteron break-up
($\bar{\nu}_e + d \rightarrow n + p + \bar{\nu}_e$)
using heavy water as a target. These processes were used to compare the
NC and CC cross sections~\cite{Reines:1980pc,Riley:1998ca}. Similar
processes involving $\nu_e$ were also used in the SNO experiment in
detecting the flavor transformation of solar neutrinos~\cite{Ahmad:2001an}.

The second method is the antineutrino-electron
elastic scattering, $\bar{\nu}_e + e^- \rightarrow \bar{\nu}_e + e^-$, which
combines the amplitudes of the charged-current (exchange of $W$ boson) and the neutral-current
(exchange of $Z$ boson). The signature of this process would be a
single electron in the final state. This process has been used to
measure the weak mixing angle $\theta_{W}$ and to constrain anomalous
neutrino electromagnetic properties~\cite{Deniz:2009mu,Reines:1976pv,Vidyakin:1992nf,Derbin:1993wy,Daraktchieva:2005kn}.
Neutrino-electron scattering is also one of the primary
approaches to detect solar neutrinos~\cite{Ahmad:2001an,Hosaka:2005um,Bellini:2008mr}.

The third method is the coherent antineutrino-nucleus interaction, in
which the signature is a tiny energy deposition by
the recoil nuclei. Although coherent elastic neutrino-nucleus scattering
was observed recently for the first time~\cite{Akimov:2017ade} using
neutrinos produced in the decay of stopped pions, the observation for this
process for less-energetic reactor neutrinos has not been
achieved. Table~\ref{tab:anue_reaction} summarizes some essential information
for these detection channels.

\begin{figure}[htp]
\begin{centering}
\includegraphics[width=0.42\textwidth]{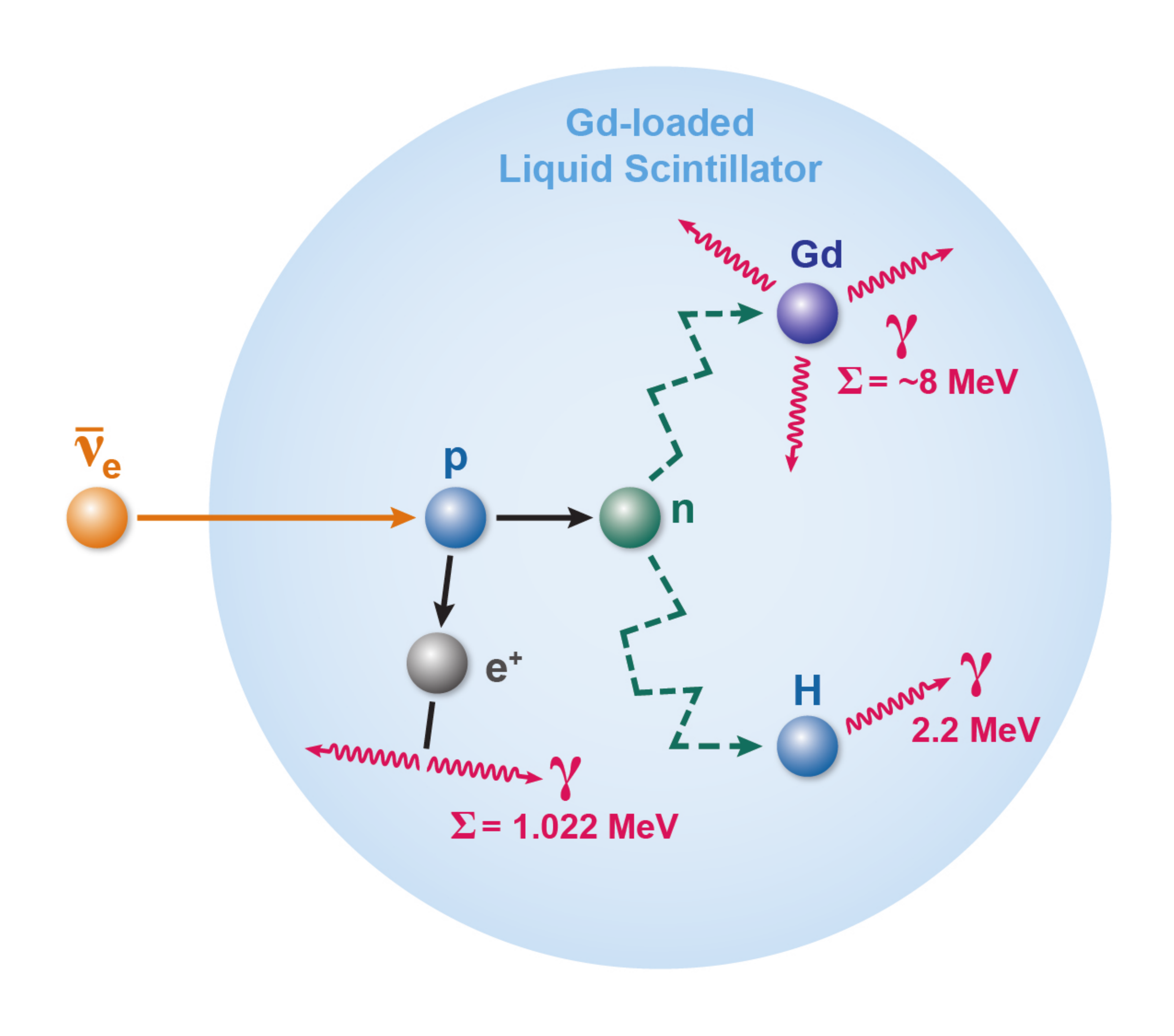}
\par\end{centering}
\caption{\label{fig:IBD_detect} Principle of the IBD detection in a Gd-loaded liquid scintillator. The electron
  antineutrino interacts with a free proton. The ionization and annihilation
  of the final-state positron form the prompt signal. The capture of the
  recoil neutron on Gd (or H) gives the delayed signal.
}
\end{figure}

\begin{table*}[!htp]
\caption{Various nuclei used in experiments to capture recoil neutrons from the
  IBD reaction. The detection channels and their cross sections~\cite{ENDF}
  for thermal neutron capture are listed. $^{157}$Gd has the highest
  thermal-neutron capture cross section of any stable nuclide.}  
\medskip
\renewcommand{\arraystretch}{1.1} \centering 
\begin{tabular}{|c|c|c|}
\hline 
Target nucleus &  process & cross section (barn) \\
 & & for thermal neutron \\\hline
H      & $n + p \rightarrow d + \gamma~(2.2~ {\rm MeV})$ & $\sim$0.33  \\ \hline
$^{3}$He & $n + ^3$He$ \rightarrow p + ^3$H$ + 0.764~ {\rm MeV}$ & $\sim$5300 \\\hline
$^{6}$Li & $n + ^6$Li$ \rightarrow \alpha + ^3$H$ + 4.6~ {\rm MeV}$& $\sim$950  \\\hline
$^{10}$B & $n + ^{10}$B$ \rightarrow \alpha + ^7$Li$ + 6.2~ {\rm MeV}$& $\sim$3,860  \\\hline
$^{108}$Cd  & $n + ^{108}$Cd$ \rightarrow ^{109m}$Cd$ \rightarrow ^{109}$Cd$ + \gamma~(0.059 ~{\rm MeV})$ & $\sim$1000~\footnotemark \\\hline
Gd & $n+^{155}$Gd$ \rightarrow ^{156}$Gd$ + \gamma s ~(8.5~ {\rm MeV})$ & $\sim$61,000 \\
   & $n+^{157}$Gd$ \rightarrow ^{158}$Gd$ + \gamma s ~(7.9~ {\rm MeV})$ & $\sim$256,000 \\\hline
\end{tabular}\label{tab:IBD_neutron_capture}
  \end{table*}

So far, the primary method to detect the reactor $\bar{\nu}_e$ is the IBD
reaction: $\bar{\nu}_e + p \rightarrow e^+ + n$.
The energy threshold of this process is about 1.8~MeV, and the cross section
is accurately known~\cite{Vogel:1999zy,Strumia:2003zx}. At the zeroth order in
1/$M$, with $M$ being the nucleon mass, the cross section can be written as:
\begin{equation}
  \sigma^{(0)} = \frac{G_F^2 \cos^2\theta_C}{\pi}\left( 1 + \Delta^R_{inner}\right) \cdot
  \left(f^2 + 3g^2 \right) \cdot E_e^{(0)} \cdot p_e^{(0)}, 
\end{equation}
with $G_F$ being the Fermi coupling constant and $\theta_C$ being the
Cabibbo angle. The vector and axial vector coupling constants are $f=1$ and $g=1.27$, respectively.
$\Delta^R_{inner}$ represents the energy independent inner radiative corrections. $E_e$ and $p_e$
are the energy and momentum of the final-state positron having $E^{(0)}=E_{\nu} - (M_n - M_p)$ after ignoring the recoil neutron kinetic energy. The IBD cross section can be linked to the neutron
lifetime $\tau_n=880.2\pm1.0$~s~\cite{Patrignani:2016xqp} as:
\begin{eqnarray}
  \sigma^{(0)} &=&  \frac{2\pi^2/m_e^2}{f^{R}\tau_n} E^{(0)}_e \times p^{(0)}_e \nonumber \\
  &\approx& 9.52\times \left( \frac{E^{(0)}_e\cdot p^{(0)}_e}{{\rm MeV}^2} \right) \times 10^{-44} {\rm cm}^2,
\end{eqnarray}
with $m_e$ being the mass of the electron and $f^R=1.7152$, representing the neutron decay phase
space factor that includes the Coulomb, weak magnetism, recoil, and outer radiative
corrections. The above formula represents the zeroth order in 1/$M$, and we should note that
the  corrections of the first order in $1/M$ are still important at reactor energies.

The various forms of extension to all orders in $1/M$, as well as the convenient
numerical form of radiative corrections of order $\alpha/\pi$ can be found in
Refs.~\cite{Vogel:1999zy,Strumia:2003zx}. Figure~\ref{fig:IBD} shows the IBD yield
obtained from the convolution of the IBD cross section and the antineutrino energy spectra.
While peak positions for the thermal neutron fission ($^{235}$U, $^{239}$Pu, and
$^{241}$Pu) occur at an energy around 3.5~MeV, the peak position for fast-neutron fission
($^{238}$U) is at a slightly higher energy, around 4~MeV. The IBD yield is also
larger for the latter.

\begin{figure*}[htp]
\begin{centering}
  \includegraphics[width=0.45\textwidth]{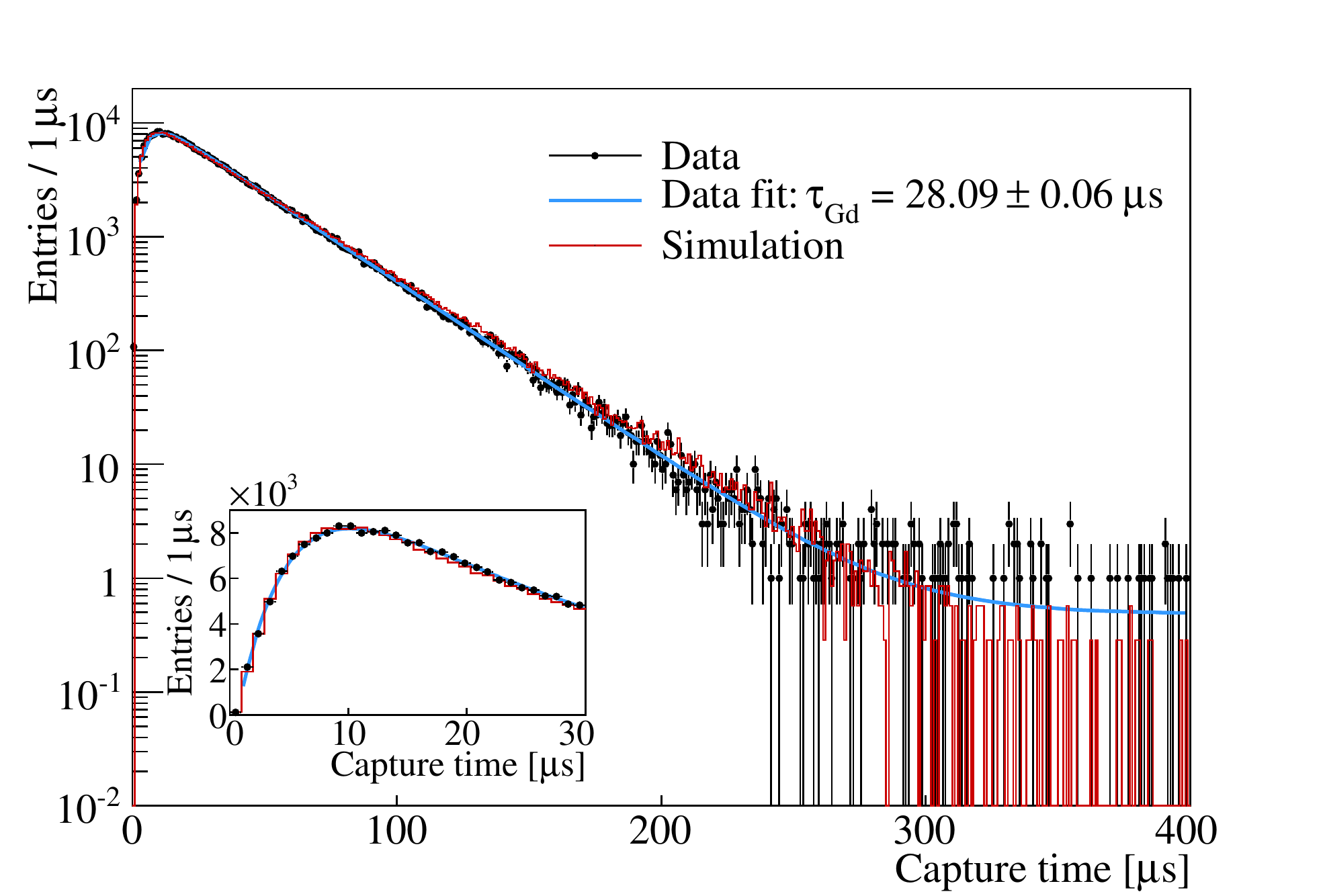}
  \includegraphics[width=0.45\textwidth]{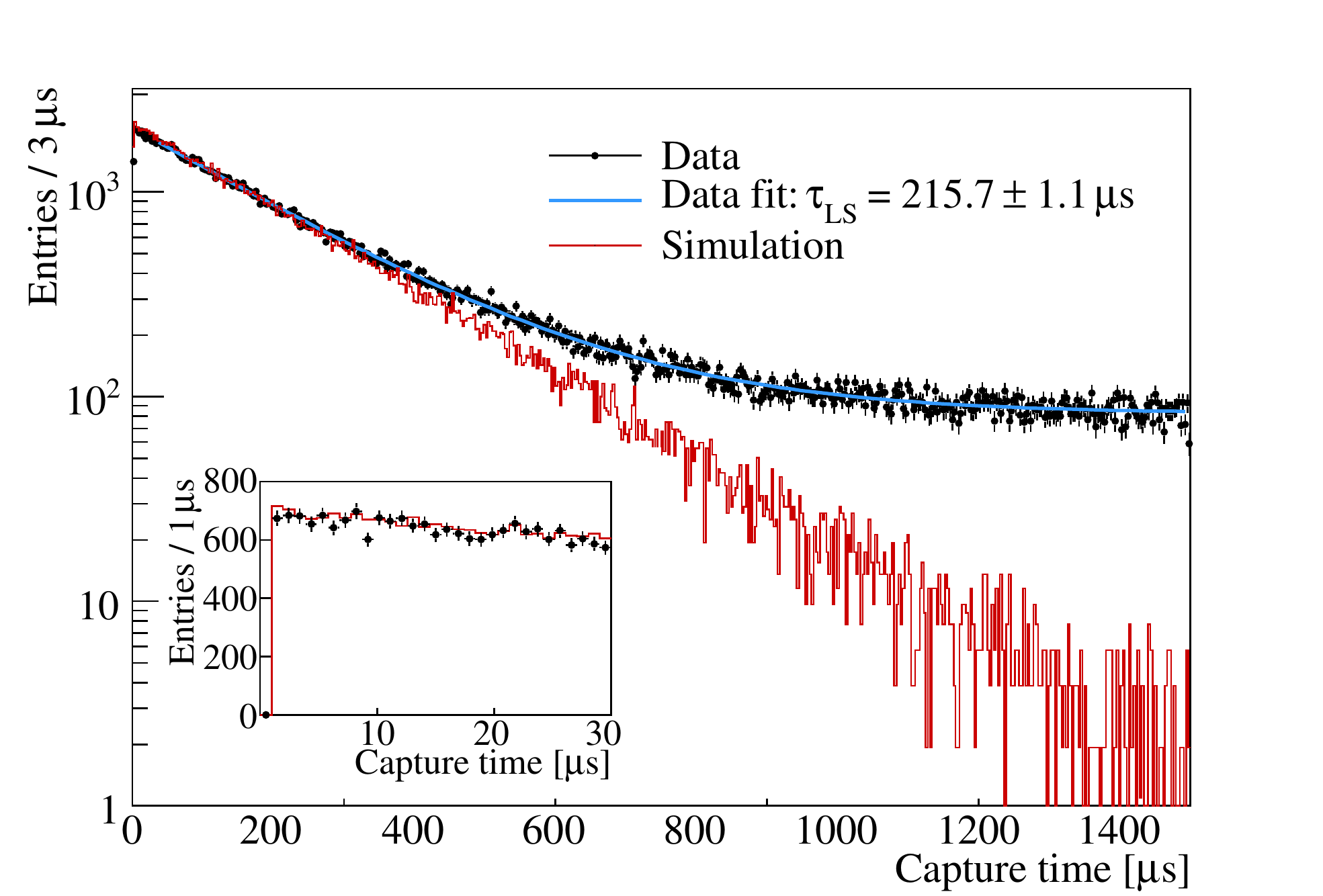}
  \Put(-335,-20){\fontfamily{phv}\selectfont \textbf{A)}}
\Put(-120,-20){\fontfamily{phv}\selectfont \textbf{B)}}
\par\end{centering}
\caption{\label{fig:IBD_capture_time} The time difference between
  prompt and delayed signals for a neutron captured on Gd (A)
  and hydrogen (B), taken from Ref.~\cite{An:2016bvr}.
  The data histograms contain backgrounds leading to non-exponential
  distributions visible at large capture times.}
\end{figure*}

\footnotetext{The cross section corresponds to the metastable resonance
  state around 0.3-keV neutron kinematic energy.}

As shown in Fig.~\ref{fig:IBD_detect}, an IBD event is indicated by a pair of coincident
signals consisting of i) a prompt signal induced by positron ionization
and annihilation inside the detector; and ii) a delayed signal produced by the
neutron captured on a proton or a nucleus (such as Gd). Because of
time correlation, IBD can be clearly distinguished from radioactive backgrounds,
which usually contain no delayed signal.

The energy of the prompt signal
is related to the neutrino energy via $E_{\bar{\nu}} \approx E_{prompt}
+ 0.78~{\rm MeV} + T_{n}$, with $T_n$ being the kinetic energy of the
recoil neutron. Since $T_n$, of the order of tens of keV, is much smaller than that of $\bar{\nu}_e$,
the neutrino energy can be accurately determined by the prompt energy,
which is a very attractive feature for measuring neutrino oscillation.

Table~\ref{tab:IBD_neutron_capture} summarizes various nuclei used in
past experiments to capture recoil neutrons from the IBD reaction. For example,
for a neutron captured on a proton, the delayed signal comes from a
single 2.2-MeV $\gamma$ ray. In comparison, for a neutron captured on Gd, the delayed
signal consists of a few $\gamma$ rays having the total energy of $\sim$8~MeV.
For a pure liquid scintillator, the average time between the prompt and delayed
signals is $\sim$210~$\mu$s. This is reduced to $\sim$30~$\mu$s for a 0.1\%
  Gd-doped liquid scintillator because of the additional contribution of neutron capture on Gd,
  which has a much higher cross section than
  that of hydrogen. The slow rise in the initial nGd capture rate, shown
  in the inset of Fig.~\ref{fig:IBD_capture_time}A, reflects the time it
  takes to thermalize neutrons from the IBD reaction. The nGd capture cross
  section is much larger for thermal neutrons than higher-energy neutrons.
  In contrast, the nH capture probability is essentially independent of
  neutron's kinetic energy. Hence, no such initial slow rise in the nH capture
  rate is observed (inset of Fig.~\ref{fig:IBD_capture_time}B).


Besides the advantages of good background rejection and
excellent reconstruction of the neutrino energy, the IBD process allows
organic (liquid) scintillators and water to be used as detector media.
These materials can be easily prepared in large volumes at low cost, which is
ideal for experiments studying neutrino properties. In addition, these
features also allow IBD to be used for non-intrusive
surveillance of nuclear reactors by providing an independent and accurate
measurement of reactor power away from the reactor core.
In addition, a precision measurement of the rate and energy
spectrum may provide a measurement of isotopic composition in the reactor core,
providing a safeguard application (i.e., to detect diversion of civilian nuclear
reactors into weapon's programs). For more details, see
Refs.~\cite{Bernstein:2001cz,Huber:2004xh,Bowden:2008gu,Christensen:2014pva},
among others.

\subsection{Detector Technology in Reactor Neutrino Experiments}\label{sec:detector}

In this section, we briefly review the detector technology used in
  reactor neutrino experiments. A recent review containing additional
  information can be found in Ref.~\cite{Cao:2017drk}.

The scintillator technology is widely used in reactor neutrino experiments.
  Given its advantage in mass production, uniformity, doping capability, and
  relatively low cost, liquid scintillator (LS) is often selected as
  the medium for large-scale reactor neutrino experiments. 
  For example, the Daya Bay, Double
  Chooz, and RENO experiments all utilized Gd-doped LS as the medium to detect
  IBD events. As discussed earlier, the coincidence between the prompt signal
  and the $\sim$8~MeV nGd-capture delayed signal provides a powerful means
  for identifying IBD events and rejecting accidental backgrounds.  
  Another example is the $^6$Li-doped LS, used in very-short-baseline
  experiments, such as Bugey-3 and PROSPECT experiments.
  The alpha and triton produced in the n$^6$Li capture (see Table~\ref{tab:IBD_neutron_capture})
  generate relatively slow scintillation light, allowing an
  effective reduction of the fast signals from $\gamma$-ray backgrounds
  via pulse-shape discrimination (PSD).

In addition to the time correlation, the spatial correlation between
  the prompt and delayed signals for IBD events can also be utilized
  for accidental background rejection. A good spational resolution can
  be obtained using a segmented detector configuration. The capability
  to reject background with finely segmented detector
  is particularly important for detectors without much overburden
  (e.g. Palo Verde) and/or situated close to the reactor
  core (e.g. very-short-baseline experiments described in
  Sec.~\ref{sec:sterile_exp}). As a result of the 
  inactive materials separating the segments, 
  its energy resolution is typically worse than that of a
  homogeneous detector with a similar scintillation light yield
  and photo-cathode coverage.

Spherical, cylindrical, and rectangular shape are typical choices of detector
  geometry. The spherical geometry has the largest volume-to-surface ratio.
  Since the light detectors are typically placed on the inner surface,
  this choice is the most cost-effective for large detectors (such as KamLAND
  and JUNO). Having the maximal symmetry, the spherical geometry also has the
  advantage in energy reconstruction.

Compared to a spherical-geometry
  detector, a cylindrical-geometry detector is much easier to construct. This
  is particularly important for the recent $\theta_{13}$ reactor experiments:
  Daya Bay, Double Chooz, and RENO, which utilized multiple functional-identical
  detectors at the same and/or different sites to limit the detector-related systematics. Besides the choice of 
  the cylindrical geometry, the recent reactor $\theta_{13}$ experiments also adopt a
  3-zone detector design with the inner, middle, and outer layers being Gd-loaded LS,
  pure LS, and mineral oil, respectively. The inner Gd-loaded LS region is the main
  target region, where IBD events with neutron captured on Gd are identified.
  The middle
  LS region is commonly referred to as the gamma catcher, which measures
  $\gamma$ rays escaping from the target region.
  The choice of two layers instead of one significantly reduced the uncertainty on the
  fiducial volume. The outer
  region serves as a buffer to suppress radioactive backgrounds from PMTs and the stainless-steel
  container. In comparison, the KamLAND detector contains two layers: the target LS region and
  the mineral oil layer. The rectangular detector shape is a typical choice for 
  segmented detectors in very-short-baseline reactor experiments. 


While the overburden is crucial for reducing cosmogenic backgrounds, additional
  passive and active shields are needed to further suppress
  radioactive backgrounds from environment.
  For example, the KamLAND, Daya Bay, RENO detectors are installed
  inside water pools, which also function as active Cerenkov detectors.
  The shieldings for very-short-baseline reactor experiments
  are typically more complicated in order to significantly reduce the
  surface neutron flux
  from cosmic rays and reactors. For example, PROSPECT experiment installed multiple layers of
  shielding including water, polyethylene, borated-polyethylene, and lead.


  Despite being the best known neutrino source with the longest history,
  there is still much to learn about the production and detection of reactor neutrinos,
  which can be crucial for future experiments. In Sec.~\ref{sec:sterile}, we will
  discuss measurements of the reactor neutrino flux and discrepancies with
  theoretical predictions, and how recent and future measurements of the
  reactor neutrino energy spectrum and the time evolution of the neutrino flux can
  shed light on these discrepancies. In Sec.~\ref{sec:additional}, we will describe how
  additional reactor neutrino detection methods beyond IBD can enable
  searches for new physics beyond the standard model.

\section{Neutrino Oscillation Using Nuclear Reactors}\label{sec:nu_osc}


We discuss in this section the recent progress of reactor experiments
in advancing our knowledge of neutrino oscillation. Following an
overview of the theoretical framework for neutrino oscillation,
a highlight of the KamLAND experiment, which was the first experiment 
to observe reactor neutrino oscillation, is presented. The recent global
effort to search for a non-zero neutrino mixing angle $\theta_{13}$, carried out
by three large reactor neutrino experiments, is then
described in some detail. We conclude this section with a discussion
of the prospects for future reactor experiments to explore other aspects
of neutrino oscillation.

\subsection{Theoretical Framework for Neutrino Oscillations}\label{sec:nu_mixing}

Neutrino oscillation is a quantum mechanical phenomenon analogous to
  $K^\circ-\bar{K}^\circ$ oscillation in the hadron sector. This phenomenon is
  only possible when neutrino masses are non-degenerate and when the flavor
  and mass eigenstates are not identical, leading to the flavor-mixing for each
  neutrino mass eigenstate. A recent review on the neutrino oscillation can
  be found in Ref.~\cite{Giganti:2017fhf}.

The standard model of particle physics posits three active neutrino
flavors, $\nu_e,~\nu_\mu,~\nu_\tau$ that participate in the weak interaction.
These active neutrinos are all left-handed in chirality and nearly all
negative in helicity~\cite{PhysRev.109.1015}, where their spin direction
is antiparallel to their momentum direction~\footnote{In the massless or
  high-energy limit, the chirality is equivalent to the helicity.}. The 
number of (light) active neutrinos,  determined
from the measurement of the invisible width of the Z-boson at LEP
to be $N_{\nu}^{LEP}=2.984\pm0.008$~\cite{ALEPH:2005ab}, is
consistent with recent measurement of the effective number of (nearly) massless
neutrino flavors $N_{\nu}^{CMB}=3.13\pm0.31$~\cite{Ade:2015xua} from the power
spectrum of the cosmic microwave background (CMB). For a long time, the masses
of neutrinos were believed to be zero, as no right-handed neutrino 
has ever been detected in experiments. However, in the past two decades,
results from several neutrino experiments can be described as neutrino
oscillation involving non-zero neutrino mass and 
mixing among the three neutrino flavors. The neutrino mixing is
analogous to the quark mixing via the
Cabibbo--Kobayashi--Maskawa (CKM) matrix~\cite{PhysRevLett.10.531,Kobayashi:1973fv}.

Although a definitive description of massive neutrinos beyond the standard
model has not yet been elucidated, the existing data firmly establishes that the three neutrino
flavors are superpositions of at least three light-mass states $\nu_1$, $\nu_2$,
$\nu_3$ having different masses, $m_1,~m_2,~m_3$:
\begin{widetext}
\begin{equation}\label{eq:pmns_matrix}
\left( \begin{array}{c} \nu_{e} \\ \nu_{\mu} \\ \nu_{\tau} \end{array} \right) = 
\left ( \begin{array}{ccc} U_{e1} & U_{e2} & U_{e3} \\
	U_{\mu1} & U_{\mu2} & U_{\mu3} \\
	U_{\tau1} & U_{\tau2} & U_{\tau3} 
\end{array} \right) \cdot \left( \begin{array}{c} \nu_{1} \\ 
\nu_{2} \\ \nu_{3} \end{array} \right).
\end{equation}
\end{widetext}
The unitary $3\times 3$ mixing matrix, $U$,  
called the Pontecorvo--Maki--Nakagawa--Sakata (PMNS) matrix~\cite{ponte1,Maki,ponte2}, 
is parameterized by three Euler angles, $\theta_{12}$, $\theta_{13}$, and $\theta_{23}$, 
plus one or three phases (depending on whether neutrinos are Dirac or Majorana types),
potentially leading to CP violation. The mixing matrix $U$ is conventionally expressed
as the following product of matrices:
\begin{eqnarray}~\label{eq:3x3_rot}
  U &=&R_{23}(c_{23},s_{23},0) \cdot R_{13}(c_{13},s_{13},\delta_{CP}) \cdot R_{12}(c_{12},s_{12},0) \nonumber \\
  & \cdot & R_{M}
\end{eqnarray}
with $R_{ij}$ being $3\times3$ rotation matrices, e.g.,
\begin{equation}\label{eq:rot_13}
R_{13} = \left( \begin{array}{ccc}
	 c_{13} & 0 & s_{13} \cdot e^{-i\delta_{CP}} \\
	0 & 1 & 0 \\
	-s_{13} \cdot e^{i\delta_{CP}} & 0 & c_{13} 
\end{array} \right), \nonumber
\end{equation}
and $R_M$ being a diagonal matrix:
\begin{equation}
  R_{M} = \left(\begin{array}{ccc}
          e^{i \alpha}& 0& 0\\
          0& e^{i \beta}& 0\\
          0& 0& 1\end{array}
          \right). \nonumber 
\end{equation}
Here $c_{ij} = \cos\theta_{ij}$, $s_{ij} = \sin\theta_{ij}$.
The Dirac phase is $\delta_{CP}$. Majorana phases are denoted by
$\alpha$ and $\beta$. Therefore, a total of seven or nine additional parameters
are required 
in the minimally extended standard model to accommodate massive Dirac or
Majorana neutrinos, respectively.

\begin{figure*}[htp]
\begin{centering}
\includegraphics[width=0.8\textwidth]{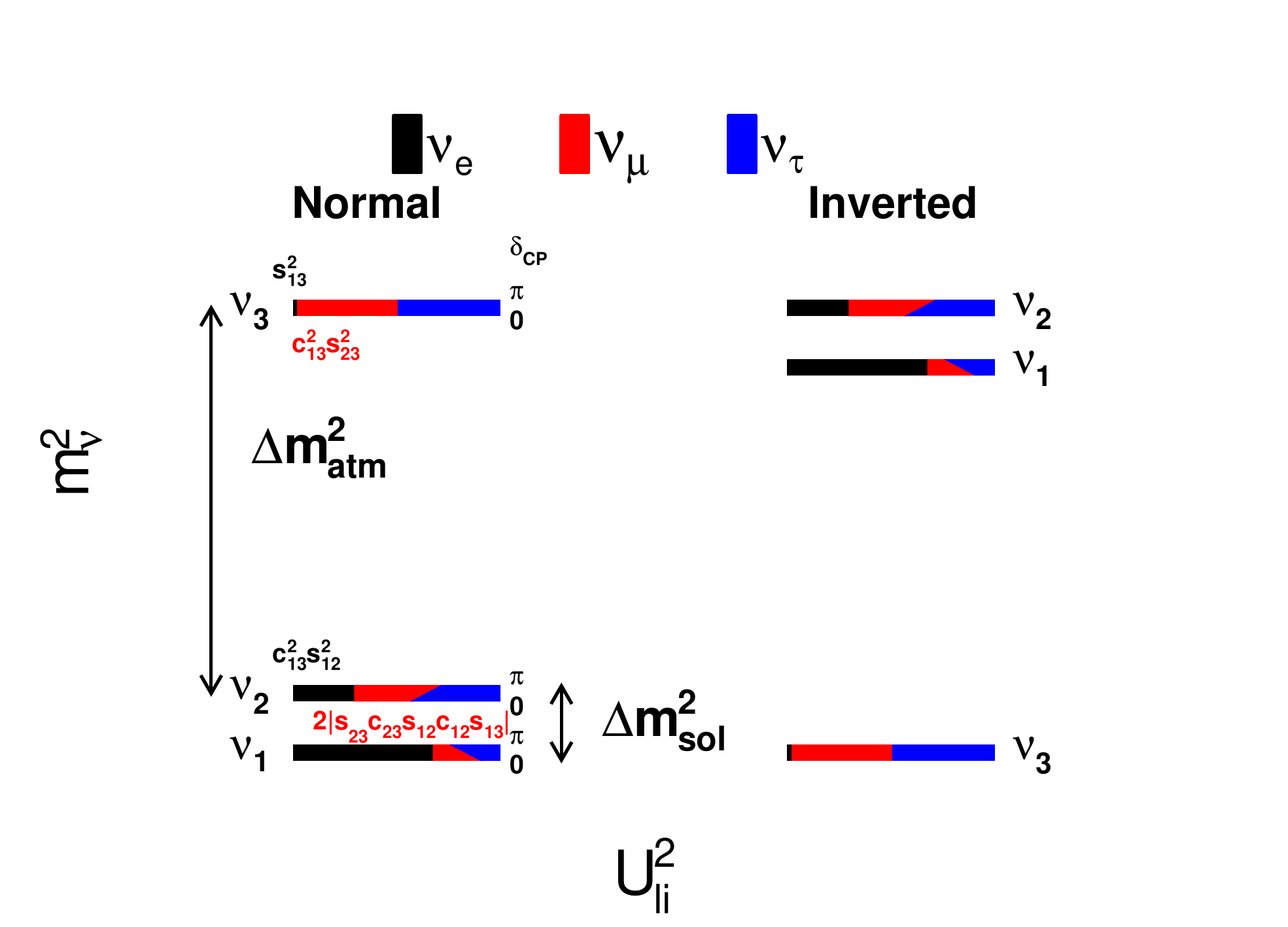}
\par\end{centering}
\caption{\label{fig:mh} 
Patterns of neutrino mass and mixing for the normal (left)
and inverted (right) hierarchy following Ref.~\cite{Mena:2003ug}.
The best-fit values of neutrino mixing parameters in
Ref.~\cite{Esteban:2016qun} are used, which results in slightly
different decompositions of the mass eigenstates in terms of
flavor eigenstates depending on the mass hierarchy.
$\Delta m^2_{sol} = \Delta m^2_{21}$ and $\Delta m^2_{atm} =
|\Delta m^2_{32}| \approx |\Delta m^2_{31}|$. The $l$ flavor 
component in the $i$th mass eigenstate is expressed as
$|U^2_{li}|$.  The magnitude
in front of $\cos\delta_{CP}$ is $2|s_{23}c_{23}s_{12}c_{12}s_{13}|$.}
\end{figure*}

\begin{table*}[h] 
\caption{Neutrino oscillation parameters taken from Ref.~\protect\cite{Esteban:2016qun}. 
  For the atmospheric mass-squared difference ($|\Delta m^2_{31}| \approx |\Delta m^2_{32}|$), the
  best fit results for both the normal (NH) and the inverted mass hierarchy (IH) are shown.
  These values are used in all the following plots, except where noted.}  
\medskip
\renewcommand{\arraystretch}{1.1} \centering 
\begin{tabular}{|c|cc|}
\hline 
parameter & best fit value $\pm~1\sigma$ & 3$\sigma$ range\\
\hline \hline
$\sin^2\theta_{12}$ & $0.306^{+0.012}_{-0.012}$ & (0.271, 0.345) \\
$\theta_{12}$ (degrees) & $33.56^{+0.77}_{-0.75}$ & (31.38, 35.99) \\
\hline 
$\Delta m^2_{21}$ $\times 10^{-5}$ eV$^2$ & $7.50^{+0.19}_{-0.17}$ & (7.03, 8.09) \\
\hline 
(NH) $\sin^2\theta_{23}$ & $0.441^{+0.027}_{-0.021}$  & (0.385, 0.635)\\
(NH) $\theta_{23}$ (degrees) & $41.6^{+1.5}_{-1.2}$ & (38.4, 52.8) \\
(IH) $\sin^2\theta_{23}$ & $0.587^{+0.020}_{-0.024}$  & (0.393, 0.640)\\
(IH) $\theta_{23}$ (degrees) & $50.0^{+1.1}_{-1.4}$ & (38.8, 53.1) \\
\hline 
(NH) $\sin^2\theta_{13}$ & $0.02166^{+0.00075}_{-0.00075}$ & (0.01934, 0.02392) \\
(NH) $\theta_{13}$ (degrees) & $8.46^{+0.15}_{-0.15}$ & (7.99, 8.90) \\
(IH) $\sin^2\theta_{13}$ & $0.02179^{+0.00076}_{-0.00076}$ & (0.01953, 0.02408) \\
(IH) $\theta_{13}$ (degrees) & $8.49^{+0.15}_{-0.15}$ & (8.03, 8.93) \\
\hline 
(NH) $\delta_{CP}$ (degrees) & $261^{+51}_{-59}$ & (0, 360) \\
(IH) $\delta_{CP}$ (degrees) & $277^{+40}_{-46}$ & (145, 391)~\footnotemark \\
\hline \hline

(NH) $\Delta m^2_{31}$ $\times 10^{-3}$ eV$^2$ & $+2.524^{+0.039}_{-0.040}$ & (+2.407, +2.643) \\
(IH) $\Delta m^2_{32}$ $\times 10^{-3}$ eV$^2$ & $-2.514^{+0.038}_{-0.041}$ & (-2.635, -2.399) \\
\hline
\end{tabular}\label{tab:PMNS}
\end{table*}

The phenomenon of neutrino flavor oscillation arises
because neutrinos are produced and detected in their flavor
eigenstates but propagate as a mixture of mass eigenstates. For example,
in vacuum, the neutrino mass eigenstates having energy $E$ would 
propagate as:
  \begin{eqnarray}\label{eq:osc_prop}
\frac{d}{dL} \left( \begin{array}{c} \nu_{1}(L) \\ \nu_{2}(L) \\ \nu_{3}(L) \end{array} \right) 
= -i \cdot V \cdot  \left( \begin{array}{c} \nu_{1}(L) \\ 
\nu_{2}(L) \\ \nu_{3}(L) \end{array} \right) \nonumber \\
= -i 
\left ( \begin{array}{ccc} \frac{m_1^2}{2E} & 0 & 0 \\
	0 & \frac{m_2^2}{2E} & 0 \\
	0 & 0 & \frac{m_3^2}{2E}
\end{array} \right) \cdot \left( \begin{array}{c} \nu_{1}(L) \\ 
\nu_{2}(L) \\ \nu_{3}(L) \end{array} \right), 
\end{eqnarray}
  after traveling a distance $L$. The above equation leads to the solution
$\nu_i(L) = e^{-i \frac{m_i^2}{2E}\cdot L}\nu_i(0)$. Therefore, for a neutrino produced
with flavor $l$, the probability of its transformation to flavor $l'$ is 
expressed as:
\begin{eqnarray}\label{eq:osc_dis}
P_{ll^\prime} \equiv P(\nu_{l}\rightarrow \nu_{l^\prime}) = |<\nu_{l^\prime}(L)|\nu_{l}(0)>|^2  \nonumber  \\
  =   \left |\sum_{j} U_{lj}U^{*}_{l'j}e^{-i(V_{jj})L} \right | ^2 \hfill \nonumber \\ 
 =  \sum_{j}|U_{lj}U^*_{l'j}|^2 +  \sum_{j} \sum_{k \neq j} U_{lj} U^{*}_{l'j} U^{*}_{lk} U_{l'k} e^{i\frac{\Delta m^2_{jk} L}{2E}},  
\end{eqnarray}
with $\Delta m^2_{jk} = m^2_j - m^2_k$. From Eq.~(\ref{eq:osc_dis}), it is obvious
that the two Majorana phases are not involved in neutrino flavor
oscillation. In other words, these Majorana phases cannot be determined
from neutrino flavor oscillation. 

\begin{figure}[h]
\begin{centering}
\mbox{ 
\includegraphics[width=0.48\textwidth]{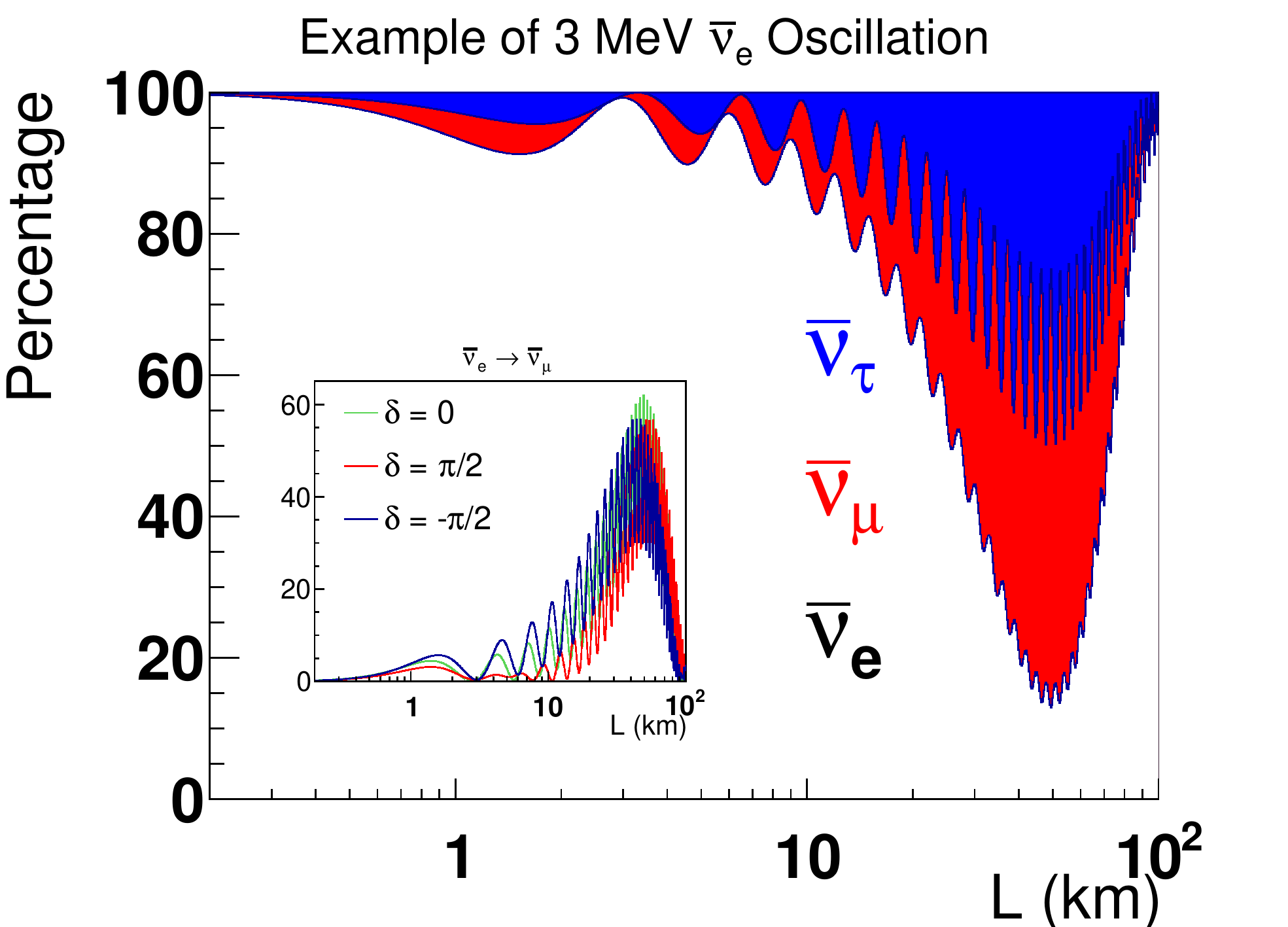}}
\par\end{centering}
\caption{\label{fig:reactor_osc} Example of a 3-MeV reactor electron
  antineutrino oscillation in the three-neutrino framework.
  The current best estimate of neutrino mixing parameters
  (tabulated in Table~\ref{tab:PMNS}) is used. The red and blue
  bands refer to the oscillation into $\bar \nu_\mu$ and
  $\bar \nu_\tau$ respectively, and the black curve is
  the $\bar \nu_e$ disappearance probability in percentages. The inner panel replots
  the $\bar{\nu}_{\mu}$ appearance probability in percentages, which is in principle
  sensitive to the unknown CP phase $\delta_{CP}$.
  However, the energy of the reactor neutrino is
  less than the $\bar{\nu}_{\mu}$ charged-current interaction threshold. The
  corresponding CPT-invariant process $\nu_{\mu} \rightarrow \nu_{e}$ is
  the primary method to measure $\delta_{CP}$ using accelerator neutrinos.}
\end{figure}

When neutrinos propagate in matter, Eq.~(\ref{eq:osc_dis}) must be modified 
because of the additional contribution originating from the interaction between 
neutrinos and matter constituents. This phenomenon is commonly referred to
as the Mikheyev--Smirnov--Wolfenstein (MSW)~\cite{Wolfenstein:1978ue,Mikheev:1986wj,Mikheev:1986gs} 
or matter effect.
The modification in oscillation probabilities is a result of the additional
contribution of charged-current interaction (W-boson exchange) between
  electrons in matter with electron neutrinos (antineutrinos). For neutrinos
  of other flavors (muon and tau), interaction with electron can only proceed
  via neutral current (Z-boson exchange). 

  Taking into account the matter effect, we have 
\begin{equation}
\frac{d}{dL} \left( \begin{array}{c} \nu_{e}(L) \\ \nu_{\mu}(L) \\ \nu_{\tau}(L) \end{array} \right) = -i 
\left ( \begin{array}{ccc} V_C & 0 & 0 \\
	0 & 0 & 0 \\
	0 & 0 & 0
\end{array} \right) \cdot \left( \begin{array}{c} \nu_{e}(L) \\ 
\nu_{\mu}(L) \\ \nu_{\tau}(L) \end{array} \right), 
\end{equation}
where $V_C = \sqrt{2} G_F N_e$ with $G_F$ being the Fermi constant and $N_e$ 
being the electron density in matter.
The sign of $V_C$ is  reversed for electron antineutrinos. The propagation matrix
$V$ in Eq.~(\ref{eq:osc_prop}) is modified as
\begin{eqnarray}\label{eq:matter_osc}
V^\prime &=& \left ( \begin{array}{ccc} \frac{m_1^2}{2E} & 0 & 0 \\
  0 & \frac{m_2^2}{2E} & 0 \\
  0 & 0 & \frac{m_3^2}{2E}
\end{array} \right)  + U^* \cdot \left ( \begin{array}{ccc} V_C & 0 & 0 \\
  0 & 0 & 0 \\
  0 & 0 & 0
\end{array} \right) \cdot U \nonumber \\
&=& U^*_{new} \cdot D \cdot U_{new},
\end{eqnarray}
where $U$ is the PMNS matrix.

The new matrix  $V^\prime$ can be expressed as
  a product of a unitarity matrix $U_{new}$, a diagonal matrix $D$, and 
$U_{new}^*$. The new energy eigenstates of neutrinos are thus $\nu^\prime_{j} = \sum_i 
U_{new}^{ij} \cdot \nu_i$, and the new mixing matrix connecting the flavor eigenstates 
and the energy eigenstates becomes $U^\prime = U\cdot U^*_{new}$. The 
oscillation probability in Eq.~(\ref{eq:osc_dis}) can be obtained by substituting
the mixing matrix $U$ by $U^\prime$ and the mass eigenstates $\nu_i$ by the
energy eigenstates $\nu^\prime_i$.
For reactor neutrino experiments, this effect is generally small because of low
neutrino energies and short baselines. For example, the changes in disappearance
probabilities are below 0.006\% and 7\% for the Daya Bay ($\sim$1.7~km baseline) and KamLAND
($\sim$180~km baseline) experiments, respectively, when the matter effect is taken into account.

\begin{figure*}[h]
  \centering
\includegraphics[width=0.34\textwidth]{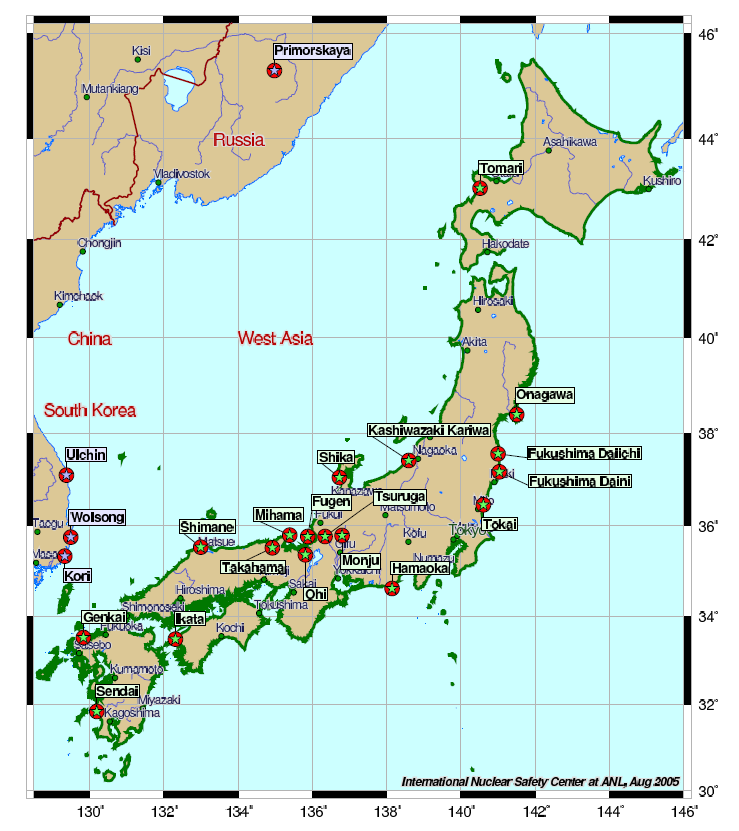}
\includegraphics[width=0.52\textwidth]{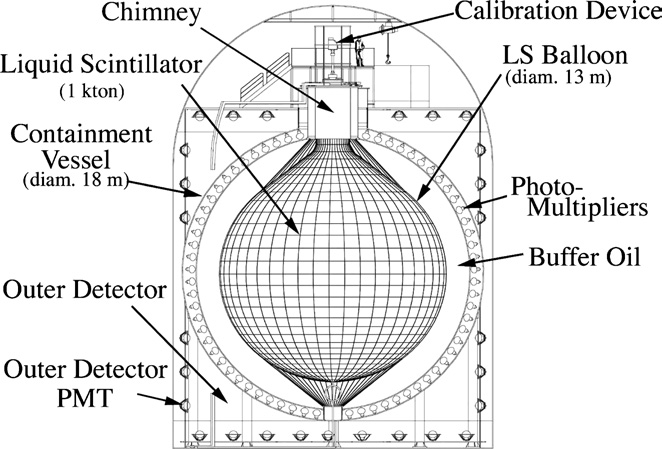}
\Put(-355,-20){\fontfamily{phv}\selectfont \textbf{A)}}
\Put(-130,-20){\fontfamily{phv}\selectfont \textbf{B)}}
\caption{\label{fig:kamland_map} A) The locations of nuclear
  power plants in Japan, Korea, and Far East Russia from International
  Nuclear Safety Center at Argonne National Laboratory
  (\url{http://www.insc.anl.gov/}). The KamLAND detector is located at
  (36.42$^\circ$ N, 137.31$^\circ$ E) in the middle of Japan.
  B) The structure of the KamLAND detector taken from Ref.~\cite{Eguchi:2002dm}.}
\end{figure*}

The  best values for the parameters obtained from a global
fit~\cite{Esteban:2016qun} to neutrino oscillation data after the
Neutrino 2016 conference~\cite{neutrino2016} are summarized
in Table~\ref{tab:PMNS}. A comparable result has also been obtained in
Ref.~\cite{deSalas:2017kay}. Incremental updates on neutrino
oscillation parameters have been presented in the Neutrino 2018 conference~\cite{neutrino2018}.
The patterns of neutrino mass and mixing
are shown in Fig.~\ref{fig:mh}. Regarding the parameters that can be accessed
through neutrino oscillation,  two crucial pieces, i) the neutrino
mass hierarchy (or the ordering of neutrino masses), which is the sign of 
$\Delta m^2_{32}=m^2_{3}-m^2_{2}$; and ii) the magnitude of the Dirac 
charge and parity (CP) phase $\delta_{CP}$, are still missing. 
Figure~\ref{fig:reactor_osc} shows an example of a 3-MeV reactor electron
antineutrino oscillation in the standard three-neutrino framework:
\begin{strip}
  \begin{eqnarray}\label{eq:3f_osc}
  P_{\bar{\nu}_e \rightarrow \bar{\nu}_e} &=& 1 - 4|U^2_{e1}||U^2_{e3}|\sin^2 \Delta_{31} -
  4|U^2_{e2}||U^2_{e3}|\sin^2\Delta_{32} - 4|U^2_{e1}||U^2_{e2}|\sin^2\Delta_{21} \nonumber\\
  &=& 1-\sin^{2}2\theta_{13}(\cos^{2}\theta_{12}\sin^{2}\Delta_{31}+\sin^{2}\theta_{12}\sin^{2}{\Delta_{32}}) 
-\cos^{4}\theta_{13}\sin^{2}2\theta_{12}\sin^{2}\Delta_{21}  \\
  P_{\bar{\nu}_e \rightarrow \bar{\nu}_{\mu}} &=& 4|U^2_{e3}||U^2_{\mu3}|\sin^2\Delta_{31} +
  4|U^2_{e2}||U^2_{\mu2}|\sin^2\Delta_{21} 
  + 8|U_{e3}||U_{\mu3}|U_{e2}||U_{\mu2}|\sin\Delta_{31}
  \sin\Delta_{21}\cos\left(\Delta_{32}-\delta_{\mu e}\right) \\
  P_{\bar{\nu}_e \rightarrow \bar{\nu}_{\tau}} &=& 4|U^2_{e3}||U^2_{\tau3}|\sin^2\Delta_{31} +
  4|U^2_{e2}||U^2_{\tau2}|\sin^2\Delta_{21} 
  + 8|U_{e3}||U_{\tau3}|U_{e2}||U_{\tau2}|\sin\Delta_{31}
  \sin\Delta_{21}\cos\left(\Delta_{32}-\delta_{\tau e}\right),
\end{eqnarray}
\end{strip}
with $\Delta_{ij} \equiv \frac{\Delta m^2_{ij}L}{4E}$ and
$\delta_{le}=-arg\left(U^*_{l3}U_{e3}U_{l2}U^*_{e2}\right)$ for lepton flavor $l$.
The fast and slow oscillation corresponds to
$|\Delta m^2_{32}|\approx |\Delta m^2_{31}|$ and $\Delta m^2_{21}$
mass squared difference, respectively.

\footnotetext{{(360,391) degrees are essentially (0,31) degrees.}}

\subsection{Observation of Neutrino Oscillations in the Solar Sector}~\label{sec:KamLAND} 


The first hint of solar neutrino flavor transformation was Ray Davis's
measurement of the solar $\nu_e$ flux using 610 tons of liquid C$_2$Cl$_4$,
through the reaction $\nu_e+^{37}$Cl$\rightarrow e^-+^{37}$Ar~\cite{Davis:1968cp}.
Compared with the prediction from the standard solar model (SSM)~\cite{Bahcall:1997ha,Bahcall:2000nu},
the measured $\nu_e$ flux was only about one-third as large~\cite{Cleveland:1994er,Cleveland:1998nv}.
This result was subsequently confirmed by SAGE~\cite{Abdurashitov:1999zd,Abdurashitov:2009tn}
and GALLEX~\cite{Anselmann:1993mh,Hampel:1998xg} using the reaction $\nu_e+^{71}$Ga$\rightarrow e^-+^{71}$Ge
and by Kamiokande~\cite{Hirata:1990fj,Fukuda:1996sz} and Super-K~\cite{Fukuda:1998fd,Fukuda:2002pe}
experiments using $\nu+e^-\rightarrow \nu+e^-$ elastic
scattering. This large discrepancy between measurements and predictions
from the SSM was commonly referred to as the \lq{solar neutrino puzzle}\rq. While many
considered this discrepancy as evidence for the inadequacy of SSM, others suggested
neutrino oscillation as the cause.

To solve the \lq{solar neutrino puzzle}\rq, the Sudbury Neutrino Observatory (SNO)
experiment was performed to measure the total flux of all neutrino flavors from the Sun
using three processes: i) the neutrino flux of all flavors from the neutral current (NC) reaction on deuterium from heavy water
$\nu_{e,\mu,\tau} + d \rightarrow \nu + p + n$; ii) the $\nu_e$ flux through
the charged current (CC) reaction $\nu_e + d \rightarrow e^- + p + p$; and iii) a combination
of $\nu_e$ and $\nu_{\mu,\tau}$ flux through the elastic scattering (ES) on electrons
$\nu + e \rightarrow \nu + e$. 
The measured flux of all neutrino flavors from the NC channel was entirely
consistent with the prediction of SSM~\cite{Ahmad:2002jz}, while
the measured $\nu_e$ flux from the CC channel clearly showed a deficit.
This result was consistent with neutrino mixing and flavor transformation
modified by the matter effect in the Sun.

The solar neutrino data allowed several solutions in the parameter space of
the neutrino mixing angle $\theta_{12}$ and the mass squared difference
$\Delta m^2_{21}$. This ambiguity was the result of several factors, including
the relatively large uncertainty of the solar $\nu_e$ flux predicted by
SSM, the matter effect inside the Sun, and the long distance neutrinos travel
to terrestrial detectors.
To resolve this ambiguity, a reactor neutrino experiment, the
Kamioka Liquid-scintillator ANtineutrino Detector (KamLAND)~\cite{Eguchi:2002dm},
was constructed in Japan to search with high precision for the $\sim$MeV reactor $\bar{\nu}_e$
oscillation at $\sim$200~km. Assuming CPT invariance, 
KamLAND directly explored the so-called \lq{large mixing angle}\rq~(LMA) parameter
region suggested by solar neutrino experiments.

As shown in Fig.~\ref{fig:kamland_map}A, the KamLAND experiment
was located at the site of the former Kamiokande experiment~\cite{Fukuda:1996sz}
under the summit of Mt. Ikenoyama in the Japanese Alps. A 2700-m water
equivalent (m.w.e.) vertical overburden was used to suppress backgrounds associated
with cosmic muons. The experimental site was surrounded by 55 Japanese nuclear reactor cores.
Reactor operation information, including thermal power and fuel burn-up,
was provided by all Japanese nuclear power plants, allowing KamLAND to calculate the
expected instantaneous neutrino flux.
The contribution to the total $\bar{\nu}_e$ flux
from Japanese research reactors and all reactors outside of Japan was about 4.5\%~\cite{Araki:2004mb}.
In particular, the contribution from reactors in Korea 
was estimated at 3.2$\pm$0.3\% and from other countries at 1.0$\pm$0.5\%. The flux-weighted
average $\bar{\nu}_e$ baseline was about 180 km, which was well suited to explore the LMA solution. 

The schematic layout of the KamLAND detector is shown in
Fig.~\ref{fig:kamland_map}B. One kiloton of highly purified LS, 80\% dodecane + 20\% pseudocumene, was enclosed in a 13-m diameter
balloon. The balloon was restrained by ropes inside a mineral-oil
buffer that was housed in a 18-m diameter stainless steel (SS) sphere.
An array of 554 20-inch and 1325 17-inch PMTs was
mounted to detect light produced by the IBD interaction. The SS vessel was
then placed inside a purified water pool, which also functioned as an active muon-veto
Cerenkov detector. The detector response was calibrated by deployments
of various radioactive sources. Resolutions of 12~cm/$\sqrt{E~({\rm MeV})}$,
6.5\%/$\sqrt{E~({\rm MeV})}$, and 1.4\% were achieved for the position,
energy, and the absolute energy scale uncertainty, respectively. 

\begin{figure}
\begin{centering}
\mbox{ 
\includegraphics[width=0.48\textwidth]{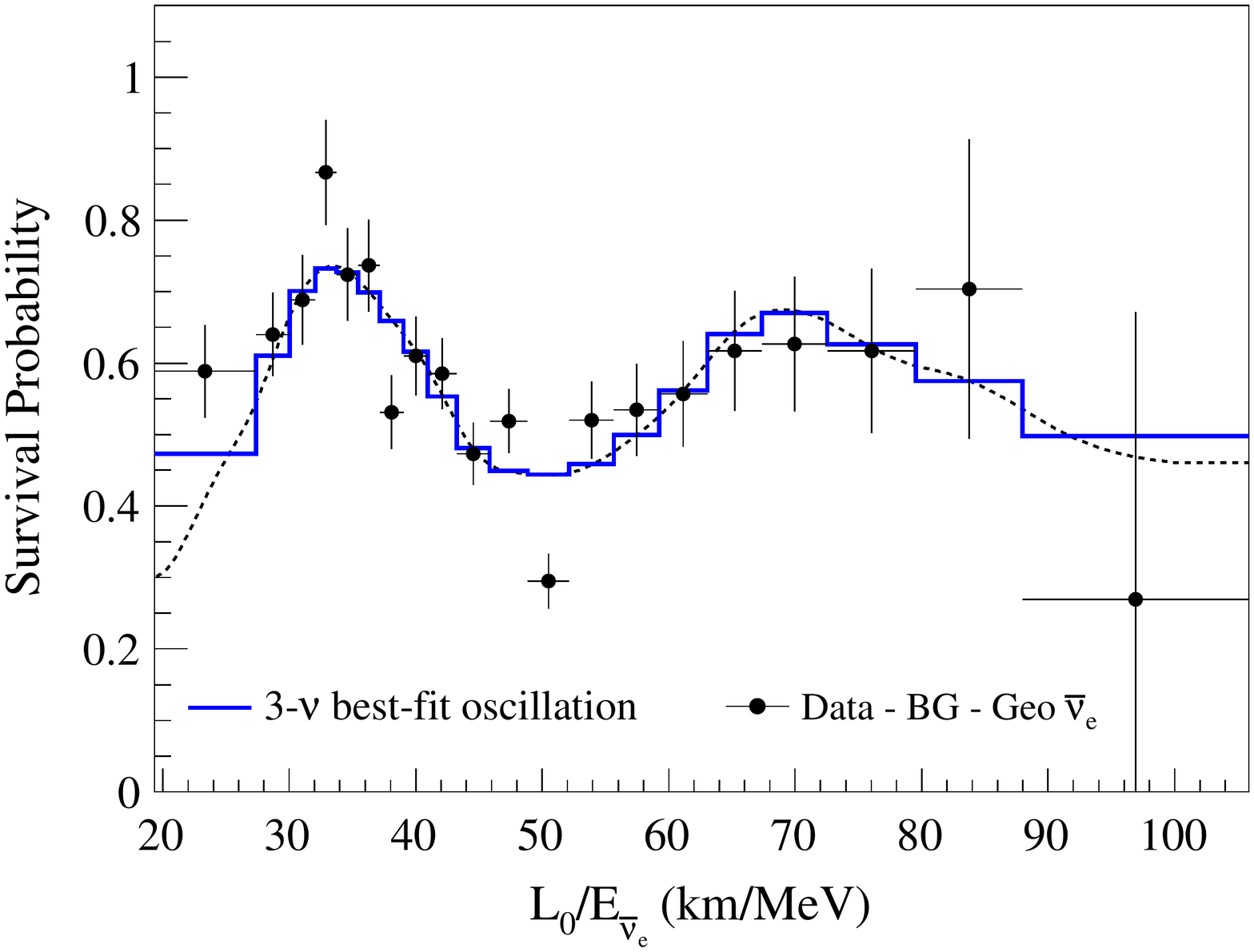}}
\par\end{centering}
\caption{\label{fig:edisap} 
  Survival probability of $\bar\nu_e$ events as a function of $L/E$
  in the KamLAND detector~\cite{Gando:2013nba}. The matter effect is included
  in the calculation of the survival probability. The dip position of
  the oscillation ($\sim$50 km/MeV) is consistent with the second
  oscillation node for  $\Delta m^2_{21}$. The size of the depletion
  is a measure of  $\sin^22\theta_{12}$.}
\end{figure}

Given the long baselines between the detector and the reactors, KamLAND expected to
observe about one reactor IBD event every day. The IBD events were selected by
requiring less than 1 ms time difference and 2-meter distance between the
prompt and delayed signals. The latter is a 2.2-MeV $\gamma$ ray from neutron
capture on hydrogen (see Table~\ref{tab:IBD_neutron_capture}). To
reduce the accidental coincidence backgrounds from external
radioactivities, the IBD selection was restricted to the innermost
6-m radius LS region. With the additional information of the event energy,
position, and time, the accidental background was suppressed to
$\sim$5\% of the IBD signal. The dominant background ($\sim$10\%) was
from the $\alpha+^{13}$C$\rightarrow n + ^{16}$O reaction ($\alpha-n$
background).  The incident $\alpha$ is from the decay of $^{210}$Po, a decay
product of $^{222}$Rn with a half-life of 3.8~days.
A decay product of uranium, $^{222}$Rn is commonly found in air and many materials
as a trace element. The prompt signal came from either a neutron scattering off
a proton or $^{16}$O de-excitation, and the delayed neutron capture signal
mimicked a $\bar{\nu}_e$ IBD event. Additional backgrounds included i) the
geoneutrinos produced in the decay chains of $^{232}$Th and $^{238}$U inside
the earth, which is an active research area by itself~\cite{Araki:2005qa,Agostini:2015cba};
ii) cosmogenic $^{9}$Li or $^{8}$He through $\beta$ decay accompanied by
a neutron emission;
iii) fast neutrons produced from muons interacting with the nearby rocks; and iv) atmospheric
neutrinos.

\begin{figure}
\begin{centering}
\mbox{ 
\includegraphics[width=0.48\textwidth,trim={0 4cm 7cm 0},clip]{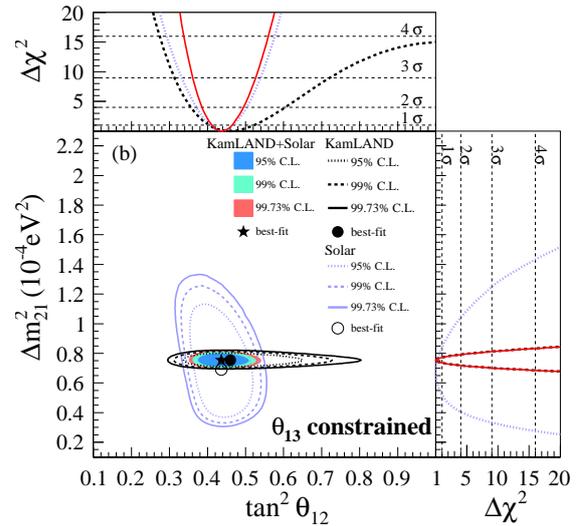}}
\par\end{centering}
\caption{\label{fig:edisap1} Allowed regions projected in the
  (tan$^2\theta_{12}$, $\Delta m^2_{21}$) plane,
  for solar and KamLAND data from a three-flavor oscillation
  analysis~\cite{Gando:2013nba}.
  The shaded regions are from a combined analysis
  of the solar and KamLAND data. The side panels show the
  $\Delta \chi^2$-profiles projected onto the tan$^2\theta_{12}$
  and $\Delta m^2_{21}$ axes. For this result,
  the value of $\theta_{13}$ is constrained by the results from
  reactor experiments with $\sim$km baselines.}
\end{figure}

The KamLAND experiment~\cite{Eguchi:2002dm,Araki:2004mb,Abe:2008aa}
clearly observed the oscillation of reactor neutrinos and
unambiguously established LMA as the solution of the
solar neutrino puzzle. The latest KamLAND result~\cite{Gando:2013nba}
is shown in Fig.~\ref{fig:edisap} as a function of $L/E_{\nu}$, where an
oscillatory pattern covering three oscillation extrema
is clearly observed.
Figure~\ref{fig:edisap1} shows $\Delta m^2_{21}$ vs.
$\tan^2 \theta_{12}$ from KamLAND and solar neutrino experiments. 

While the solar neutrino experiments are more sensitive to the mixing
angle $\theta_{12}$, KamLAND measures the mass-squared difference
$\Delta m^2_{21}$ more accurately through fitting the spectral
distortions. The observation of consistent mixing
parameters with two distinct neutrino sources (solar vs. reactor neutrinos)
and two different physics mechanisms (flavor transformation with the
matter effect vs. flavor oscillation in vacuum) provides compelling evidence
for non-zero neutrino mass and mixing.

Besides contributing to the measurement of neutrino mass and mixing parameters
  in the solar sector, the KamLAND data also gave an early hint of a non-zero
  $\theta_{13}$~\cite{Fogli:2008jx}. With $\theta_{13}=0$, the data from
  KamLAND~\cite{Abe:2008aa} favors a larger value of $\theta_{12}$ as compared
  to that from the SNO solar neutrino data~\cite{Aharmim:2005gt}. This small difference in
  $\theta_{12}$ can be reduced for a non-zero value of $\theta_{13}$
  ($\theta_{13}>0$ at $\sim$1.2$\sigma$ level)~\cite{Fogli:2008jx}.
  In the next section,
  we review the discovery of a non-zero $\theta_{13}$.

\subsection{Discovery of a Non-zero $\theta_{13}$}\label{sec:DYB}

\begin{table*}[htp] 
  \caption{Key parameters of five past and present reactor $\theta_{13}$
    experiments, including the reactor thermal power (in giga-watts),
    distance to reactors, target mass and material of the detectors,
    and overburden of the underground site (in meter-water-equivalent).
    PC, PXE, and LAB stands for Pseudocumene, Phenylxylylethane, and
    Linear Alkybenzene for liquid scintillator (LS) materials, respectively.}  
\medskip
\renewcommand{\arraystretch}{1.1} \centering 
\begin{tabular}{|c|c|c|c|c|c|}
\hline 
Experiment & Power     & Baseline  & Target Material & Mass  & Overburden \\
& (GW$_{th}$) & (m)      &   Gd-doped LS       & (tons) & (m.w.e.)     \\
\hline
CHOOZ      & 8.5        &1050     & paraffin-based       & 5     & 300         \\\hline
Palo Verde & 11.6       & 750-890 &   (segmented) PC-based  & 12    & 32          \\
\hline
Double Chooz & 8.5      & 400     & PXE-based  & 8     & 120         \\
             &          & 1050    &     & 8     & 300        \\\hline
RENO         & 16.8     & 290     &  LAB     & 16    & 120        \\
&          & 1380    &   & 16    & 450         \\\hline
Daya Bay  & 17.4   & 360      & LAB    & $2\times20$  & 250   \\
&         & 500    &   & $2\times20$  & 265    \\
          &        &  1580    & & $4\times20$  & 860    \\
\hline
\end{tabular}\label{tab:reactor_experiments}
\end{table*}

\subsubsection{History of Searching for a Non-zero $\theta_{13}$}\label{sec:theta13_history}


As introduced in Sec.~\ref{sec:nu_mixing}, three mixing angles,
one phase, and two independent mass-squared differences govern the
phenomenon of neutrino flavor oscillation. KamLAND and solar neutrino
experiments determined $\theta_{12}\approx$~33$^{\circ}$ and
$\Delta m^2_{21}\approx$~7.5$\times10^{-5}~{\rm eV}^2$.
Meanwhile, the results
$\theta_{23}\approx 45^{\circ}$ and $|\Delta m^2_{32}|\approx$~2.3$\times10^{-3}~{\rm eV}^2$
came from atmospheric
neutrino experiments such as Super-K~\cite{Fukuda:1998mi} and long-baseline
disappearance experiments, including K2K~\cite{Ahn:2006zza}, MINOS~\cite{Ahn:2006zza},
T2K~\cite{Abe:2013fuq}, and NO$\nu$A~\cite{Adamson:2017qqn}. In particular, the zenith-angle dependent
deficit of the upward-going atmospheric muon neutrinos reported by the Super-K
experiment~\cite{Fukuda:1998mi} in 1998 was the first compelling evidence of
neutrino flavor oscillation. Given that both the $\theta_{23}$ and $\theta_{12}$
angles are large, it is natural to expect that the third mixing angle
$\theta_{13}$ is also sizable.

There are at least two ways to access $\theta_{13}$. The first is to use reactor
neutrino disappearance $P\left(\bar{\nu}_e\rightarrow\bar{\nu}_e\right)$
(see Eq.~\ref{eq:3f_osc}). For a detector located at a distance $L$ near the first
maximum of $\sin^2\Delta_{31}$, the amplitude of the oscillation gives
$\sin^22\theta_{13}$. The second method is to use accelerator muon neutrinos
to search for electron neutrino
appearance $P\left(\nu_\mu \rightarrow \nu_e\right) \equiv P\left(\bar{\nu}_e \rightarrow \bar{\nu}_{\mu}\right)$ (see Eq.~\ref{eq:3f_osc}). In this case, the amplitude of the
oscillation depends not only on $\theta_{13}$, but also on several parameters,
including $\theta_{23}$, the unknown CP phase $\delta_{CP}$, and neutrino mass hierarchy
(through the matter effect in Earth). While the second method can access several
important neutrino parameters, the first method provides a direct and unambiguous
measurement of $\theta_{13}$.

Historically, the CHOOZ~\cite{Chooz1,Chooz2} and Palo Verde~\cite{PaloVerde} experiments made the
first attempts to determine the value of $\theta_{13}$ in the late 1990s to early 2000s.
Both experiments utilized reactor neutrinos to search for oscillation of $\bar{\nu}_e$ at
baselines of $\sim$1~km using a single-detector configuration.
The CHOOZ experiment was located at the CHOOZ power plant in the Ardennes region of France.
The CHOOZ detector mass was about 5~tons, and the distance to reactor cores was about 1050~m.
The data-taking started in April 1997 and ended in July 1998.

The Palo Verde experiment
was located at the Palo Verde Nuclear Generating Station in the Arizona desert of the United
States. The Palo Verde detector mass was about 12~tons, and the distances to three reactor cores
were 750~m, 890~m, and 890~m. The data-taking started in October 1998 and ended in July 2000. 
No oscillation were observed in either experiment, and a better upper limit of
sin$^22\theta_{13}<0.12$ was set at 90\% confidence level (C.L.) by CHOOZ.

\begin{figure*}[htp]
\centering
\includegraphics[width=0.37\textwidth]{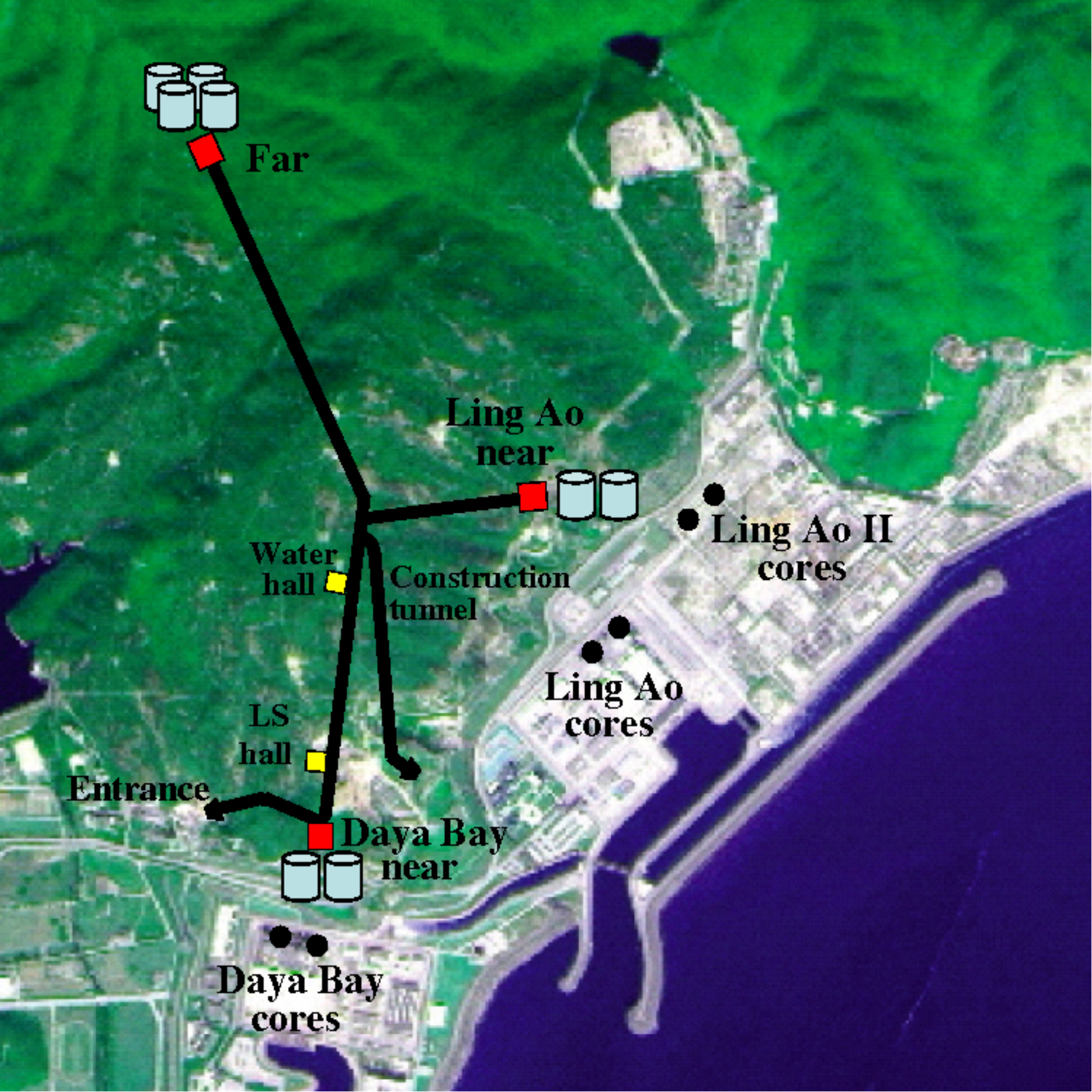}
\includegraphics[width=0.37\textwidth]{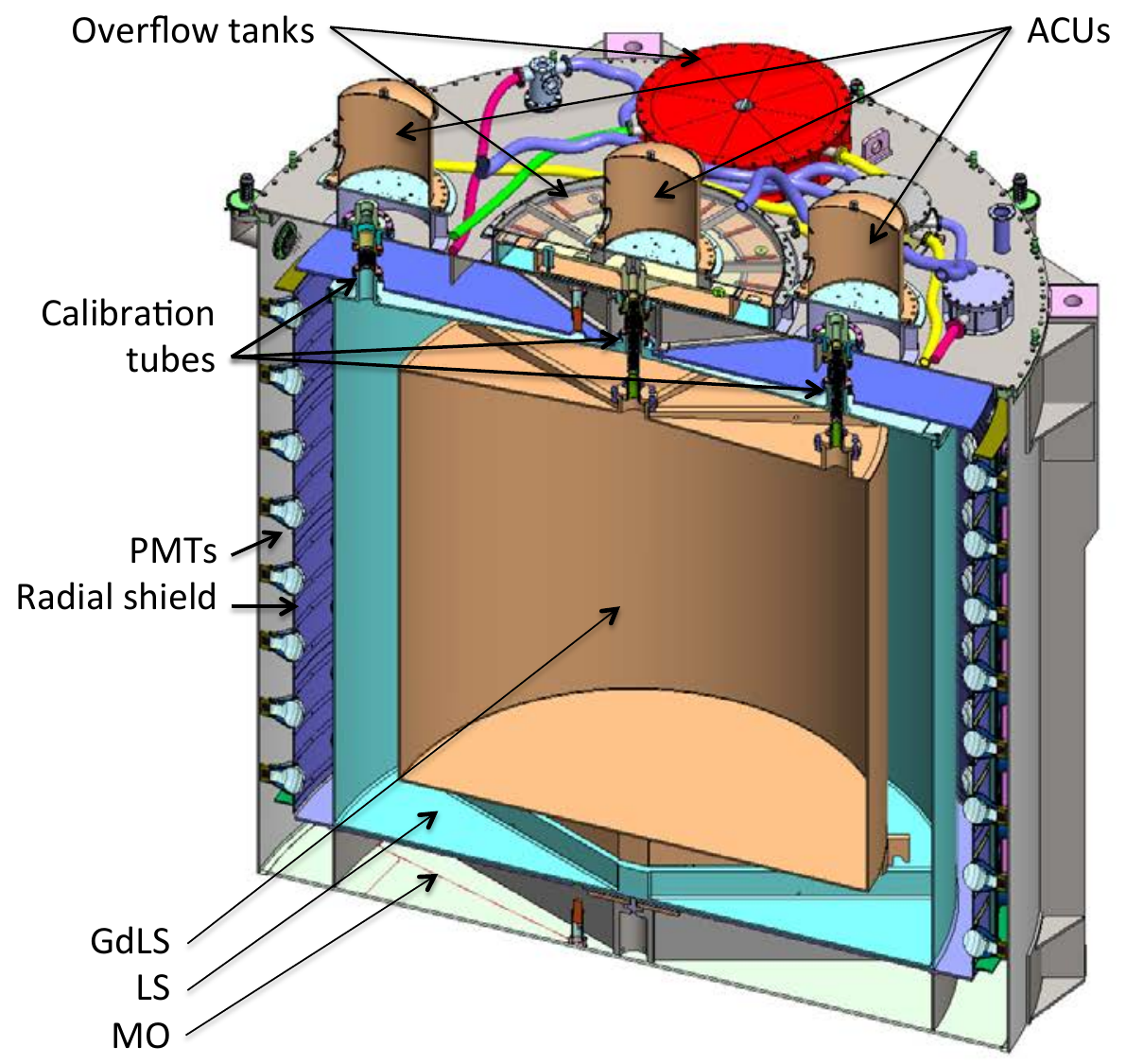}
\Put(-305,-20){\fontfamily{phv}\selectfont \textbf{A)}}
\Put(-90,-20){\fontfamily{phv}\selectfont \textbf{B)}}
\caption{\label{fig:dyb_map}
  A) The layout and the map of the Daya Bay experiment and
  the hosting Daya Bay plant campus. B) The structure of the
  Daya Bay antineutrino detector~(AD), taken from Ref.~\cite{Qian:2014xha}.
  The Daya Bay ADs were equipped with three automated calibration
  units~(ACUs), two for the Gd-LS volume and one for the LS volume. }
\end{figure*}

Given the measured values of $\theta_{12}$ and $\theta_{23}$ and the null $\theta_{13}$ results
from CHOOZ and Palo Verde, several phenomenological models of neutrino mixing
patterns, such as bimaximal and tribimaximal mixing~\cite{Harrison:2002er,Altarelli:2010gt},
became popular. In these models, the neutrino mass matrix in the flavor basis,
\begin{equation}
  M_{\nu} = U \cdot M_{\nu}^{diag} \cdot U^{\dag},
\end{equation}
is constructed based on flavor symmetries~\footnote{Here, $M_{\nu}^{diag}$ is a
  diagonal matrix with eigenvalues being the three neutrino masses $m_{1,2,3}$.},
and $\theta_{13}$ was predicted to be either zero or very small. 
Therefore, a new generation of reactor experiments (Double Chooz, Daya Bay, and RENO)
was designed to search for a small non-zero $\theta_{13}$. To suppress reactor-
  and detector-related systematic uncertainties, all three experiments
  adopted the ratio method advocated in Ref.~\cite{ratio}
  , which required placing multiple identical detectors at different baselines.
Table~\ref{tab:reactor_experiments} summarizes the key
parameters for past and present reactor $\theta_{13}$ experiments.

In 2011, almost 10 years after CHOOZ and Palo Verde, several hints
collectively suggested a non-zero $\theta_{13}$~\cite{Fogli:2011qn}.  The first one was based
on a small discrepancy between KamLAND and the solar neutrino
measurements~\cite{Fogli:2008jx}. Subsequently, accelerator neutrino
experiments MINOS~\cite{minos1} and T2K~\cite{t2k} reported their search
for $\nu_{\mu}$ to $\nu_e$.
In particular, T2K disfavored the $\theta_{13}$ = 0 hypothesis at
2.5$\sigma$~\cite{t2k}.

In early 2012, the Double Chooz reactor experiment
reported that the $\theta_{13}$ = 0 hypothesis was disfavored at 1.7$\sigma$,
based on their far-detector measurement~\cite{dc}. These hints of
a non-zero $\theta_{13}$ culminated in March 2012, when the Daya Bay reactor
neutrino experiment reported the discovery of a non-zero $\theta_{13}$
with a 5.1$\sigma$ significance~\cite{An:2012eh}.

About one month later,
RENO confirmed Daya Bay's finding of a non-zero $\theta_{13}$ with a
4.9$\sigma$ significance~\cite{Ahn:2012nd}. Later in 2012, Daya Bay
increased the significance to 7.7$\sigma$ using a larger
data set~\cite{An:2013uza}. A non-zero $\theta_{13}$ was firmly established.
In the following, we review three reactor $\theta_{13}$ experiments: Daya Bay, RENO, and Double Chooz.
Since these three experiments had many similarities in their design and
  physics analysis, we use Daya Bay to illustrate some common
  features.

\subsubsection{The Daya Bay Reactor Neutrino Experiment}

\begin{figure}[htp]
\begin{centering} 
\includegraphics[width=0.48\textwidth]{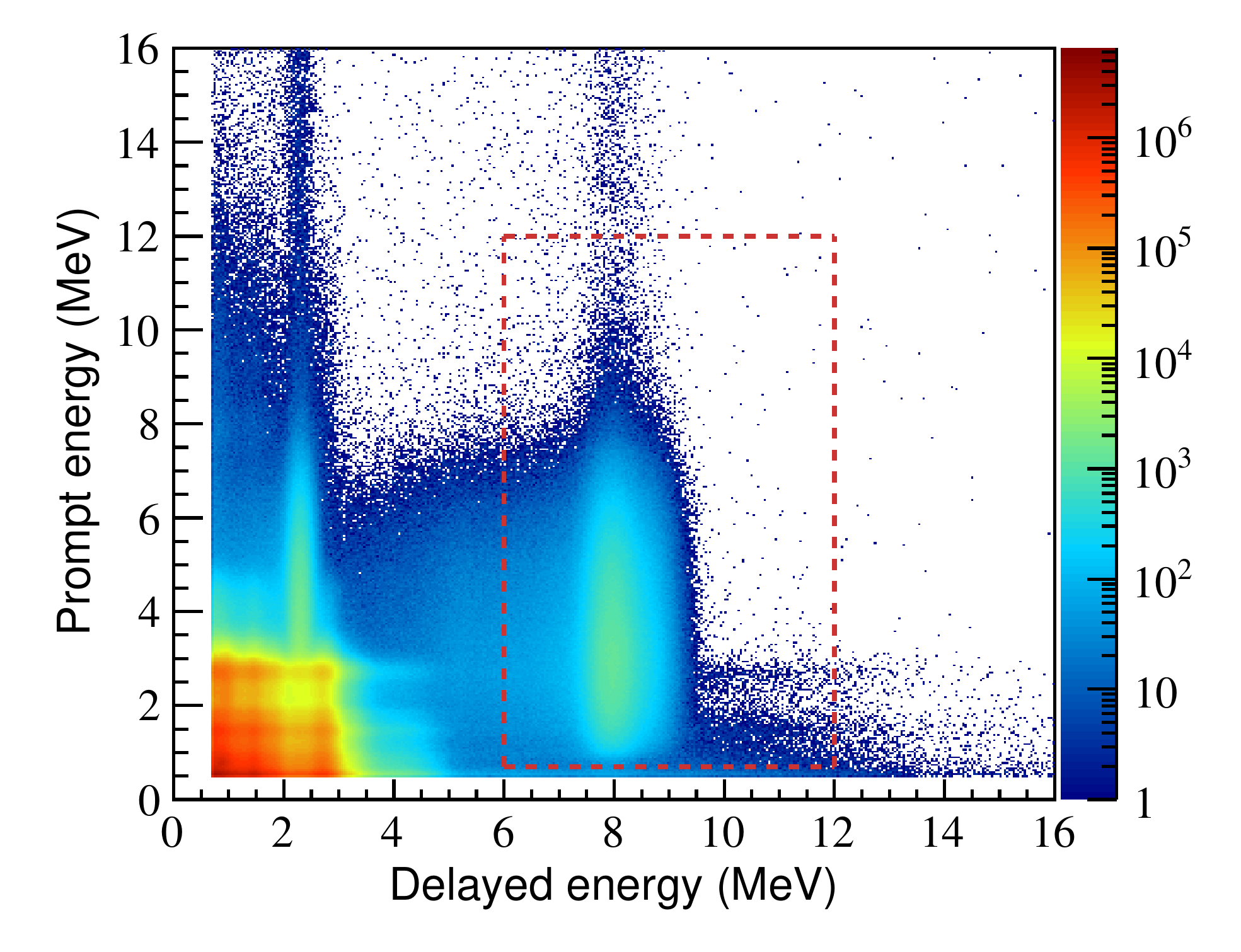}	
\par\end{centering}
\caption{\label{fig:ibd_selection} The distribution of prompt versus 
delayed energy for signal pairs which satisfied the $\bar{\nu}_e$ inverse 
beta decay selection criteria, taken from Ref.~\protect\cite{An:2016ses}. A
few-percent contamination from accidental backgrounds (symmetric
under interchange of prompt and delayed energy) and $^9$Li decay and
fast neutron backgrounds (high prompt and $\sim$8 MeV delayed energy)
are visible within the selected region. Inverse beta decay interactions
where the neutron was captured on hydrogen provided an additional signal
region with delayed energy around 2.2 MeV, albeit with much higher background.}
\end{figure}

\begin{figure*}[htp]
\begin{centering} 
\includegraphics[width=0.85\textwidth]{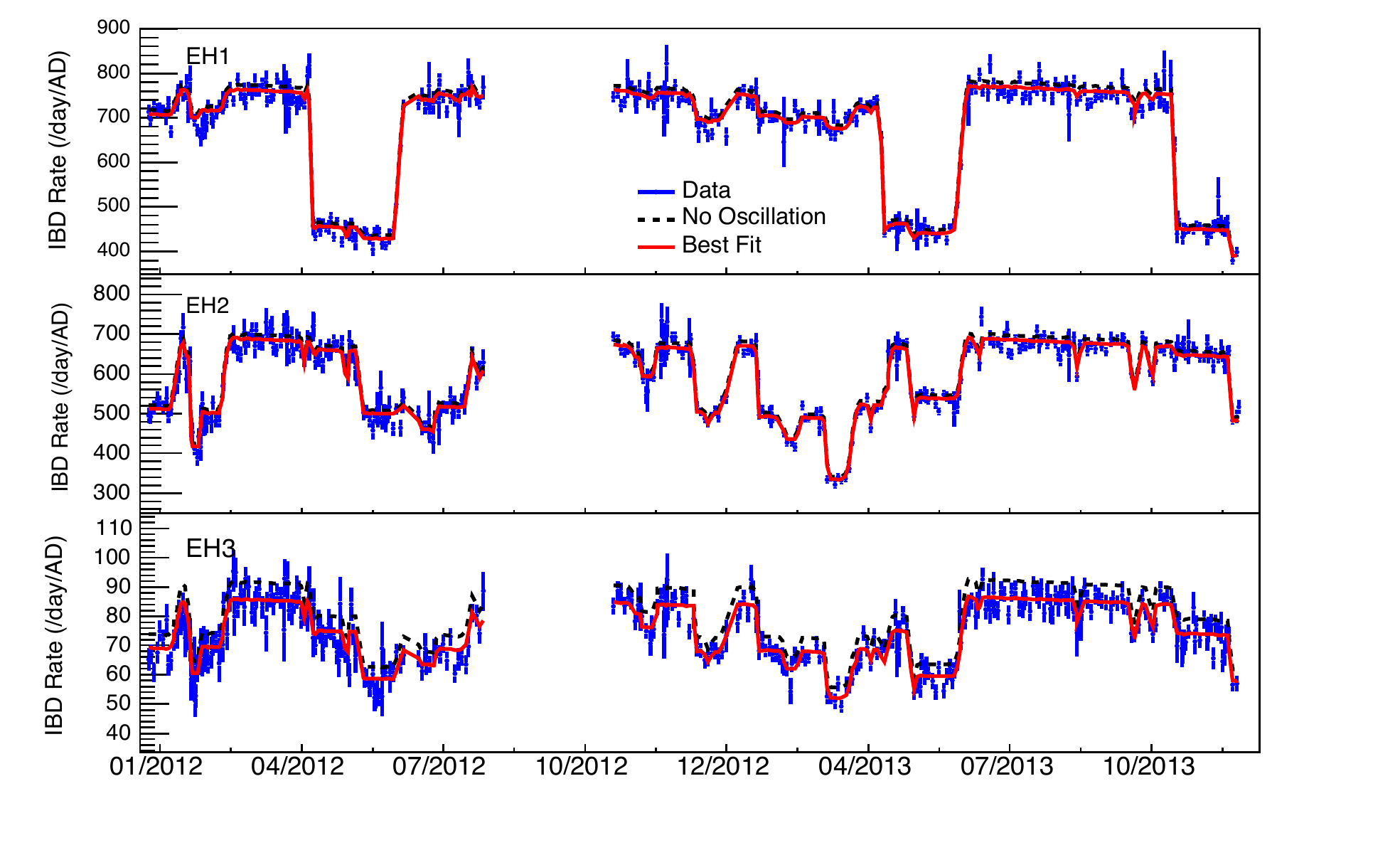}	
\par\end{centering}
\caption{\label{fig:dyb_rate} Daily averaged rates of IBD candidate
  events per detector in three experimental halls of Daya Bay as
  a function of time. The dotted curves represent no-oscillation
  predictions. The rates predicted with the best-fit non-zero
  $\sin^22\theta_{13}$ are shown as the red solid curves.
  The plot is taken from Ref.~\cite{An:2016srz}.} 
\end{figure*}

The Daya Bay Reactor Neutrino Experiment was located on the campus of the
Daya Bay nuclear reactor power plant in southern China.
As shown in Fig.~\ref{fig:dyb_map}A, the plant hosted six reactor cores whose
locations were grouped into three clusters: the Daya Bay, 
Ling Ao, and Ling Ao II clusters. The total thermal power was about 17.4~GW.
To monitor antineutrino flux from the
  three reactor clusters, near-detector sites were implemented.
  Two near-detector sites: the Daya Bay site ($\sim$363~m
  from the Daya Bay cluster) and the Ling Ao site ($\sim$500~m from the Ling Ao
  and Ling Ao II clusters) were constructed. The locations
  of the near and far sites were chosen
  to maximize the sensitivity to $\theta_{13}$. In particular, the
  Ling Ao near site
  and the far site were both located at approximately equal distances from 
  the Ling Ao and Ling Ao II clusters, largely reducing the effect of
  antineutrino flux uncertainties from these two clusters. 
  The average baseline of the far site was $\sim$1.7~km. 

Each near underground site hosted two antineutrino
detectors (ADs). The far site hosted four ADs that pair with the four ADs of the two
near sites, providing a maximal cancellation of detector effects.
The effective vertical overburdens were 250, 265, and 860 m.w.e.
for the Daya Bay site (EH-1), the Ling Ao site (EH-2), and the far site (EH-3),
respectively. With the near- and far-sites configuration, the contribution from 
reactor flux uncertainties was suppressed by a factor of 20~\cite{An:2013uza}, which
was the best among the reactor $\theta_{13}$ experiments.

Figure~\ref{fig:dyb_map}B shows the schematic view of
an AD~\cite{Band:2013osa,An:2015qga}. The innermost region was filled with
20~tons of Gd-doped linear-alkylbenzene-based liquid scintillator (LAB GdLS).
An array of 192 8-inch PMTs was installed on each AD.
Three automated calibration units (ACUs)~\cite{Liu:2013ava}
were equipped to periodically calibrate the detector response. Similar
to KamLAND,  ADs were placed inside high-purity water pools to
reduce radioactive backgrounds from the environment.
With PMTs installed, the water pool was also operated as an
independent water Cerenkov detector to veto cosmic
muons~\cite{Dayabay:2014vka,Hackenburg:2017syz}. Each water pool was further
split into two sub-detectors, so that the efficiency in each sub-detector could
be cross calibrated. A plane of resistive plate
chambers (RPC) was installed on the top of each water pool as an active muon
veto.


\begin{figure}[htp]
\begin{centering} 
\includegraphics[width=0.48\textwidth]{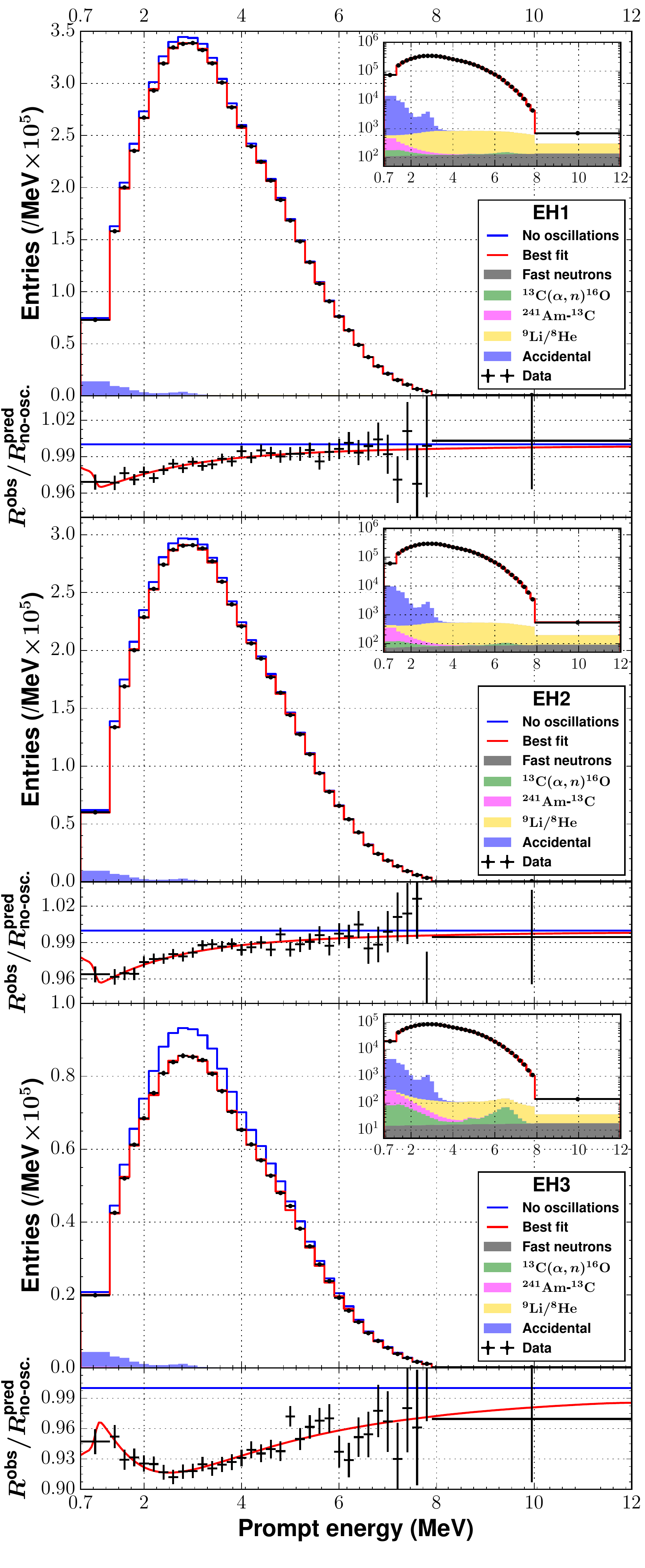}	
\par\end{centering}
\caption{\label{fig:dyb_spectrum}
  Reconstructed positron energy spectra for the $\bar{\nu}_e$ candidate
  interactions (black points) from Daya Bay{~\protect\cite{An:2016ses}}.
  The spectra of the detectors in each
experimental hall are combined: EH1 (top), EH2 (middle), and
EH3 (bottom). The measurements are compared with the prediction
assuming no oscillation (blue line) and the best-fit three-flavor
neutrino oscillation model (red line). The inset in semi-logarithmic
scale shows the backgrounds. The ratio of the background-subtracted
spectra to prediction assuming no oscillation is shown in the panel
beneath each energy spectrum.}
\end{figure}

Figure~\ref{fig:ibd_selection} shows the distribution of prompt versus 
  delayed energy for signal pairs satisfied the $\bar{\nu}_e$
  selection criteria, which included a crucial cut on the time difference
  between the prompt and delayed signals (1 $< \Delta t <$ 200 $\mu$s).  
  Five sources of backgrounds were identified. Ordering them in terms of
  their magnitudes at the near halls, they were accidental coincidence
  background, $\beta$-n decays from cosmogenic $^9$Li and $^8$He, fast
  neutrons produced by untagged muons, correlated $\gamma$-rays from
  Am-C neutron calibration units~\cite{Liu:2015cra}, and background from
  the $(\alpha,n)$ reactions~\cite{An:2016ses}. The accidental coincidence
  background was
  evaluated with high precision. Two of the three Am-C sources were
  removed during the 8-AD period for background reduction. Using information
  from the muon veto system, the fast neutron background rate was well
  determined. The total backgrounds accounted for $\sim$3\% (2\%) of the IBD
  candidate sample in the far (near) sites before the background subtraction.

\begin{table*}[htp] 
  \caption{Summary of major systematic uncertainties included in the Daya Bay oscillation
  analysis~\protect\cite{An:2016ses}. }  
\medskip
\renewcommand{\arraystretch}{1.1} \centering 
\begin{tabular}{|c|c|c|}
  \hline
  \hline
  Source & Uncertainty & Correlation \\\hline
  {\bf Reactor flux} & & \\
  Fission fractions & 5\%  & Correlation among isotopes from Ref.~\cite{Djurcic:2008ny},\\
  & & correlated among reactors\\
  Average energy per fission & Uncertainties from Ref.~\cite{Ma:2012bm}& Correlated among reactors\\
  $\bar{\nu}_e$ flux per fission & Huber--Mueller model\cite{Huber:2011wv,Mueller:2011nm} & Correlated among reactors \\
  Non-equilibrium $\bar{\nu}_e$ emission & 30\% (rel.)& Uncorrelated among reactors\\
  Spent nuclear fuel & 100\% (rel.)& Uncorrelated among reactors\\
  Reactor power & 0.5\% & Uncorrelated among reactors\\\hline
  {\bf Detector response } & & \\
  Absolute energy scale & $<$1\% & Correlated among detectors \\
  Relative energy scale & 0.2\%  & Uncorrelated among detectors \\
  Detector efficiency   & 0.13\% & Uncorrelated among detectors \\
  & & partial correlated (0.54 correlation coefficient) \\
  & & with relative energy scale \\
  IAV thickness & 4\% below  1.25 MeV (rel.)& Uncorrelated among detectors\\
  & 0.1\% above 1.25 MeV & \\\hline
  {\bf Background} &  &\\
  Accidental rate  & 1\% (rel.) & Uncorrelated among detectors \\
  $^9$Li-$^8$He rate & 44\% (rel.) & Correlated among same-site detectors \\
  Fast neutron rate & 13--17\% (rel.)& Correlated among same-site detectors \\
  $^{241}$Am-$^{13}$C rate & 45\% (rel.) & Correlated among detectors \\
  ($\alpha$,n) rate & 50\% (rel.) & Uncorrelated among detectors \\
  \hline
\end{tabular}\label{tab:dyb_uncertainties}
\end{table*}

Since the measurement of oscillation effect was obtained through the comparison
  of rate and spectra between near and far detectors, the identically designed
  detectors facilitated a near complete cancellation of the correlated detector systematic
  uncertainties. The accuracy of the oscillation parameters was thus governed by
  the uncertainties uncorrelated among detectors.
  Table~\ref{tab:dyb_uncertainties} summarizes the systematic uncertainties
  included in the Daya Bay oscillation analysis~\cite{An:2016ses}. In particular,
  the nature of each uncertainty (correlated or uncorrelated among reactors or
  detectors) is explicitly listed. 
  For the $\theta_{13}$ determination,
  an uncorrelated 0.1\% uncertainty from the hydrogen-to-Gd neutron capture ratio,
  which was related to the Gd concentrations in GdLS for all detectors,
  and an uncorrelated 0.08\% uncertainty from the 6-MeV cut on the delayed signal,
  which depended on the energy scale established in all detectors,
  were the major uncorrelated uncertainties. 

In earlier reactor neutrino experiments, measurements with reactor
power on and off provided a powerful tool to separate neutrino
signals from backgrounds. While this tool is not applicable in
Daya Bay, a clear correlation between the rates of IBD candidate
events and the reactor power was observed. Figure~\ref{fig:dyb_rate} shows the 
daily averaged rates of IBD candidate events at the three experimental
halls versus time. The IBD rates exhibit patterns that track well
with the variation of effective reactor power viewed at each hall.
These data show that the IBD candidate events originate predominantly
from the reactors rather than from cosmogenic and radioactive 
backgrounds.

Based on $\bar{\nu}_e$ data from all
eight detectors collected in 1230 days, Daya Bay determined
$\sin^22\theta_{13}=0.0850\pm0.0030~(stat.)\pm0.0028~(syst.)$ in a rate-only
analysis~\cite{An:2016ses}, with $|\Delta m^2_{32}|$ constrained
by atmospheric and accelerator neutrino experimental results. The
measured non-zero value of $\sin^22\theta_{13}$ was
only about 30\% below the upper limit set by the previous
CHOOZ experiment.

Prior to the discovery of a non-zero $\theta_{13}$, the only method to
measure the mass-squared difference $|\Delta m^2_{32}|$ was through
muon (anti)neutrino disappearance in atmospheric or accelerator
neutrino experiments. Given the IBD spectrum covering the antineutrino
energy range from 1.8 MeV to $\sim$8 MeV, the ``large'' value of
$\theta_{13}$ offered an alternative way to precisely measure this
quantity.

The first-ever extraction of
$|\Delta m^2_{ee}|:=|\cos^2\theta_{12}\Delta m^2_{31}+\sin^2\theta_{12}\Delta m^2_{32}|$~\cite{Nunokawa:2005nx}
was made by Daya Bay~\cite{An:2013zwz} through probing the relative spectral
distortion measured between the near and far detectors. In addition to the various
systematic uncertainties in the previous rate analysis, the absolute
detector energy response was another important ingredient to extract
$|\Delta m^2_{ee}|$, since the spectral distortion depended on $\Delta m^2_{ee} \frac{L}{E_{\nu}}$.
A physics-based energy model was constructed and constrained by calibrations using
various $\gamma$-ray sources and the well-known $^{12}$B beta
decay spectrum~\cite{An:2016ses}.

\begin{figure}[htp]
\begin{centering} 
\includegraphics[width=0.48\textwidth]{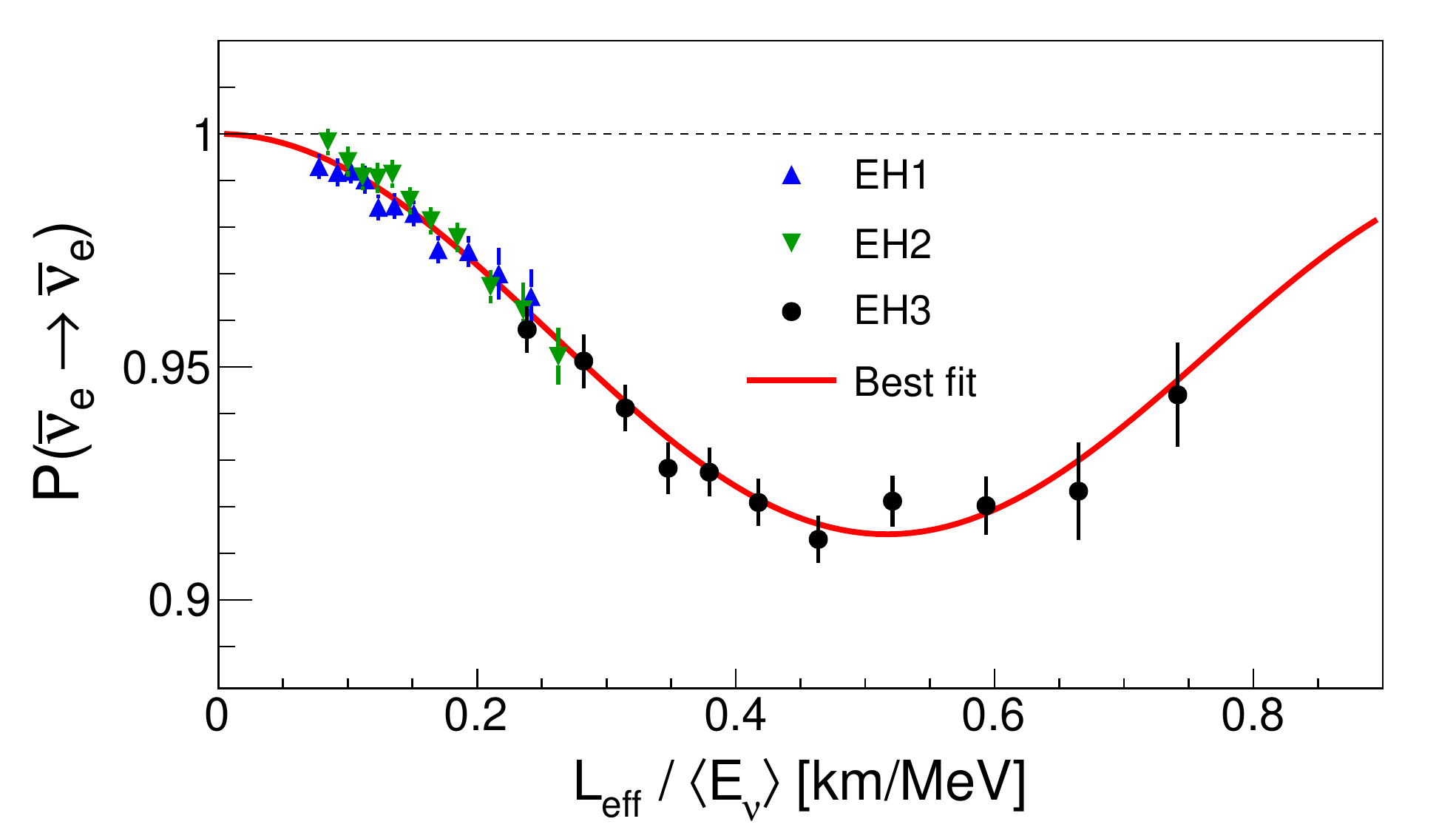}	
\par\end{centering}
\caption{\label{fig:dyb_loe}  The measured $\bar{\nu}_e$ disappearance
probability as a function of $L/E$ from Daya Bay 
{\protect\cite{An:2016ses}}. The oscillation amplitude corresponds to
$\sin^22\theta_{13}=0.0841\pm0.0027~(stat.)\pm0.0019~(syst.)$. 
The oscillation frequency corresponds to 
$|\Delta m^2_{ee}|=2.50\pm0.06~(stat.)\pm0.06~(syst.)\times10^{-3}$ eV$^2$. } 
\end{figure}

Figure~\ref{fig:dyb_spectrum} shows reconstructed positron energy spectra for the IBD candidate events
from Daya Bay~\cite{An:2016ses}. The best fit curve corresponds to
$\sin^22\theta_{13}=0.0841\pm0.0027~(stat.)\pm0.0019~(syst.)$ and
$|\Delta m^2_{ee}|=2.50\pm0.06~(stat.)\pm0.06~(syst.)\times10^{-3}$ eV$^2$~\cite{An:2016ses}.
Figure~\ref{fig:dyb_loe} shows the measured $\bar{\nu}_e$ disappearance
probability as a function of $L/E_{\nu}$.
As shown in Fig.~\ref{fig:theta13_dm2_global}, improved measurements were reported
  at the Neutrino 2018 conference~\cite{neutrino2018}.
Another benefit of the \lq{large}\rq~value of $\theta_{13}$ is that a different sample of
the IBD events using neutron capture on hydrogen (nH) in both the GdLS and LS regions can
also be employed to independently measure $\theta_{13}$.
Since the oscillation signal is large, many systematic associated
with the nH channel, which are generally larger than those of the nGd channel,
become less important. The details of extracting $\theta_{13}$
using the nH channel from Daya Bay can be found in Ref.~\cite{An:2014ehw,An:2016bvr}.

\subsubsection{The RENO and Double Chooz Experiments}

The Reactor Experiment for Neutrino Oscillation (RENO) was a short-baseline
reactor neutrino experiment built near the Hanbit nuclear power plant in South
Korea. Like the Daya Bay experiment, RENO was designed to measure the
mixing angle $\theta_{13}$. The six reactor cores in RENO had a total
thermal power of 16.4 GW. The reactor cores were equally spaced in a straight
line, with the near and far detector sites located along a line perpendicular
to and bisecting the reactor line. The near site was $\sim$290 m from the
geometric center of reactor cores, while the far site, located on the opposite side
of the reactor line, was at a distance of  $\sim$1380 m. Because of the large
variation in the distances between the near detector and various reactor cores,
the suppression of the uncertainty in the reactor neutrino flux was
less than ideal.
Taking a similar approach as Daya Bay, RENO adopted a three-zone LS antineutrino
detector nested in a muon veto system. The central target zone contained 16~tons
of 0.1\% Gd-doped LAB LS. A total of 354 10-inch PMTs were mounted on the
inner wall and the top and bottom surfaces of a stainless steel container. Unlike
Daya Bay, RENO had one detector in each experimental site.

RENO started data taking
in both the near and far detectors in the summer of 2011, ahead of all competing
experiments.  The first RENO $\theta_{13}$ result was published in
Ref.~\cite{Ahn:2012nd} in 2012. This result was in agreement with
Daya Bay's finding of a non-zero $\theta_{13}$~\cite{An:2012eh} with a
near-5$\sigma$ confidence
level. The observation of a 4~MeV--6~MeV anomaly in the prompt energy spectrum,
which is discussed in detail in Sec.~\ref{sec:RAA}, was first
reported by RENO~\cite{neutrino2014}. Most recently, RENO also reported a
measurement of $|\Delta m^2_{ee}|$ from the antineutrino energy spectral
distortion~\cite{RENO:2015ksa}, which was consistent with world measurements.
Figure~\ref{fig:theta13_dm2_global} shows RENO's latest results on
$\sin^22\theta_{13}$ and $|\Delta m^2_{ee}|$, reported at the Neutrino 2018
conference~\cite{neutrino2018}. In particular, the first measurement of
$|\Delta m^2_{ee}|$ using the nH channel was performed.


Double Chooz built upon the former CHOOZ experiment that set
the best upper limit of $\sin^22\theta_{13}$ prior to the discovery of
a non-zero $\theta_{13}$. It added a near site detector at a distance
of $\sim$410~m with a 115-m.w.e. overburden. The far site was the
original CHOOZ detector site, having a 1067 m baseline and 
a 300-m.w.e. overburden. The total thermal power of the two Double
Chooz reactors was 8.7 GW. Based on the original CHOOZ design, Double Chooz
adopted the three-zone design.  Instead of LAB-based LS,
Double Chooz's central target region was a 10-ton PXE-based LS. For each detector,
390 low-background 10-inch PMTs were mounted on the inner surfaces of
the stainless steel container. Unlike Daya Bay, Double Chooz had one detector
in each experimental site. Because of a construction delay,
the first result of Double Chooz~\cite{dc,Abe:2012tg},
a 1.7$\sigma$ hint of a non-zero $\theta_{13}$, included only the far-site data. 
To constrain the reactor neutrino flux uncertainty, Double Chooz
used the Bugey-4 measurement~\cite{Declais:1994ma} to normalize the flux. 
The systematic uncertainties of the first result were subsequently improved, as reported in
Ref.~\cite{Abe:2014lus}, with backgrounds constrained by the reactor-off data.
An improved measurement of $\theta_{13}$ with about twice the antineutrino flux
exposure was reported in Ref.~\cite{Abe:2014bwa}. Double Chooz carried out
the first independent $\theta_{13}$ analysis using the neutron-capture-on-hydrogen
data~\cite{Abe:2013sxa,Abe:2015rcp}. The Double Chooz
near detector started taking data in 2014. The latest Double Chooz
result using both near and far detector data yielded
$\sin^22\theta_{13}=0.105\pm0.014$~\cite{neutrino2018}. 

\subsubsection{Impacts of a Non-zero $\theta_{13}$}

\begin{figure*}[htp]
  \begin{centering}
    \includegraphics[width=0.4\textwidth]{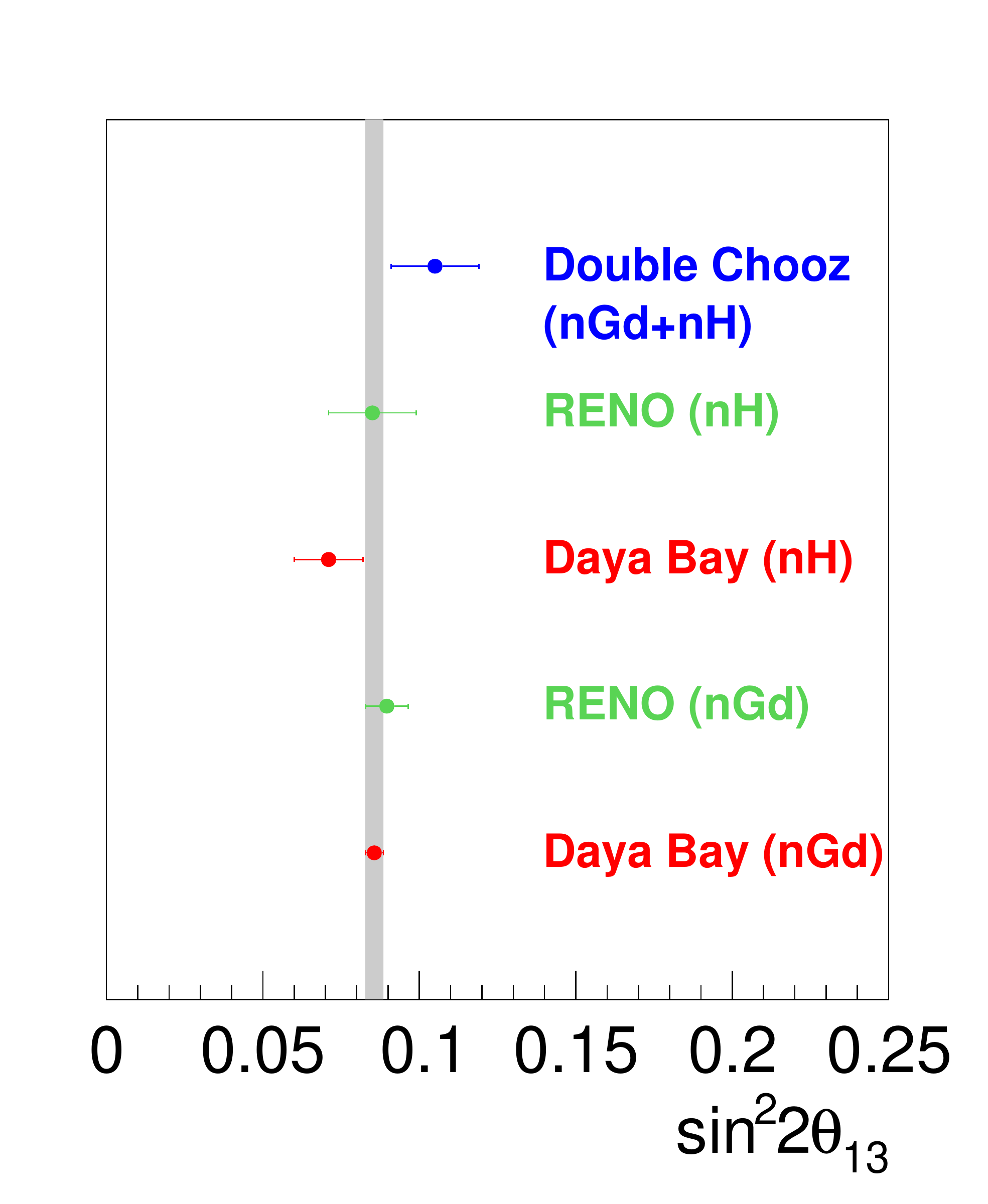}
    \includegraphics[width=0.4\textwidth]{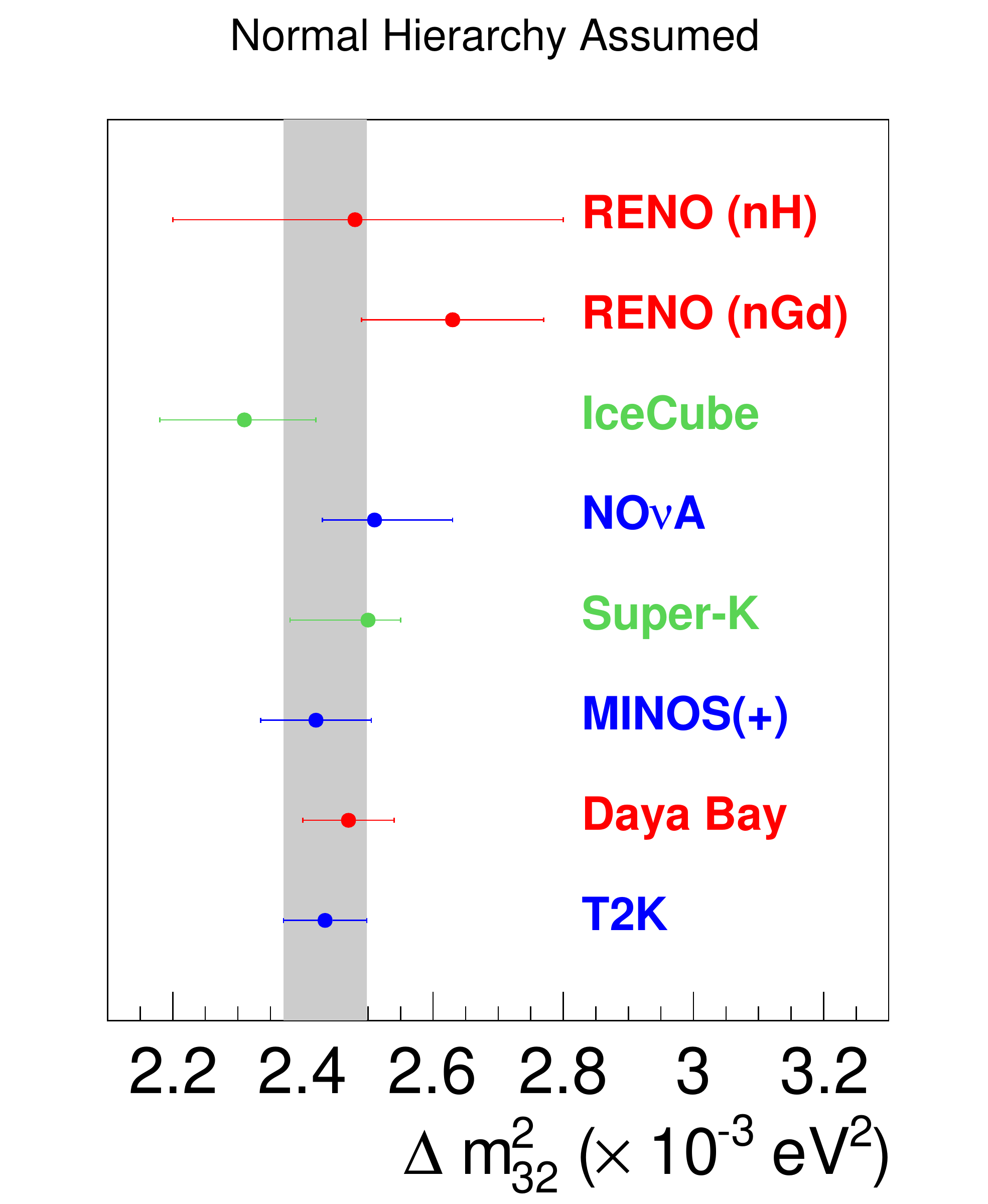}
    \Put(-305,-20){\fontfamily{phv}\selectfont \textbf{A)}}
    \Put(-90,-20){\fontfamily{phv}\selectfont \textbf{B)}}
    \par\end{centering}
    \caption{\label{fig:theta13_dm2_global} Global results on $\theta_{13}$ (A)
      and $\Delta m^2_{32}$ (B) taken from the results presented at the Neutrino 2018 conference~\protect{\cite{neutrino2018}}. For $\Delta m^2_{32}$, only results of the normal hierarchy
    are shown. }
    \end{figure*}

Figure~\ref{fig:theta13_dm2_global} summarizes the status
of $\theta_{13}$ and $|\Delta m^2_{32}|$ after the Neutrino 2018
conference~\cite{neutrino2018}. The precision of $\sin^22\theta_{13}$
from Daya Bay was better than 3.5\%, making it the best measured mixing
angle. Given the relatively \lq{large}\rq~ value of $\theta_{13}$, the
$|\Delta m^2_{32}|$ was measured precisely using reactor neutrinos,
given the well-controlled systematics for the detector and the antineutrino
flux. In particular, the precision of $|\Delta m^2_{32}|$ from Daya Bay had
reached a similar precision as those from accelerator neutrino and
atmospheric neutrino experiments, as shown in Fig.~\ref{fig:theta13_dm2_global}.

Besides the precision measurement of $|\Delta m^2_{32}|$, a non-zero
$\theta_{13}$ also opens up many opportunities for future
discoveries. In particular, it allows for a determination of the
neutrino mass hierarchy in a medium-baseline reactor neutrino
experiment, which is elaborated in Sec.~\ref{sec:mh}. In addition,
it enables the search for CP violation in the leptonic sector, as
well as the determination of the neutrino mass hierarchy through
precision (anti-)$\nu_{\mu}\rightarrow$~(anti-)$\nu_e$ oscillation
in accelerator neutrino experiments (see Ref.~\cite{Diwan:2016gmz} for a
recent review). To leading order in
$\alpha=\Delta m^2_{21}/\Delta m^2_{31}$, the probability of the
$\nu_{\mu}\rightarrow\nu_e$ oscillation can be
written as~\cite{Freund:2001pn}:
\begin{eqnarray}
\label{eq:nueapp} 
P(\nu_\mu \rightarrow \nu_e)  &  = & \sin^2\theta_{23} \frac{\sin^22\theta_{13}}{(A-1)^2} \sin^2[(A-1)\Delta_{31}]  \hfill \nonumber   \\
&+ &  \alpha^2\cos^2\theta_{23}\frac{\sin^22\theta_{12}}{A^2}\sin^2(A\Delta_{31})  \hfill \nonumber  \\
&- &   \alpha \frac{\sin2\theta_{12}\sin2\theta_{13}\sin2\theta_{23}\cos\theta_{13}\sin\delta_{CP}}{A(1-A)} \nonumber \\
&\times& \sin\Delta_{31}\sin(A\Delta_{31}) \sin[(1-A)\Delta_{31}]  \hfill \nonumber \\ 
&+ &  \alpha \frac{\sin2\theta_{12}\sin2\theta_{13}\sin2\theta_{23}\cos\theta_{13}\cos\delta_{CP}}{A(1-A)} \nonumber \\
& \times& \cos\Delta_{31}\sin(A\Delta_{31})\sin[(1-A)\Delta_{31}],  \hfill 
\end{eqnarray}
where
\begin{eqnarray}
  \Delta_{ij} &=& \Delta m^2_{ij}L/4E_{\nu}, \nonumber \\
  A &=& \sqrt{2}G_{F}N_{e}2E_{\nu}/\Delta m^2_{31}. 
\end{eqnarray}

For antineutrinos, the signs of $\delta_{CP}$ and $A$ are reversed.
The sensitivity to the mass hierarchy (i.e., the sign of $A$) mainly
comes from the first term in Eq. (\ref{eq:nueapp}), which becomes
non-zero for a non-zero
$\theta_{13}$. In addition, the sensitivity to the mass hierarchy is larger
for a larger value of $\theta_{13}$. Similarly, the sensitivity to CP
violation (i.e., a non-zero value for $\sin\delta_{CP}$) comes from the
last two terms, which are in play for a non-zero $\theta_{13}$. In contrast
to the mass hierarchy sensitivity, the sensitivity to CP violation
is approximately independent of the value of $\theta_{13}$~\cite{Marciano:2001tz}. To illustrate
this point, we use the fractional asymmetry
\begin{equation}
A^{\mu e}_{CP} \equiv  \frac{({P(\nu_\mu\to \nu_{e}) -
 P(\bar\nu_\mu \to \bar\nu_{e})})}{({P(\nu_\mu\to \nu_{e}) +
  P(\bar\nu_\mu \to \bar\nu_{e})})}. 
\end{equation}
At larger values of $\theta_{13}$, $A^{\mu e}_{CP} \sim$1/$\sin2\theta_{13}$
becomes smaller for a given value of CP phase. However,
the increase in the number of events leads to a better measurement
of $A^{\mu e}_{CP}$, with statistical uncertainties $\delta A^{\mu e}_{CP}
\sim$1/$\sin2\theta_{13}$. These two effects approximately cancel each
other. In real experiments, a larger value of $\theta_{13}$ is actually
favored, as the impact of various backgrounds on the
$\nu_{\mu}\rightarrow\nu_e$ signal is reduced with larger signal
strength. 

By 2020, the precision of $\sin^22\theta_{13}$ and $\Delta m^2_{32}$
in Daya Bay is projected to be better than 3\%. The comparison of the
$\theta_{13}$ measurement from reactor $\bar{\nu}_e \rightarrow \bar{\nu}_e$
disappearance and that from the accelerator $\nu_{\mu}\rightarrow\nu_e$
appearance in the future DUNE~\cite{Adams:2013qkq} and
Hyper-K~\cite{Kearns:2013lea} experiments
will provide one of the best unitarity tests of the PMNS matrix~\cite{Qian:2013ora}.


\subsection{Future Opportunities}\label{sec:mh}

\begin{figure*}[htp]
\begin{centering}
  \includegraphics[width=0.85\textwidth]{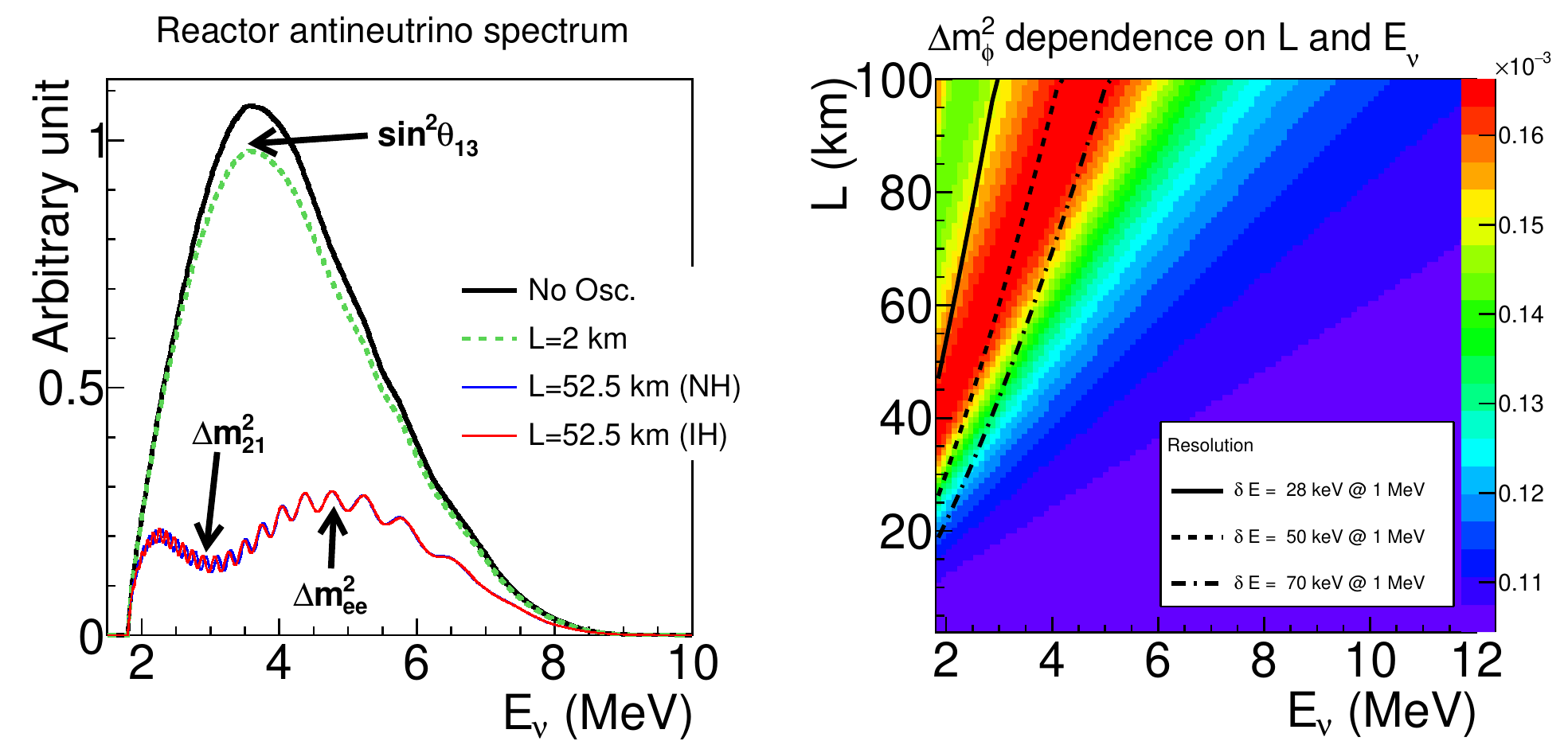}
  \Put(-325,-0){\fontfamily{phv}\selectfont \textbf{A)}}
\Put(-100,-0){\fontfamily{phv}\selectfont \textbf{B)}}
\par\end{centering}
\caption{\label{fig:reactor_mh} A) Expected antineutrino energy spectra at
  different baselines with $\Delta m^2_{ee}=2.41\times10^{-3}$ eV$^2$. 
  The effects of a non-zero $\sin^22\theta_{13}$ and different
  MHs are emphasized. B) $\Delta m^2_{\phi}$ is shown as a function of the
  neutrino energy and the baseline.
  At 50~km--60 km, the $\Delta m^2_{\phi}$ shows a clear dependence on the neutrino
  energy. Such
  a dependence is the key to determine the MH. The plot is taken from
  Ref.~\cite{Qian:2012xh}.}
\end{figure*}

\subsubsection{Determination of the Neutrino Mass Hierarchy}

The neutrino mass hierarchy (MH), i.e., whether the third generation neutrino mass 
eigenstate is heavier or lighter than the first two, is one of the remaining
unknowns in the minimal extended $\nu$SM (see Ref.~\cite{Qian:2015waa} for a
recent review)~\footnote{The other two unknowns are the CP phase $\delta_{CP}$
  and the absolute neutrino mass. In addition, the octant of $\theta_{23}$,
  i.e., whether $\theta_{23}$ is larger or smaller than 45$^\circ$, is also an
  interesting question.}. The determination of the MH,
together with searches for neutrinoless double beta decay, may reveal whether
neutrinos are Dirac or Majorana fermions, which could significantly advance
our understanding of the Universe. 

The precise measurement of $\sin^{2}2\theta_{13}$ by the current generation of 
short-baseline reactor neutrino experiments has provided a unique opportunity
to determine the MH in a medium-baseline ($\sim$55 km) reactor neutrino
experiment~\cite{Qian:2012xh,Petcov:2001sy, Learned:2006wy, Zhan:2008id,Zhan:2009rs, Ciuffoli:2012iz,Ge:2012wj,Li:2013zyd}. 
The oscillation from the atmospheric mass-squared difference manifests
itself in the energy spectrum as multiple cycles that contain the MH information, 
as shown in the following formula derived from Eq. (\ref{eq:3f_osc}): 
\begin{eqnarray}\label{eq:osc}
P_{\bar{\nu}_{e}\rightarrow\bar{\nu}_{e}} 
& = & 1-2s_{13}^{2}c_{13}^{2}-4c_{13}^{2}s_{12}^{2}c_{12}^{2}\sin^{2}\Delta_{21}  \\
&+&2s_{13}^{2}c_{13}^{2}\sqrt{1-4s_{12}^{2}c_{12}^{2}\sin^{2}\Delta_{21}}\cos(2|\Delta_{32}|\pm\phi), \nonumber
\end{eqnarray}
where $\Delta_{21}\equiv\Delta m_{21}^{2}L/4E$, $\Delta_{32}\equiv\Delta
m_{32}^{2}L/4E$, and 
\[
\sin\phi=\frac{c_{12}^{2}\sin2\Delta_{21}}{\sqrt{1-4s_{12}^{2}c_{12}^{2}\sin^{2}\Delta_{21}}}\,,
\]
\[
\cos\phi=\frac{c_{12}^{2}\cos2\Delta_{21}+s_{12}^{2}}{\sqrt{1-4s_{12}^{2}c_{12}^{2}\sin^{2}\Delta_{21}}}.
\]
The $\pm$ sign in the last term of Eq. (\ref{eq:osc}) depends on the MH: the plus sign 
indicates the normal hierarchy (NH) and the minus sign indicates the inverted hierarchy (IH).
The principle of determining MH through spectral distortion can be understood from
Fig.~\ref{fig:reactor_mh}B, which shows the energy and baseline dependent
$\Delta m^2_{\phi}:=4E\cdot \phi/L$, based on Eq. (\ref{eq:osc}). The three lines
represent three different choices of energy resolution. In the region left of the line,
the measurement of $\Delta m^2_{\phi}$ is compromised.
Above $\sim$40~km,
$\Delta m^2_{\phi}$ possesses a clear energy dependence. In particular, at
$\sim$50~km, $\Delta m^2_{\phi}$ at low-energy region (2~MeV--4~MeV) is larger than that
at high-energy region (4~MeV--8~MeV). This distinction provides an excellent opportunity to determine
the MH. 
For NH, the $\Delta m^2_{eff} :=  2|\Delta m^2_{32}| + \Delta m^2_{\phi}$ measured
in the low-energy region (2~MeV--4~MeV) would be higher than that measured in the
high-energy region (4~MeV--8~MeV).
In comparison, for the IH, the $\Delta m^2_{eff}:=2|\Delta m^2_{32}| - \Delta m^2_{\phi}$
measured in the low-energy region would be lower than that measured at high energy.
Figure~\ref{fig:reactor_mh}A shows the reactor neutrino energy spectra at a baseline of 
52.5~km for both NH and IH.  The choice of MH leads to a shift in the oscillation pattern
at low-energy region relative to that at high-energy region.

The Jiangmen Underground Neutrino Observatory (JUNO)~\cite{An:2015jdp} is a
next-generation (medium-baseline) reactor neutrino experiment under construction
in Jiangmen City, Guangdong Province, China. It consists of a
20-kton underground LS detector having a
1850~m.w.e. overburden and two reactor complexes at
baselines of $\sim$53~km, with a total thermal power of 36~GW. With $\sim$100k IBD
events from reactor neutrinos (about six years data-taking),
JUNO aims to determine the MH at 3$\sigma$ sensitivity.~\footnote{The MH
  determination involves two non-nested hypotheses. The statistical
  interpretation of MH sensitivity can be found in
  Ref.~\cite{Qian:2012zn,Blennow:2013oma}.} This goal in sensitivity relies on
an unprecedented 3\%/$\sqrt{E~\rm (MeV)}$ energy resolution, which requires a $\sim$80\%
photo-cathode coverage, an increase in both LS light yield
and attenuation length, and an increase in PMT quantum efficiency. In addition,
excellent control of the energy-scale uncertainty~\cite{Qian:2012xh,Li:2013zyd,Minakata:2006gq}
is crucial. 

\subsubsection{Precision Measurements of Neutrino Mixing Parameters}

In addition to determining the MH, JUNO will access four fundamental neutrino
mixing parameters: $\theta_{12}$, $\theta_{13}$, $\Delta m^2_{21}$, and $|\Delta m^2_{32}|$.
JUNO is expected to be the first experiment to observe 
neutrino oscillation simultaneously from both atmospheric and solar neutrino
mass-squared differences and will be the first experiment to observe more 
than two oscillation cycles of the atmospheric mass-squared difference.
Moreover, JUNO is expected to achieve better than 1\% precision measurements
of $\sin^22\theta_{12}$, $|\Delta m^2_{32}|$, and $\Delta m^2_{21}$,
which provides very powerful tests of the standard three-flavor neutrino model.
In particular, the precision measurement of $\sin^22\theta_{12}$ will
lay the foundation for a future sub-1\% direct unitarity test of the
PMNS matrix $U$.

The combination of short-baseline reactor neutrino
experiments (such as Daya Bay, RENO, and Double Chooz), medium-baseline
reactor neutrino experiments (such as KamLAND and JUNO), and solar
neutrino experiments (such as SNO) enable the first direct unitarity test
of the PMNS matrix~\cite{Qian:2013ora,Antusch:2006vwa}: $|U_{e1}|^2 +
|U_{e2}|^2 + |U_{e3}|^2 \stackrel{?}{=} 1$. When combined with results from
Daya Bay and SNO, JUNO's precision measurement will test this unitarity condition
to 2.5\%~\cite{Qian:2013ora}. An accurate value of $\sin^22\theta_{12}$
will also allow for testing model predictions of neutrino mass and
mixing~\cite{deGouvea:2013onf}, which could guide us towards a more
complete theory of flavor~\cite{Antusch:2013ti}.
Furthermore, the precision measurement of $\sin^22\theta_{12}$
will constrain the allowed region, in particular the minimal value, of
the effective neutrino mass
$|m_{ee}|:=|\sum U_{ei}^2 m_i|$~\cite{Dueck:2011hu,Ge:2015bfa}, to which the decay
width of neutrinoless double beta decay is proportional.

As shown in Ref.~\cite{Nunokawa:2005nx},
the measurements of muon neutrino disappearance and electron antineutrino
disappearance are effectively measuring $|\Delta m^2_{\mu\mu}|$ and $|\Delta m^2_{ee}|$
(two different combinations of $\Delta m^2_{31}$ and $\Delta m^2_{32}$), respectively.
When combined with the precision $|\Delta m^2_{\mu\mu}|$ measurements from muon
neutrino  disappearance, the precision measurement of $|\Delta m^2_{ee}|$
will allow a test of the sum rule $\Delta m^2_{13} + \Delta m^2_{21} +
\Delta m^2_{32} = 0$, which is an important prediction of the $\nu$SM, and
will reveal additional information regarding the neutrino MH.

Using the convention of Ref.~\cite{Qian:2012xh}, we have
$|\Delta m^2_{ee,\mu\mu}| \approx |\Delta m^2_{23}| \pm \Delta m^2_{\phi~ee,\mu\mu}/2$,
in which the plus/minus sign depends on the MH. Since
$\Delta m^2_{\phi~ee}$ ($\sim$10$^{-4}$~eV$^{2}$) is larger than $\Delta m^2_{\phi~\mu\mu}$ 
($\sim$5$\times 10^{-5}$ eV$^{2}$), the precision measurements of both
$|\Delta m^2_{\mu\mu}|$ and $|\Delta m^2_{ee}|$ would provide new information
about the neutrino MH~\cite{Nunokawa:2005nx,Minakata:2006gq}. Furthermore,
the comparison of $\Delta m^2_{32}$ extracted from the reactor electron antineutrino
disappearance and that extracted from the accelerator muon neutrino disappearance
can be a stringent test of CPT symmetry~\cite{deGouvea:2017yvn}.

In addition to the sub-percent precision measurements of solar-sector oscillation
parameters, the atmospheric mass-squared difference, and the MH determination,
the 20-kton target mass offers a rich physics program of proton decay,
geoneutrinos, supernova neutrinos, and many exotic neutrino physics
topics~\cite{An:2015jdp}. For the
$p\rightarrow \bar{\nu} + K^+$ channel, which is favored by a number of supersymmetry
  grand unified theories~\cite{Nath:2006ut}, JUNO would be competitive
  relative to
Super-K and to-be-built experiments such as DUNE~\cite{Adams:2013qkq}
and Hyper-K~\cite{Kearns:2013lea}. Besides JUNO, there is a proposal
in Korea (RENO-50)~\cite{Kim:2014rfa} that has a similar physics reach.

Reactor neutrinos have played crucial roles in the discoveries of the
  non-zero neutrino mass and mixing and the establishment of the standard
  three-neutrino framework. While the current-generation reactor
  experiments continue to improve the precision of $\theta_{13}$ and
  $|\Delta m^2_{ee}|$, the next-generation reactor experiments will aim to
  determine the neutrino MH and precision
  measurements of neutrino mass and mixing, which are crucial steps towards
  completing the neutrino standard model.

\section{The Reactor Antineutrino Anomaly and Search for a Light Sterile Neutrino}\label{sec:sterile}

The majority of neutrino oscillation data can be successfully
explained by the three-neutrino framework described in Sec.~\ref{sec:nu_mixing}. Despite
this success, the exact mechanism by which neutrinos acquire their mass remains
unknown. In addition, the fact that the mass of electron neutrino is at least 5 orders
of magnitude smaller than that of electron~\cite{Otten:2008zz} also presents a puzzle. 
The possible existence of additional neutrino flavors beyond the known three
may provide a natural explanation of the smallness of neutrino mass~\cite{King:2003jb}.

In accord with precision electroweak measurements~\cite{ALEPH:2005ab},
these additional neutrinos are typically considered to be sterile~\cite{ponte2},
i.e., non-participating in any fundamental interaction of the standard model, which leaves
no known mechanism to detect them directly. Nonetheless, an unambiguous signal of their
existence can be sought in neutrino oscillation experiments, where sterile neutrinos
could affect the way in which the three active neutrinos oscillate if they mix with sterile
neutrinos.

Besides theoretical motivations in searching for sterile neutrinos,
several experimental anomalies could also be explained by additional light sterile
neutrinos at the $\sim$eV mass scale. Among them are the LSND~\cite{Aguilar:2001ty} and
MiniBooNE~\cite{miniboone,Aguilar-Arevalo:2018gpe} anomalies for (anti-)$\nu_{\mu}\rightarrow $(anti-)$\nu_e$
oscillation and the anomalies observed by GALLEX~\cite{Kaether:2010ag} and SAGE~\cite{Abdurashitov:2009tn}
when calibrated $\nu_e$ sources ($^{51}$Cr for GALLEX, $^{51}$Cr and $^{37}$Ar for SAGE)
produced lower rates of detected $\nu_e$ than expected. 


The reactor antineutrino anomaly~\cite{anom} suggests $\bar{\nu}_e \rightarrow \bar{\nu}_e$
disappearance oscillation from an observed deficit in the measured
antineutrino events relative to the expectation based on the latest reactor
antineutrino flux calculations~\cite{Huber:2011wv,Mueller:2011nm}.
In this section, we focus our discussion on the search for a light sterile neutrino in
reactor experiments and 
the reactor antineutrino anomaly. For other recent reviews on the search for light
sterile neutrinos, see Refs.~\cite{Conrad:2016sve,Conrad:2013mka}.

\subsection{Theoretical Framework for a Light Sterile Neutrino}\label{sec:frame_sterile}
Adding one light sterile neutrino into the current three-neutrino model would
lead to an expansion of the 3$\times$3 unitary matrix $U$ (Eq.~\ref{eq:pmns_matrix})
into a $4\times4$ unitary matrix:
\begin{equation}\label{eq:pmns_matrix_sterile}
  \left( \begin{array}{c} \nu_{e} \\ \nu_{\mu} \\ \nu_{\tau} \\ \nu_{s}\end{array} \right) = 
\left ( \begin{array}{cccc} U_{e1} & U_{e2} & U_{e3} & U_{e4}\\
	U_{\mu1} & U_{\mu2} & U_{\mu3} & U_{\mu4}\\
	U_{\tau1} & U_{\tau2} & U_{\tau3} & U_{\tau4} \\
        U_{s1}   & U_{s2}    & U_{s3}    & U _{s4}
\end{array} \right) \cdot \left( \begin{array}{c} \nu_{1} \\ 
\nu_{2} \\ \nu_{3} \\ \nu_4 \end{array} \right),
\end{equation}
where subscript $s$ stands for the added light sterile neutrino. This expansion
would introduce three additional mixing angles $\theta_{14}$, $\theta_{24}$,
$\theta_{34}$ and two additional phases $\delta_{24}$, $\delta_{34}$.
Similar to Eq. (\ref{eq:3x3_rot}), the matrix $U$ can be parameterized~\cite{Harari:1986xf}
as:
\begin{eqnarray}\label{eq:4x4_rot}
  U = R_{34} \left(c_{34},s_{34},\delta_{34} \right) \cdot R_{24}\left( c_{24}, s_{24}, \delta_{24} \right) \cdot R_{14}\left(c_{14},s_{14},0 \right) \nonumber \\
  \cdot R_{23}\left(c_{23},s_{23},0 \right) \cdot R_{13} \left( c_{13}, s_{13}, \delta_{CP} \right) \cdot
  R_{12}\left( c_{12}, s_{12}, 0 \right),
\end{eqnarray}
where $R$s are $4\times4$ rotation matrices. For example, Eq. (\ref{eq:rot_13}) is
expanded to 
\begin{equation}
R_{13} = 
\left ( \begin{array}{cccc} c_{13} & 0 & s_{13} \cdot e^{-i\delta_{CP}} & 0 \\
  0 & 1 & 0 & 0 \\
  -s_{13} \cdot e^{i\delta_{CP}} & 0 & c_{13} & 0 \\
  0   & 0    & 0    & 1
\end{array} \right).
\end{equation}

Given Eq. (\ref{eq:pmns_matrix_sterile}), the neutrino oscillation probabilities can be
calculated following the procedure described in Sec.~\ref{sec:nu_mixing}.
Following Eq. (\ref{eq:osc_dis}), the neutrino oscillation probability is 
written as:
\begin{equation}
  P_{\nu_l\rightarrow \nu_{l'}}(L/E) =  \left |\sum_{i=1}^4 U_{li}U^{*}_{l'i}e^{-i(m_{i}^2/2E)L} \right | ^2.
\end{equation}
More specifically, we have 
\begin{eqnarray}\label{eq:PueFullnoCP}  
  P_{\nu_\mu\rightarrow\nu_e}(L/E) &=&  \left |\sum_{i=1}^4 U_{\mu i}U^{*}_{ei}e^{-i(m_{i}^2/2E)L} \right | ^2,  \\
  P_{\nu_\mu\rightarrow \nu_\mu}(L/E) &\equiv &  P_{\bar{\nu}_\mu\rightarrow \bar{\nu}_\mu}(L/E) \nonumber \\ 
  & =& \,1-4\sum_{k>j}|U_{\mu k}|^2|U_{\mu j}|^2\sin^2\left(\frac{\Delta m^2_{kj}L}{4E}\right), \nonumber \\ 
  P_{\overline\nu_e\rightarrow\overline\nu_e}(L/E) &\equiv & P_{\nu_e\rightarrow \nu_e}(L/E) \nonumber \\
  & = & \,1-4\sum_{k>j}|U_{ek}|^2|U_{ej}|^2\sin^2\left(\frac{\Delta m^2_{kj}L}{4E}\right).  \nonumber
\end{eqnarray}
Given Eq. (\ref{eq:4x4_rot}), in which the definition of mixing angles
depends on the specific ordering of the matrix
multiplication, we have
\begin{eqnarray}\label{eq:DisapToApp}
|U_{e4}|^2 &=& \,\,s^2_{14}, \nonumber \\
|U_{\mu4}|^2 &=& \,\,s^2_{24}c^2_{14}, \nonumber \\
4|U_{e4}|^2|U_{\mu4}|^2 &=&\,\, 4 s^2_{14}c^2_{14}s^2_{24}\equiv \sin^22\theta_{\mu e}. 
\end{eqnarray}
The last line in Eq. (\ref{eq:DisapToApp}) is crucial in the region
where $\Delta m^2_{41}$ $\gg$ $|\Delta m^2_{32}|$ and for short baselines
($\Delta_{32} \equiv \frac{\Delta m^2_{32}L}{4E} \sim 0$). 
Equation (\ref{eq:PueFullnoCP}) can then be simplified to
  \begin{eqnarray}\label{eq:PueApproxMtrx}
    P_{\nu_\mu\rightarrow\nu_e}(L/E) &\approx& P_{\bar{\nu}_\mu\rightarrow \bar{\nu}_e}(L/E) 
    \approx  \sin^22\theta_{\mu e} \sin^2 \Delta_{41} , \nonumber \\
    P_{\nu_\mu\rightarrow \nu_\mu}(L/E) &\equiv& P_{\bar{\nu}_\mu\rightarrow \bar{\nu}_\mu}(L/E)  \nonumber \\
    &\approx&  1 - \sin^2 2\theta_{24} \sin^2\Delta_{41} \nonumber \\
    & & -\sin^22\theta_{23} \cos2\theta_{24} \sin^2 \Delta_{31} , \nonumber \\
    P_{\overline\nu_e\rightarrow\overline\nu_e}(L/E) & \equiv & P_{\nu_e\rightarrow\nu_e}(L/E) \nonumber \\
    &\approx & 1 - \sin^2 2\theta_{14} \sin^2 \Delta_{41} \nonumber \\
    & & - \sin^22\theta_{13} \sin^2 \Delta_{31},
  \end{eqnarray}
  in which the values of additional CP phases are irrelevant.
  This is no longer true if there are two sterile neutrino flavors.
We kept the $\sin^2\Delta_{31}$ terms in the disappearance formulas, since
they are important in some of the disappearance experiments to be discussed
in the next section. We should note that at a given $\Delta_{41}$, the
three oscillations in Eq. (\ref{eq:PueApproxMtrx}) depend on only two unknowns,
namely, $\theta_{14}$ and $\theta_{24}$. Hence, from a measurement of any
two oscillations, the third one can be deduced.

\subsection{Search for a Light Sterile Neutrino from Reactor Experiments}\label{sec:sterile_exp}

In this section, we review the searches for a light sterile neutrino from
the Bugey-3~\cite{Declais:1994su}, Daya Bay~\cite{An:2014bik,An:2016luf},
NEOS~\cite{Ko:2016owz}, DANSS~\cite{Alekseev:2018efk}, PROSPECT~\cite{Ashenfelter:2018iov},
and STEREO~\cite{Almazan:2018wln} experiments.

\begin{figure}[htp]
  \begin{centering}
    \includegraphics[width=0.45\textwidth]{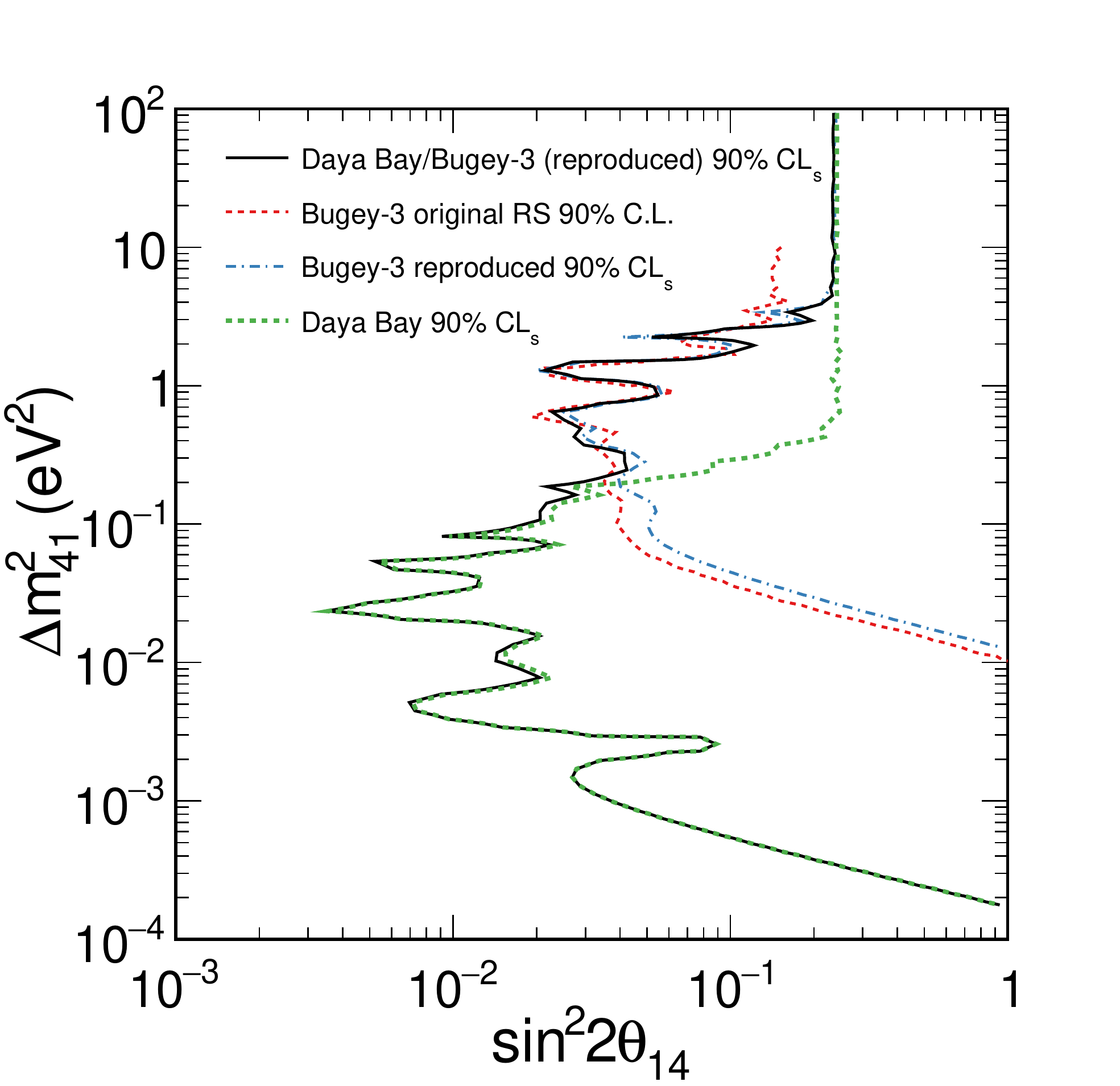}
    \par\end{centering}
    \caption{\label{fig:dyb_sterile_search}    
      Excluded regions for the combined Daya Bay and reproduced
      Bugey-3 results~\cite{Adamson:2016jku}.
      The region to the right of the curve is excluded at the 90\% CL$_s$.
      The original Bugey-3 result~\cite{Declais:1994su} using a raster
      scan (RS)~\cite{Lyons:2014kta}, the reproduced Bugey-3 result
      with adjusted fluxes, and Daya Bay result~\cite{An:2016luf} are
      shown as well.}
\end{figure}

The Bugey-3 experiment was performed in the early 1990s at the Bugey
Nuclear Power Plant located in the Saint-Vulbas commune in France, about 65 km from
the Swiss border. The main goal was to search for neutrino
oscillation. In this experiment, two LS detectors having a total of three detector modules
measured $\bar{\nu}_e$ generated from two reactors
(reactor 4 and 5) at three different baselines (15~m, 40~m, and 95~m)~\cite{Declais:1994su}.
Each detector module was a 600-liter $^6$Li-doped LS having dimensions of 
122$\times$62$\times$85 cm$^3$~\cite{Abbes:1996nc}.
Each module was optically divided into independent cells having dimensions of
$8\times8\times85$ cm$^3$. Every cell was instrumented on each side by a
PMT. The pressurized water reactor was approximated as a
cylinder of $\sim$1.6~m radius and $\sim$3.7~m height. 
Bugey-3 detected IBD interactions with recoil neutrons captured
by $^6$Li (see Table~\ref{tab:IBD_neutron_capture}).
The energy resolution was about 6\% at 4.2~MeV. The ratios of the measured positron
energy spectrum to the Monte Carlo prediction at all three distances did not
show any signature of oscillation, and exclusion contours were made in the phase space
of $\sin^22\theta_{14}$ and $\Delta m^2_{41}$ (see
Fig.~\ref{fig:dyb_sterile_search}).

\begin{figure}[htp]
  \begin{centering}
    \includegraphics[width=0.43\textwidth]{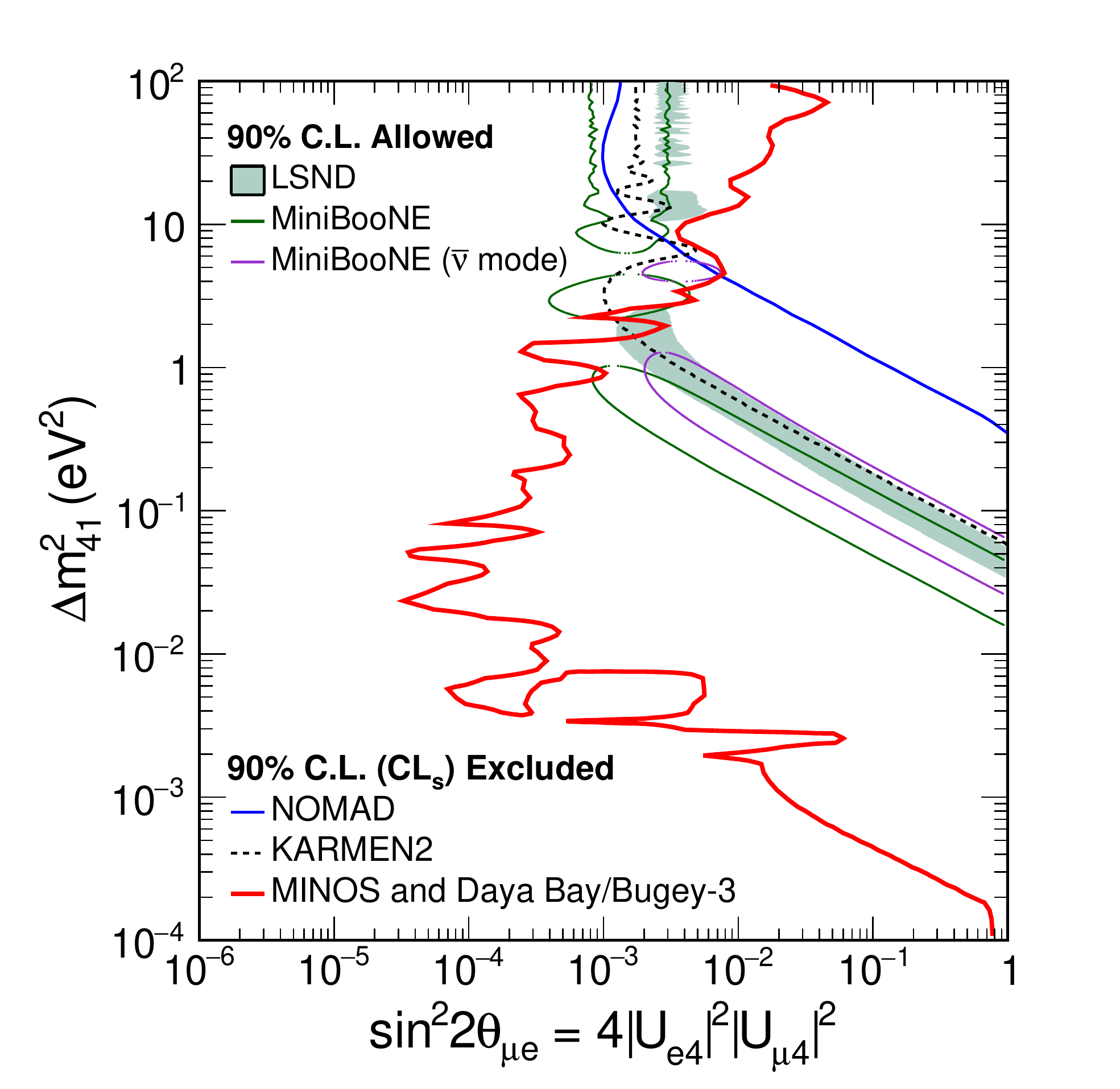}
    \par\end{centering}
    \caption{\label{fig:combined_sterile_search}
      MINOS and Daya Bay/Bugey-3 combined 90\% CL$_s$ limit on
      $\sin^22\theta_{\mu e}$~\cite{Adamson:2016jku}
      are compared to the LSND and MiniBooNE 90\% C.L. allowed regions.
      Regions of parameter space to the right of the red contour are excluded.
      The regions excluded at 90\% C.L. by KARMEN2~\cite{Armbruster:2002mp} 
      and NOMAD~\cite{Astier:2003gs} are also shown. }
\end{figure}


The main motivation of the Daya Bay experiment (described in Sec.~\ref{sec:DYB})
was to perform precision measurements of $\sin^22\theta_{13}$ and $\Delta m^2_{ee}$.
Given its unique configuration of multiple baselines to three groups of nuclear reactors,
the Daya Bay experiment also allowed a search for sterile neutrinos through
relative spectral distortions obtained at three experimental sites. With a baseline longer
than that of Bugey-3, Daya Bay was sensitive to the sterile neutrino
mixing parameter $\sin^22\theta_{14}$ at smaller $\Delta m^2_{41}$ values.

Similar to that of Bugey-3, no oscillation signature attributable to an additional sterile
neutrino was found, and exclusion contours were set in Refs.~\cite{An:2014bik,An:2016luf}
using the Feldman--Cousins~\cite{Feldman:1997qc} and CL$_s$~\cite{Read:2000ru,Junk:1999kv} approaches.
Figure~\ref{fig:dyb_sterile_search} shows the combined results of Daya Bay and
Bugey-3~\cite{Adamson:2016jku} using the Gaussian CL$_s$ method~\cite{Qian:2014nha}.
The exclusion contour combining both experiments covered about 5 orders
of magnitude in $\Delta m^2_{41}$. This result was further combined with
results from the MINOS experiment~\cite{MINOS:2016viw} to constrain the anomalous
(anti-)$\nu_{\mu}\rightarrow$(anti-)$\nu_e$ oscillation~\cite{Adamson:2016jku} using
the CL$_s$ method~\cite{Read:2000ru,Junk:1999kv,Read:2002hq}. As shown in
Fig.~\ref{fig:combined_sterile_search}, the combined result from
Daya Bay, Bugey-3, and MINOS excluded most of regions allowed by LSND
and MiniBooNE. Together with the search results from the IceCube experiment
using the matter effect~\cite{TheIceCube:2016oqi}, this result
significantly reduced the allowed parameter space for future searches. 


The NEOS~\cite{Ko:2016owz} experiment searched for a light sterile neutrino at
reactor unit 5 (2.8-GW thermal power) located at the Hanbit nuclear power complex
in Yeonggwang, South Korea, which is the same reactor complex used by the
RENO experiment~\cite{RENO:2015ksa}. The active core size was 3.1~m in diameter
and 3.8~m in height.  In this experiment, the search was performed with 1~ton
of 0.5\% Gd-loaded LS at a distance of about 24~m from
the reactor core. The LS was contained in a horizontal cylindrical
stainless-steel tank of 103~cm in diameter and 121~cm in length. Each end of the
target vessel was exposed to 19 8-inch PMTs that were packed inside mineral oil.
The energy response of the NEOS detectors was calibrated with various radioactive
sources. The energy resolution was measured to be about 5\% at 1~MeV. With 20-m
m.w.e. overburden and active muon veto counters made from 5-cm thick
plastic scintillators surrounding the detector, NEOS achieved a 22:1
signal-to-background ratio after all cuts.

\begin{figure}[htp]
  \begin{centering}
    \includegraphics[width=0.48\textwidth]{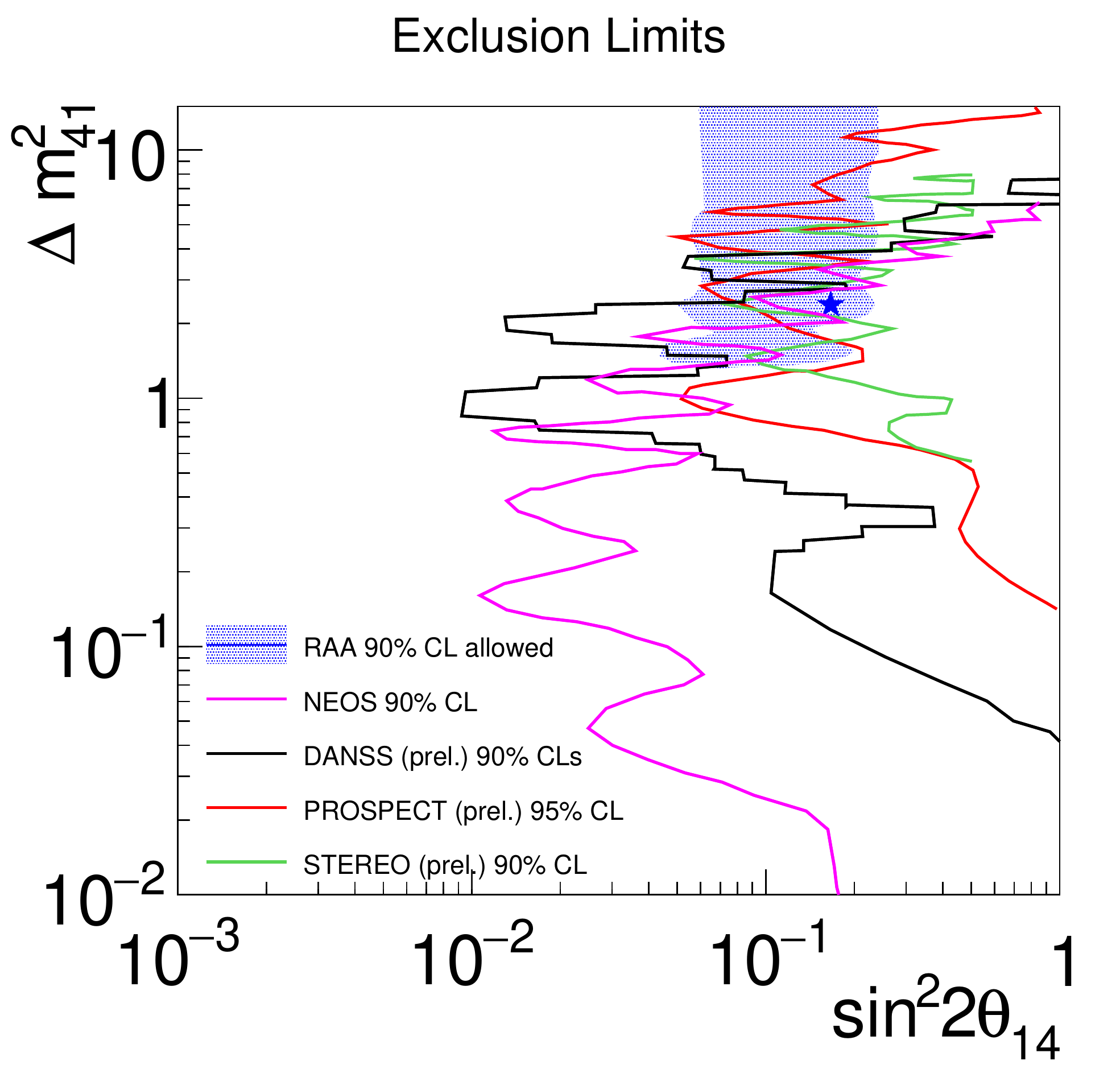}
    \par\end{centering}
    \caption{\label{fig:preliminary_sterile_search}
      Exclusion limits reported at the Neutrino 2018 conference~\cite{neutrino2018}
      from the new generation of very--short--baseline reactor neutrino
      experiments. The results from DANSS~\cite{Alekseev:2018efk} (90\% CL$_s$), PROSPECT~\cite{Ashenfelter:2018iov} (95\% CL),
      and STEREO~\cite{Almazan:2018wln} (90\% CL) are preliminary. 
      The allowed region from the reactor antineutrino anomaly (RAA) is
      compared. The star represents the best-fit point. 
      }
\end{figure}

With a single detector, NEOS relied on
external constraints on the neutrino spectrum to search for spectral distortion.
In comparison with the neutrino spectrum measured from the Daya Bay
experiment~\cite{An:2015nua}, NEOS observed no significant spectral distortion
caused by oscillation, and the exclusion limit was set using the raster-scan
method~\cite{Lyons:2014kta}. As shown in Fig.~\ref{fig:preliminary_sterile_search},
  stringent exclusion limits were set in the mass range of $0.2$ eV$^2$ $<\Delta m^2_{41}<$
  3 eV$^2$.

\begin{table*}[h] 
\caption{Major parameters of very--short--baseline reactor neutrino
  experiments that are in operation, under construction, or being planned.
  Diameter, radius, and height are indicated by d, r, and h, respectively. For the energy resolution, the unit
  of the energy \lq{E}\rq~is MeV. For signal-to-background ratios, the achieved performances (A.)
are separated from the expected performance (E.). \lq{Seg.}\rq~stands for segmentation. }  
\medskip
\renewcommand{\arraystretch}{1.1} \centering 
\begin{tabular}{|c|c|c|c|c|c|c|}
\hline 
Experiment & Reactor & Distance & Mass & Resolution & Seg. & S/B \\\hline
DANSS & LEU 3.1 GW$_{th}$& 10.7-12.7 m& 1.1 Ton & 17\%/$\sqrt{E}$ & 2D & 0.6 (A.) \\
Ref.~\cite{Alekseev:2018efk,Alekseev:2016llm} & 1.5 m r $\times$ 3.5 m h& & & & & \\\hline
NEOS                         & LEU 2.8 GW & 24 m & 1 Ton & 5\%/$\sqrt{E}$ & 1D & 21 (A.) \\
Ref.~\cite{Ko:2016owz} & 3.1 m d $\times$ 3.8 m h & & & & & \\\hline
NEUTRINO-4                   & HEU 100 MW & 6-12 m& 0.3 Ton & N/A & 2D & 0.25-0.3 (A.)\\
Ref.~\cite{Serebrov:2012sq,Serebrov:2017nxa} &   0.35$\times$0.42$\times$0.42 m$^{3}$ &       &         & &    & \\\hline
Nucifer & HEU 70 MW           & 7.2 m& 0.6 Ton & 10\%/$\sqrt{E}$& 1D & 0.06 (A.) \\
Ref.~\cite{Cucoanes:2012jv,Boireau:2015dda} & 0.3 m r$\times$ 0.6 m h& & & & &\\\hline
PROSPECT & HEU 85 MW  & 7-12 m& 1.5 Ton & 4.5\%/$\sqrt{E}$& 2D & 0.8 (A.)\\
Ref.~\cite{Ashenfelter:2015uxt,Ashenfelter:2018zdm} & 0.2 m r $\times$0.5 m h & & & & & \\\hline
STEREO          & HEU 58~MW & 8.9-11.1 m& 1.6 Ton & 8\%/$\sqrt{E}$& 2D & 0.9 (A.)\\
Ref.~\cite{Manzanillas:2017rta,Allemandou:2018vwb}  & 0.4 m d $\times$ 0.8 m h & & & & &\\\hline
SOLID & HEU 75 MW  & 6-9 m& 1.6 Ton & 14\%/$\sqrt{E}$ & 3D & 1.0 (E.) \\
Ref.~\cite{Abreu:2017bpe,Abreu:2018pxg}& 0.25 m r & & & & & \\\hline
NuLAT & HEU 20 MW  & 4 m& 1 Ton& 4\%/$\sqrt{E}$& 3D & 3 (E.)\\
Ref.~\cite{Lane:2015alq}  &  1 m d   & & & & &\\\hline
CHANDLER & HEU 75 MW & 5.5-10 m & 1 Ton & 6\%/$\sqrt{E}$& 3D & 3 (E.)\\
Ref.~\cite{chandler}  &  0.25 m r   & & & & &\\
\hline
\end{tabular}\label{tab:sb_exp}
\end{table*}

A new generation of very--short--baseline reactor neutrino experiments
  to search for an eV-mass-scale sterile neutrino
  are under construction or in operation. Table~\ref{tab:sb_exp} summarizes the
  major parameters of these experiments. The primary challenges for
  these experiments include the cosmogenic backgrounds resulting from the
  limited amount of overburden, and reactor-related backgrounds caused by the
  proximity of the detector to the reactor core. 
  A segmented detector design is generally required to achieve a
  desired signal-to-noise ratio.

  The sensitivity of a light sterile neutrino typically depends
  on the distance between the detector and the reactor core,  statistics (target mass,
  reactor power, and signal to noise ratio), sizes of reactor core and detector (smearing
  in distance), and energy resolution (smearing in energy). A comparison of
  measurements at different distances is crucial for finding evidence of
  a sterile neutrino.
  At the Neutrino 2018 conference~\cite{neutrino2018},
  three of these experiments: DANSS~\cite{Alekseev:2018efk}, PROSPECT~\cite{Ashenfelter:2018iov}, and
  STEREO~\cite{Almazan:2018wln}, reported preliminary exclusion limits shown in
  Fig.~\ref{fig:preliminary_sterile_search}.

The DANSS experiment is located at the Kalinin nuclear power plant in
  Russia. The detector was placed in a room below the reactor with an
  overburden of $\sim$50 m.w.e. 
  Polystyrene-based plastic scintillator strips (1~cm$\times$4~cm$\times$1~m) with a thin
  Gd-containing coating were arranged with two orientations in different layers. A total
  of 2500 strips were coupled to 2500 silicon photomultipliers and 50
  PMTs~\cite{Alekseev:2016llm}. Data were taken at three vertical
  detector positions with baseline varying from 10.7~m to 12.7~m.
  With about 1 million IBD events after background subtraction, DANSS
  observed no significant spectral distortion when comparing the positron
  energy spectrum measured at different detector
  positions~\cite{Alekseev:2018efk}. As shown in
  Fig.~\ref{fig:preliminary_sterile_search}, DANSS excluded the best-fit point of the RAA with a confidence level higher than 5$\sigma$.

The PROSPECT experiment is located at the 85-MW high flux isotope
  reactor (HFIR) at Oak Ridge National Laboratory in the United States.
  With a compact
  reactor core and short baselines (7~m -- 9~m), PROSPECT had good
  sensitivities for $\Delta m^2_{41}$ above 3~eV$^2$.
  The detector consisted of 154 segments (119~cm$\times$15~cm$\times$15~cm) filled with $^6$Li-doped
  EJ-309 LS. Each segment was read from two PMTs at each end.
  The $^6$Li-doped LS
  allowed a good pulse shape discrimination for the delayed signal~\cite{Ashenfelter:2018cli}, which was
  essential for rejecting cosmogenic and reactor-related backgrounds. Using multiple layers of shielding, PROSPECT achieved an overall signal to background
  ratio ($\sim$0.8).
  With a total 25k IBD events after background subtraction, energy spectra
  from six baselines were compared. No oscillation signal
  was observed~\cite{Ashenfelter:2018iov} and exclusion limits were set.
  As shown in Fig.~\ref{fig:preliminary_sterile_search}, the best-fit point of
  the RAA was excluded by PROSPECT with a confidence level of 2.2$\sigma$.

The STEREO experiment is located at a 58-MW research reactor at
  Institut Laue--Langevin (ILL) in Grenoble,
  France. Similar to PROSPECT, the research reactor core is compact
  and the baseline ranges from 9~m to 11~m.
  The target (dimensions 2.2~m$\times$0.9~m$\times$1.2~m) was longitudinally divided into six identical
  and optically separated cells filled with Gd-loaded LS.
  With about 15~m.w.e. overburden, the
  STEREO detector was further shielded by a combination of lead, polythylene,
  and boron-loaded rubber. A water
  Cerenkov muon veto was installed on top of the detector.
  About 400 IBD events were detected per day when reactor was on
  and a signal to background ratio of 0.9 was achieved~\cite{Allemandou:2018vwb}. With 66 (138) days of
  reactor on (off) data, no oscillation signal was observed when the measured spectra from six
  cells were compared~\cite{Almazan:2018wln}.
  As shown in Fig.~\ref{fig:preliminary_sterile_search}, the best-fit point of
  the RAA was excluded by STEREO with a confidence level of 97.5\%.

In the next few years, more precise results are expected from the new generation of
  very--short--baseline reactor
  neutrino experiments. Together with searches for a light sterile neutrino
  with atmospheric neutrinos~\cite{TheIceCube:2016oqi},
accelerator neutrinos~\cite{Antonello:2015lea}, pion/kaon decay-at-rest (DAR) neutrinos,
and radioactive neutrino sources~\cite{Borexino:2013xxa}, these reactor
neutrino experiments are expected to give a definitive answer regarding
the existence of 
a eV-mass-scale light sterile neutrino.

\subsection{Reactor Antineutrino Anomaly}~\label{sec:RAA}

\begin{figure}[btp]
  \begin{centering}
    \includegraphics[width=0.45\textwidth]{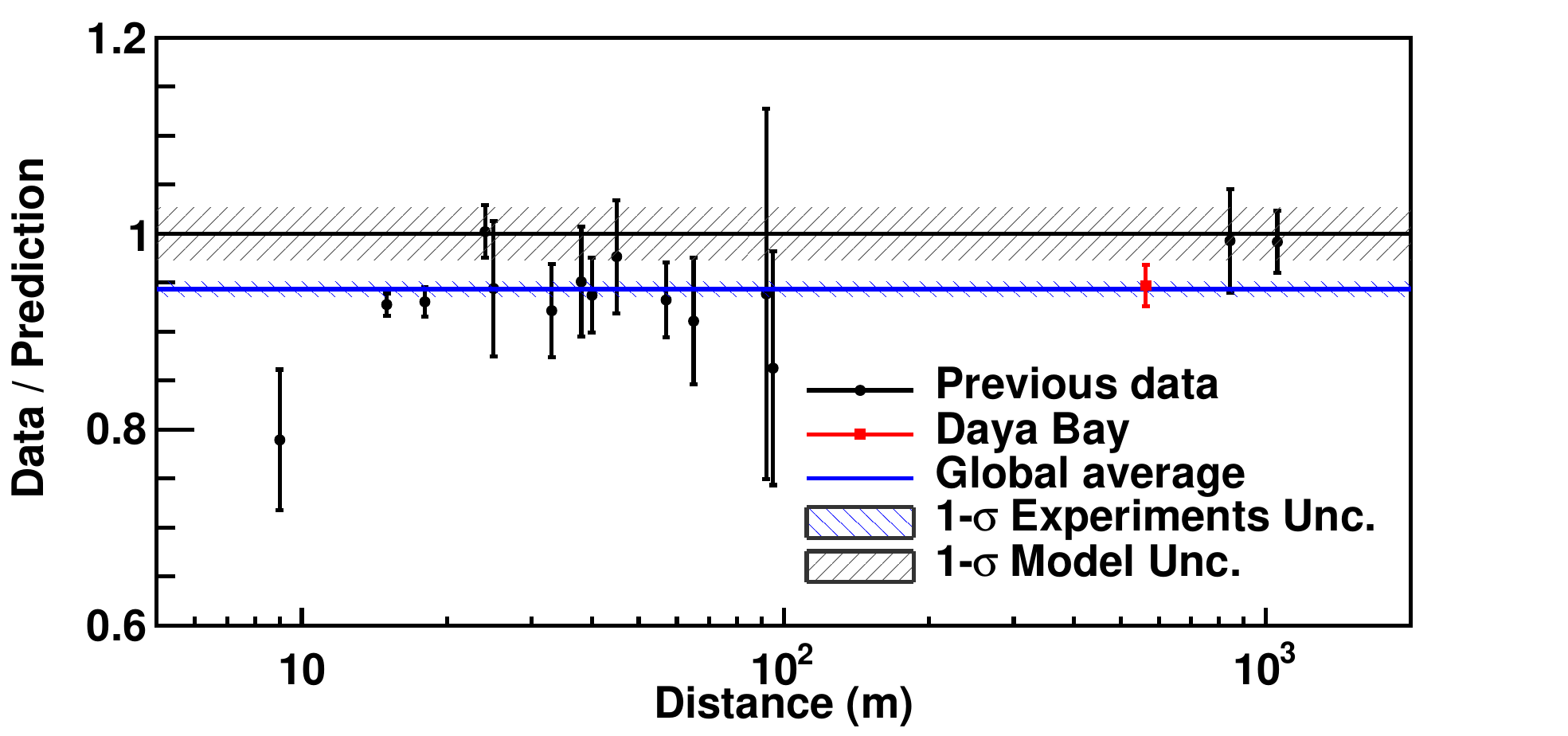}
    \par\end{centering}
    \caption{\label{fig:dyb_global_deficit}
      The measured reactor $\bar{\nu}_e$ rate as a function of the distance
      from the reactor, normalized to the theoretical prediction of the
      Huber--Mueller model~\cite{Mueller:2011nm,Huber:2016xis},
      taken from Ref.~\cite{An:2015nua}. The rate is
      corrected for three-flavor neutrino oscillation at each baseline. The blue shaded
      region represents the global average and its 1$\sigma$ uncertainty. The 2.7\%-model
      uncertainty is shown as a band around unity. Measurements at the same
      baseline are combined for clarity. The Daya Bay measurement is shown at the
      flux-weighted baseline (573 m) of the two near halls.}
\end{figure}

The reactor antineutrino anomaly~\cite{anom} refers to a deficit of the measured antineutrino
rate in short-baseline reactor experiments ($L < 2$ km) with respect to the latest
calculations of the antineutrino flux~\cite{Huber:2011wv,Mueller:2011nm}, which are
about 5\% higher than previous
calculations~\cite{VonFeilitzsch:1982jw,Schreckenbach:1985ep,Hahn:1989zr,Vogel:1980bk}.
The initial calculation of this deficit in Ref.~\cite{anom} is biased towards
a larger value by about 1.5\%~\cite{Zhang:2013ela} because of an improper treatment
of flux uncertainties in the covariance matrix, as demonstrated in
Ref.~\cite{DAgostini:1993arp}. Figure~\ref{fig:dyb_global_deficit}
displays the updated global fit, showing a data-over-prediction
ratio of $0.943\pm0.008$, excluding uncertainties associated with
the flux prediction. 

\begin{figure}[htp]
  \begin{centering}
    \includegraphics[width=0.48\textwidth]{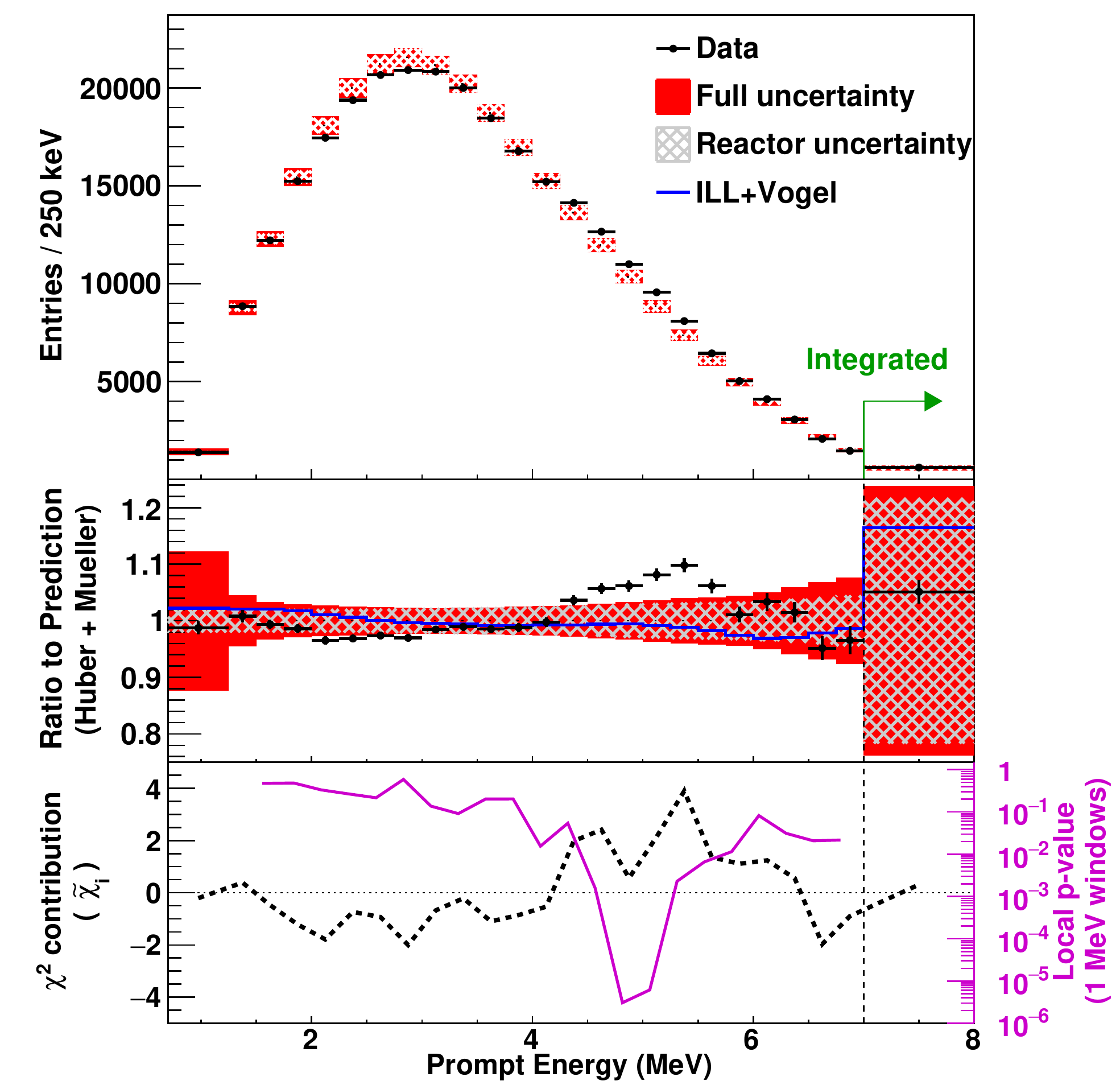}
    \par\end{centering}
    \caption{\label{fig:dyb_bump}
      Predicted and measured prompt-energy spectra, taken from
      Ref.~\cite{An:2015nua}. The prediction is based on the Huber--Mueller
      model~\cite{Mueller:2011nm,Huber:2016xis} and normalized
      to the number of measured events. The highest energy bin contains
      all events above 7~MeV. The gray hatched and red filled bands represent
      the square-root of diagonal elements of the covariance matrix
      for the reactor-related and the full (reactor, detector, and background)
      systematic uncertainties, respectively. The error bars on the
      data points represent the statistical uncertainty.
      The ratio of the measured prompt-energy spectrum to the predicted spectrum
      (Huber--Mueller model) is shown in the middle panel. The blue curve
      shows the ratio of the
      prediction based on the ILL+Vogel~\cite{Vogel:1980bk,VonFeilitzsch:1982jw,Schreckenbach:1985ep,Hahn:1989zr}
      model to that based on the Huber--Mueller model.
      The defined $\chi^2$ distribution
      of each bin (black dashed curve) and local p-values for 1-MeV energy
      windows (magenta solid curve) are shown in the bottom panel. }
\end{figure}

\begin{figure*}[htp]
  \begin{centering}
    \includegraphics[width=0.5\textwidth,angle=-90,trim={0 0 5cm 0},clip]{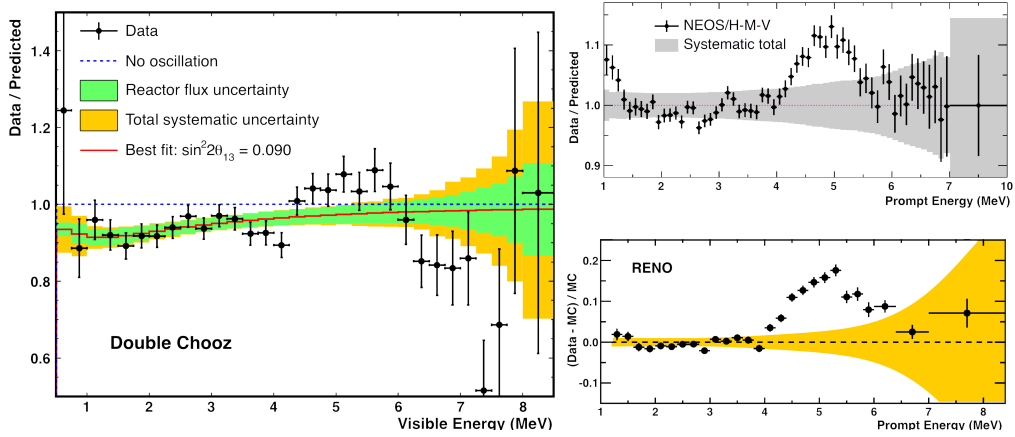}
    \par\end{centering}
    \caption{\label{fig:other_bumps} Observations of the 5-MeV prompt energy
      excess with respect to the model prediction~\cite{Huber:2011wv,Mueller:2011nm}
      from RENO~\cite{RENO:2015ksa}, Double Chooz~\cite{Abe:2014bwa},
      and NEOS~\cite{Ko:2016owz}.}
\end{figure*}

The calculated deficit cannot be explained by the quoted uncertainties of the reactor
flux model~\cite{Huber:2011wv,Mueller:2011nm}, which is around 2\%. One potential
explanation of this deficit is the existence of a sterile neutrino with its corresponding
mass eigenstate heavier than or equal to $\sim$1~eV. Recently, the foundation of
this explanation was challenged by authors of Ref.~\cite{anom2}, who carefully
examined the flux spectrum calculation and concluded that the uncertainties of the
flux calculation should be larger than 5\%. Their conclusion was supported by the
recent measurements of the reactor neutrino energy spectrum from the Daya
Bay~\cite{An:2015nua}, RENO~\cite{RENO:2015ksa}, Double Chooz~\cite{Abe:2014bwa}, and
NEOS~\cite{Ko:2016owz} experiments. Figure~\ref{fig:dyb_bump} shows the measured prompt
energy spectrum from Daya Bay~\cite{An:2015nua} in comparison with the model prediction
and its associated uncertainties.

An excess between the 4~MeV and 6~MeV prompt energy
beyond the model uncertainties can be clearly seen, which indicates an
underestimation of the model uncertainties. Taking into account the entire
energy range, this result disfavors the model prediction~\cite{Huber:2011wv,Mueller:2011nm}
at about 2.6$\sigma$. For the 2-MeV window between 4~MeV and 6~MeV, the p-value in testing the
compatibility between the measurement and calculation reaches 5$\times10^{-5}$, corresponding
to a 4.0$\sigma$ deviation.

Such an excess having a similar degree of deviation
was also observed when compared with the ILL+Vogel~\cite{Vogel:1980bk,VonFeilitzsch:1982jw,Schreckenbach:1985ep,Hahn:1989zr}
model calculation. Figure~\ref{fig:other_bumps}
compiles the observations of this excess from recent reactor neutrino experiments:
RENO~\cite{RENO:2015ksa}, Double Chooz~\cite{Abe:2014bwa}, and NEOS~\cite{Ko:2016owz}.
In addition, a re-analysis of positron spectrum from the G\"{o}sgen experiment, which was
  performed with a nuclear power plant at Switzerland in the 1980's~\cite{Zacek:1986cu}, also revealed
  a similar excess~\cite{Zacek:2018bij}.
The observation of this 5-MeV prompt energy excess has motivated many studies attempting to explain
its origin (See ~\cite{Huber:2016xis,Dwyer:2014eka,Hayes:2015yka,Sonzogni:2016yac},
among others).  At the moment, the exact origin of the 5-MeV prompt energy excess
is still not clear. Nevertheless, it indicates that the original 2\% quoted
model uncertainty was underestimated.

\begin{figure}[htp]
  \begin{centering}
    \includegraphics[width=0.45\textwidth]{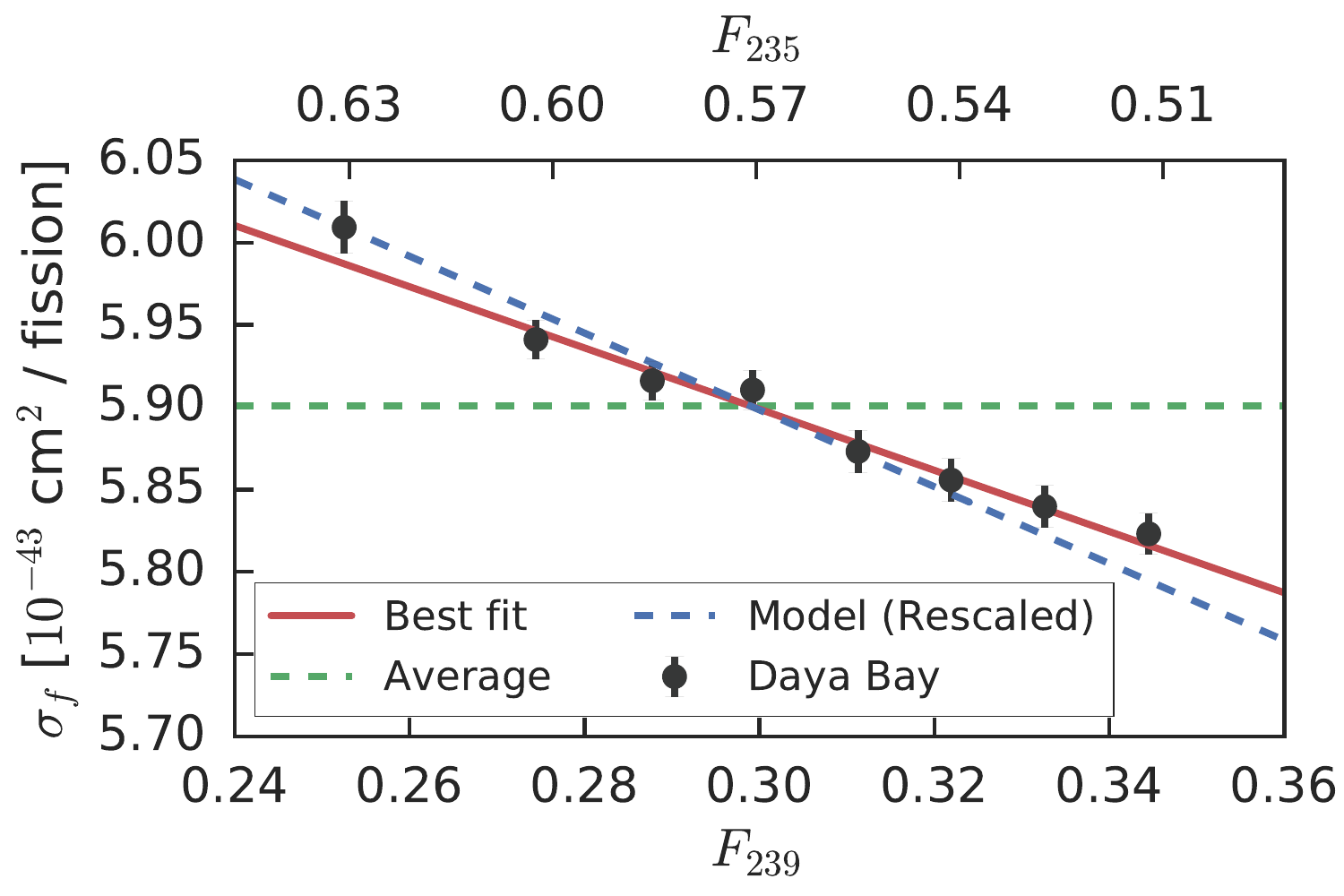}
    \par\end{centering}
    \caption{\label{fig:dyb_evolution}
      Measurements of IBD yield per fission, $\sigma_f$ (black), versus
      effective $^{239}$Pu (lower axis) or $^{235}$U (upper axis) fission fractions,
      taken from Ref.~\cite{An:2017osx}.
      The predicted yields from the Huber--Mueller model~\cite{Huber:2011wv,Mueller:2011nm} (blue),
      scaled to account
      for the difference in total yield between data and prediction, are shown. A
      clear discrepancy is seen between measurements and model predictions.}
\end{figure}

\begin{figure}[htp]
  \begin{centering}
    \includegraphics[width=0.45\textwidth]{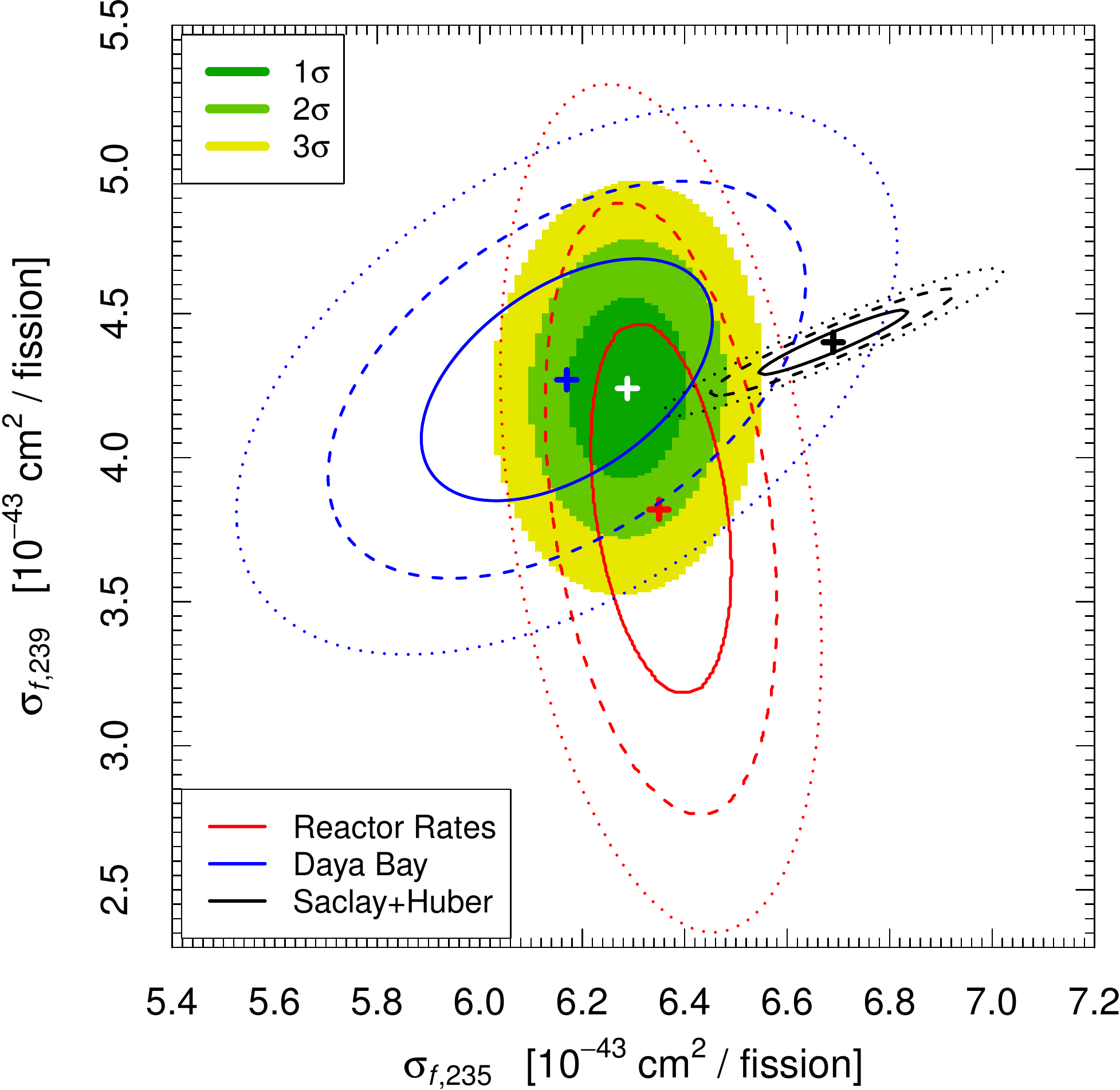}
    \par\end{centering}
    \caption{\label{fig:global_evolution} 
      Allowed regions (filled colored contours) in the $\sigma_{f,235}$-$\sigma_{f,239}$ plane obtained
      from the combined fit of the reactor rates~\cite{Giunti:2016elf} and
      the Daya Bay measurement of $\sigma_{f,235}$ and
      $\sigma_{f,239}$~\cite{An:2017osx}, taken from Ref.~\cite{Giunti:2017nww}.
      The red, blue and black curves enclose, respectively, the allowed
      regions obtained from the fit of the reactor rates~\cite{Giunti:2016elf}, the allowed
      regions corresponding to the Daya Bay measurement~\cite{An:2017osx},
      and the theoretical Huber--Mueller model~\cite{Huber:2011wv,Mueller:2011nm}
      allowed regions.}
\end{figure}


In addition to the measured reactor neutrino energy spectra, evidence also indicates
the underestimation of the model uncertainties from the extracted
antineutrino flux of $^{235}$U and $^{239}$Pu.  Figure~\ref{fig:dyb_evolution}
shows the measured IBD yield per fission, $\sigma_f$, as a function of the effective
$^{239}$Pu fission fraction from Daya Bay~\cite{An:2017osx}. The data from Daya Bay after an overall normalization correction to account for the rate deficit
still
deviated from the prediction of the Huber--Mueller model~\cite{Huber:2011wv,Mueller:2011nm}.
Taking into account the original model uncertainty as well as the measurement
uncertainties, the Huber--Mueller model prediction was disfavored at $\sim$3.1$\sigma$. 

These data were further used to extract the IBD yield per $^{235}$U fission, $\sigma_{235}$, and
the IBD yield per $^{239}$Pu fission, $\sigma_{239}$. The IBD yield per $^{241}$Pu ($^{238}$U)
fission, $\sigma_{241}$ ($\sigma_{238}$), which contributes about 5\% (10\%)
to the antineutrino flux, was conservatively constrained to 10\% uncertainty.

The 2D confidence interval for $\sigma_{235}$ vs. $\sigma_{239}$ from Daya Bay
is shown in Fig.~\ref{fig:global_evolution}. In comparison, the results from
Ref.~\cite{Giunti:2017nww} are shown after analyzing the measured rates from all the short-baseline
reactor experiments with various average fission fractions. In the latter analysis,
the uncertainties of $\sigma_{238}$ and $\sigma_{241}$ were conservatively taken to be 15\% and
10\%, respectively.

In comparison, with the predictions from the
Huber--Mueller model~\cite{Huber:2011wv,Mueller:2011nm}, both results showed a clear deficit
in $\sigma_{235}$. The uncertainty of $\sigma_{235}$ from the rate analysis was smaller than
that of the Daya Bay fuel-evolution analysis, as some of the short-baseline experiments were
performed with highly-enriched $^{235}$U. In contrast, the uncertainty of $\sigma_{239}$
from the Daya Bay fuel-evolution analysis was smaller than that of the rate analysis.
Within experimental uncertainties, both measurements of $\sigma_{239}$ were consistent with
that from Huber--Mueller model.

In summary, the analysis of measured reactor neutrino energy spectra and fuel
evolution from Daya Bay suggests an underestimation of the calculated reactor neutrino flux,
which has shaken the foundation of the light-sterile-neutrino explanation of the reactor antineutrino
anomaly. On the other hand, an increase of the reactor neutrino flux uncertainties also
enlarges the allowed phase space for sterile neutrino couplings
(i.e., $\sin^22\theta_{14}$ and $\Delta m^2_{41}$). Additional measurements are thus
necessary to fully address this question.



\section{Additional Physics Topics Using Reactor Neutrinos}\label{sec:additional}

The high statistics data acquired by reactor neutrino experiments,
together with the accurate determination of the antineutrino energy
using the IBD reaction, have prompted various searches for new effects
within or beyond the paradigm of three-flavor neutrino oscillation.
The search for a light sterile neutrino, discussed in the previous section, is 
a prime example. In this section, we discuss examples of other
searches for new effects, including the search for the neutrino magnetic moment, 
the attempt to constrain characteristics of the wave-packet approach for
neutrino oscillation, the test of the Leggett--Garg inequality, and the search for the
breaking of Lorentz and CPT invariance.

\subsection{Search for the Neutrino Magnetic Moment via Neutrino-electron
  Scattering}

A natural extension to the standard model is the potential existence of
neutrino electromagnetic interactions with virtual
photons~\cite{Kayser:1982br,Nieves:1981zt,Shrock:1982sc}, which can be
described at low-momentum transfer by two phenomenological parameters,
the anomalous magnetic moment $\mu_{\nu}$ and the mean-square charge
radius $\langle r^2 \rangle$~\cite{Vogel:1989iv}. A non-zero $\mu_{\nu}$ would enable
left-handed neutrinos to flip into sterile right-handed neutrinos
in a magnetic field. In the minimal standard model, neutrinos
are massless and have no magnetic moment. A non-zero moment can be
generated through radiative corrections~\cite{Lee:1977tib,Marciano:1977wx} for
massive Dirac neutrinos in a simple extension~\cite{Fujikawa:1980yx}:
\begin{equation}
  \mu_{\nu} = \frac{3G_F m_e m_{\nu}}{4\sqrt{2}\pi^2} = 3.2\times 10^{-19} \left(
  \frac{m_{\nu}}{1~eV}\right) \cdot \mu_{B},
\end{equation}
with $m$ representing the mass and $\mu_B\equiv e/2m_e$ being the electron
Bohr magnetons.
In comparison, $\langle r^2 \rangle$ conserves helicity in
interactions between a neutrino and a charged particle. The
interpretation of $\langle r^2 \rangle$ is still under
debate. On one hand, authors of Refs.~\cite{Lee:1977tib,Fujikawa:2003tz,Fujikawa:2003ww}
showed that a straightforward definition of $\langle r^2 \rangle$
was gauge-dependent and thus unphysical. On the other hand,
authors of Refs.~\cite{Bernabeu:2000hf,Bernabeu:2002nw,Bernabeu:2003xj}
 interpreted $\langle r^2 \rangle$ as a physical observable,
and $\langle r_{\bar{\nu}_e}^2 \rangle=0.4\times10^{-32}$~cm$^2$ was
predicted within the standard model framework.

For reactor neutrinos, both $\mu_{\nu}$ and $\langle r^2 \rangle$  can be
accessed through
the neutrino-electron elastic scattering having a cross
section~\cite{Vogel:1989iv}:
\begin{eqnarray}
  \frac{d\sigma}{dT} = \frac{G_F^2m_e}{2\pi} ( (g_V + x + g_A)^2 \nonumber \\
  + (g_V + x - g_A)^2 \left(1-\frac{T}{E_\nu}\right)^2) + (g^2_A - (g_V+x)^2) \frac{m_e T}{E_\nu^2} ) \nonumber \\
  + \frac{\pi\alpha^2\mu_{\nu}^2}{m_e^2} \frac{1-T/E_\nu}{T},
\end{eqnarray}
where $E_{\nu}$ is the neutrino energy and 
\begin{eqnarray}
  g_V &=& 2\sin^2\theta_W + 1/2 \nonumber \\
  g_A &=& -1/2 \nonumber \\
  x &=& \frac{\sqrt{2}\pi\alpha}{3G_F} \langle r^2 \rangle
\end{eqnarray}
for $\bar{\nu}_e$. Here, $\theta_W$ is the weak mixing angle
and  $T$ stands for the kinetic energy of the scattered electron.
In particular, the $1/T$ term associated with $\mu_{\nu}$ leads to
a significant increase of the cross section at low kinetic energies.
Therefore, the most sensitive direct limit,
$\mu_{\nu}<3.2\times10^{-11}\mu_B$, came from high-purity germanium
detectors at about a 10-keV
threshold~\cite{Beda:2009kx,Li:2002pn,Wong:2006nx}. The
$\mu_{\nu}$ contribution at the present limit are still orders of magnitude
higher than the standard model prediction.  Other technologies,
such as time projection chamber~\cite{Daraktchieva:2005kn},
organic scintillator~\cite{Vidyakin:1992nf}, and scintillating
crystal~\cite{Deniz:2009mu}, were also used to set direct limits
on $\mu_{\nu}$. A relaxed indirect limit on $\mu_{\nu}$ was set by
KamLAND's search for solar $\bar\nu_e$~\cite{Eguchi:2003gg}. 
In addition, limits on $\langle r_{\bar{\nu}_e}^2 \rangle$
were set at a few times
$10^{-32}$ cm$^2$~\cite{Deniz:2009mu,Vidyakin:1992nf}.
Neutrino-electron elastic scattering from reactor
neutrinos can also be used to perform (precision) measurements of
the weak mixing angle $\theta_W$ at low momentum
transfer~\cite{Reines:1976pv,Deniz:2009mu}.

\subsection{Wave Packet and Neutrino Oscillation}\label{sec:decoh}
       
The phenomenon of neutrino oscillation is usually formulated as a quantum
mechanical effect using a plane-wave approximation. While successful in
describing many neutrino oscillation results,
the plane-wave approach can lead to apparent paradoxes~\cite{Giunti:2003ax,
Akhmedov:2009rb}. The necessity of a wave-packet treatment 
for neutrino oscillation
has been considered since the 1970s~\cite{Nussinov:1976uw,Kayser:1981ye}. 
The wave-packet
models of neutrino oscillation contain a quantity
$\sigma_p$ that effectively describes the momentum dispersions of
all particles involved in the production and detection of neutrinos.
A consequence of a non-zero value of $\sigma_p$ is the \lq{decoherence}\rq~
of the quantum superposition of mass eigenstates, leading to
a modification or diminishing of the neutrino oscillation pattern.
Moreover, the width of the wave packet would also broaden as time elapses,
as a result of the momentum dispersion.

Despite many theoretical advances in formulating
the wave packet models, within quantum mechanical or field-theoretical 
approaches, no quantitative estimates for $\sigma_p$ or
the related spatial width $\sigma_x = \left(2\sigma_p\right)^{-1}$ are available. A treatment
of the decoherence length for neutrinos produced in pion decays using
density matrix formalism was recently performed~\cite{Jones:2014sfa}.
For antineutrinos
produced in reactors, estimates for $\sigma_x$ vary from $\sim$10$^{-12}$~cm
(the size of the uranium nucleus) to $\sim 10^{-7}$ cm (atomic scale), corresponding
to $\sigma_p \sim$10~MeV to $\sigma_p \sim$100~eV~\cite{An:2016pvi}.

The recent high-statistics reactor neutrino oscillation data have 
provided an opportunity to compare these data against the wave-packet
approach and to set a constraint on the momentum
dispersion of the wave packet for the first time~\cite{An:2016pvi}.
In particular, a search for possible decoherence effects in neutrino 
oscillation was performed
using Daya Bay data. The good energy resolution, together with 
large statistics collected at multiple baselines, allowed a meaningful
study of quantum decoherence effects based on these data.

In the wave-packet approach, the probability of a neutrino's oscillating from
flavor $\alpha$ to $\beta$ at a distance $L$, $P_{\alpha \beta} (L)$,
can be written as~\cite{An:2016pvi}
\begin{equation}
P_{\alpha\beta}  =\sum_{k,\,j=1}^3\frac{ V^*_{\alpha k}
V^{\phantom *}_{\beta k}V^{\phantom\dagger}_{ \alpha j}  V^*_{\beta j} }
{\sqrt[4]{1 +\left(L/L^{\mathrm{d}}_{kj}\right)^2}}
\mathrm{e}^{- \frac{\left(L/L^\mathrm{coh}_{kj}\right)^2}
{1+\left(L/L^{\mathrm{d}}_{kj}\right)^2} -\mathrm{D}^2_{kj}-i\widetilde{\varphi}_{kj}},
\label{eq:ossc}
\end{equation}
where $V_{\alpha k}$ is the usual neutrino mixing matrix element. Three
length scales appear in Eq.~(\ref{eq:ossc}):
\begin{equation}
L^\mathrm{osc}_{kj}  = \frac{4\pi p}{\Delta m^2_{kj}},~~
L^\mathrm{coh}_{kj}  =\frac{L^\mathrm{osc}_{kj}}{\sqrt 2 
\pi\sigma_\mathrm{rel}}, ~~
L^\mathrm{d}_{kj}  = \frac{L^\mathrm{coh}_{kj}}
{2\sqrt{2}\sigma_{\mathrm{rel}}},
\label{lengthts-vacuum_1}
\end{equation}
where the relative momentum spread, $\sigma_{\mathrm{rel}} = \sigma_p/p$, is a 
Lorentz invariant quantity. $L^{\mathrm{osc}}_{kj}$ refers to the usual 
oscillation length where maximal oscillation occurs for the neutrino 
mass-squared difference $\Delta m^2_{kj}$. The neutrino
coherence length, $L^{\mathrm{coh}}$, corresponds to the distance at which the
wave packet splits into non-overlapping components, diminishing the interference
between neutrino mass eigenstates $k$ and $j$. The dispersion
length, $L^{\mathrm{d}}$, 
characterizes the distance when the spatial 
widths of the wave packets for $k$ and $j$ mass eigenstates differ 
sufficiently because of momentum dispersion, and oscillation is suppressed. The 
quantity $\mathrm{D}_{kj}$ in Eq. (\ref{eq:ossc}) is given as
\begin{equation}
\mathrm{D}_{kj} = \frac{\sqrt{2}\pi\sigma_x}{L_{kj}^\mathrm{osc}}, 
\label{eq:D_factor}
\end{equation}
which suppresses the oscillation when the spatial width, $\sigma_x$, of the
wave packet is large compared with the oscillation width, 
$L^{\mathrm{osc}}_{kj}$. The expression for the phase $\widetilde{\varphi}_{kj}$, 
which is the sum of the usual plane-wave phase 
$\varphi_{kj} = 2\pi L/L^\mathrm{osc}_{kj}$ and another correction term
arising from the wave packet, can be found in Ref.~\cite{An:2016pvi}.

From Eq.~(\ref{eq:ossc}) and Eq.~(\ref{eq:D_factor}), in the
limits of $\sigma_p \to 0$ or $\sigma_p \to \infty$, the oscillation probability
in Eq.~(\ref{eq:ossc}) becomes
\begin{equation}
P_{\alpha\beta}=\sum_k|V_{\alpha k}|^2|V_{\beta k}|^2.
\label{eq:prob_decoherent}
\end{equation}
The interference terms with $k \ne j$ in Eq.~(\ref{eq:ossc}) now all vanish. 
Thus $P_{\alpha\beta}$ is now independent of distance, and the oscillation 
pattern disappears. This result can be understood
intuitively. As $\sigma_p \to 0$, the spatial width of the wave packet
approaches infinity, washing out any oscillation pattern having a 
finite oscillation length. Similarly,
an infinite $\sigma_p$ gives zero coherence and dispersion 
lengths, preventing any interference effects. 
Observation of oscillation behavior in reactor neutrino experiments
clearly shows that $\sigma_p$ must lie somewhere between these 
two extremes. 

The Daya Bay Collaboration has performed~\cite{An:2016pvi} a fit 
to the neutrino oscillation
data utilizing the wave packet oscillation expression 
of Eq.~(\ref{eq:ossc}).
The allowed region for $\sigma_{\mathrm{rel}}$ at a 95\% C.L. was found to be
$2.38 \times 10^{-17} < \sigma_{\mathrm{rel}} < 0.23$. Adding the constraints
of the sizes of the reactor cores and detectors, the upper limit reduces
to 0.20, corresponding to 
$10^{-11}\mathrm{ cm } \lesssim\sigma_x \lesssim 2$~m. It is worth noting
that the lower limit in $\sigma_x$ is roughly 10 times the size of the uranium
nucleus.

With additional data from Daya Bay, the sensitivity on
the upper limit of $\sigma_{\mathrm{rel}}$ is expected to be improved
by $\sim$30\%. Nevertheless, a decoherence effect from the wave-packet
approach was found to be insignificant for the Daya Bay 
experiment~\cite{An:2016pvi}. 
Thus, the neutrino oscillation parameters $\sin^2 2\theta_{13}$ and 
$\Delta m^2_{32}$ extracted from the plane-wave approach are
entirely reliable.

\subsection{Leggett--Garg Inequality and Neutrino Oscillation}

The phenomenon of neutrino oscillation is fundamentally a quantum mechanical
effect. It originates from the principle of superposition, which
allows a neutrino flavor eigenstate to be expressed as a 
coherent superposition of neutrino mass eigenstates. As discussed
in Sec.~\ref{sec:decoh}, decoherence effects would lead to the disappearance
of neutrino oscillation. 

The superposition principle remains an enigmatic and nonintuitive
ingredient of the quantum mechanics. At the macroscopic level, a system's being
able to coexist in different states led to the famous paradox
of Schr\"odinger's cat~\cite{Schrodinger:1935zz}. At 
the microscopic level, the celebrated
Bell's inequality~\cite{Bell:1964kc} was proposed as a quantitative means to
probe quantum mechanical coherence, or entanglement, within a spatially
separated system. While Bell's inequality has been extensively tested,
a loophole-free test of this inequality remains an elusive goal. 

In 1985, Leggett and Garg~\cite{Leggett:1985zz} proposed a new test 
of quantum coherence
not only for microscopic systems, for which Bell's inequality
applies, but also for macroscopic systems. To facilitate such a test
for macroscopic systems, Leggett and Garg considered the correlations
of a single system measured at different times.

The Leggett--Garg inequality
(LGI) is derived based on two principles: macroscopic realism (MR) and
non-invasive measurability (NIM). Realism, often encoded in hidden-variable
theories, implies that a measurement on a system reveals a pre-existing
value. Under realism, systems prepared identically can be distinguished
via a set of hidden variables, and a measurement would uncover a pre-existing
value. NIM stipulates that a measurement could be performed without 
disturbing the system. While MR and NIM are consistent with classical
mechanics, they certainly contradict quantum mechanics. 
The LGI provides a method to test the applicability of quantum mechanics
to macroscopic systems, and LGI is often regarded as the time analogue
of Bell's inequality~\cite{PhysRevLett.71.3235}. A recent review 
on LGI can be found in Ref.~\cite{1304.5133}.

The LGI involves the two-time correlation function $C_{ij} = \langle
Q(t_i)Q(t_j) \rangle$, where $Q$ is a dichotomic observable with 
$Q = \pm 1$. The value of $C_{ij}$ is obtained by summing over the four
possible values of $Q(t_i)Q(t_j)$ (namely, +1, -1, -1, +1) weighted by the
corresponding probability $P_{ij}(Q_i,Q_j)$. From $C_{ij}$ 
the quantity $K_n$ could be defined from measurements performed at $n$ distinct
times:
      
\begin{equation}
K_n = C_{21} + C_{32} + C_{43} + \cdot \cdot + C_{n (n-1)} - C_{n1}.
\label{eq:kn}
\end{equation}
Under the assumptions of MR and NIM, Leggett and Garg obtained the
inequality $K_n \le n-2$ for $n \ge 3$. 

Twenty-five years after the work of Leggett and Garg, the
first observation of the violation of LGI was 
reported~\cite{doi:10.1038/nphys1641},
followed by many other LGI tests~\cite{1304.5133}. However, most of
the tests suffer from the \lq{clumsiness loophole}\rq~\cite{1001.1777}, 
for which the LGI
violation could be attributed to unintentional disruption of the system
during measurements. This loophole could be avoided by using weak or
indirect measurements. 

The idea of testing LGI using neutrino oscillation was proposed several years
ago~\cite{Gangopadhyay:2013aha}, and the first test was performed 
recently~\cite{Formaggio:2016cuh}. As an example, consider the case of reactor
neutrino oscillation with an electron antineutrino at $t=0$. If at time $t$,
a measurement finds an electron antineutrino, then $Q=+1$. Otherwise, $Q=-1$. The key idea
is to mimic a series of measurements at various times on a single neutrino by
measurements made on an ensemble of neutrinos of various energies at a given 
time. Details of this method can be found in 
Refs.~\cite{Gangopadhyay:2013aha,Formaggio:2016cuh}. One unique 
feature of this method is the
long coherence length for neutrino oscillation, unlike other LGI tests
involving much shorter coherence lengths. This method is also free from the
\lq{clumsiness loophole}\rq. Using the MINOS muon neutrino oscillation data
at a baseline of 735 km, the LGI for $K_3$ and $K_4$ was found to be 
violated at a level greater than $6\sigma$~\cite{Formaggio:2016cuh}. 
A recent analysis of the Daya Bay data also showed a very similar
result~\cite{Fu:2017hky}.

\subsection{Lorentz Violation and Neutrino Oscillation}

The standard model  and general relativity (GR) are believed to be the
low-energy limit of a theory that unifies quantum physics and 
gravity at the Planck scale, $M_P \approx 10^{19}$~GeV. An effective
field theory at lower energies, called the standard-model 
extension (SME)~\cite{Colladay:1996iz,Colladay:1998fq,Kostelecky:2003fs},
extends the GR-coupled SM by including Lorentz-violating terms constructed
from SM and GR fields. 
The Lorentz and CPT violations in the SME are caused by background Lorentz tensor fields of the Universe. These
background fields are fixed in spacetime, implying rotation and boost
dependence of physics in a specific frame.
While suppressed at presently accessible energy $E$ by an order 
of $\sim$E/M$_P$,
the predicted violations of Lorentz and CPT symmetries might be revealed
in sensitive measurements. 

Quantum interference phenomena such as neutral-meson 
oscillation~\cite{Kostelecky:1997mh} and
neutrino oscillation~\cite{Kostelecky:2003cr} might provide 
sensitive searches for the Lorentz
and CPT violations predicted by the SME. A small coupling between neutrinos
and a Lorentz-violating field can conceivably alter the pattern of neutrino
oscillation~\cite{Kostelecky:2003cr}. In the SME, the effective 
Hamiltonian for neutrino oscillation
is given as~\cite{Kostelecky:2003cr}
\begin{equation}
(h^\nu_{eff})_{ab} \sim \frac{(m^2)_{ab}}{2E} + \frac {1}{E} [(a_L)^\mu p_\mu
-(c_L)^{\mu \nu} p_\mu p_{\nu}]_{ab},
\label{eq:lvh}
\end{equation}  
where $a$ and $b$ refer to the neutrino flavors and $E$ and $p_\mu$ are the
energy and the energy-momentum 4-vector of the neutrino, respectively. The first term
on the right-hand-side of
Eq.~(\ref{eq:lvh}) is the SM contribution from massive neutrinos. The
coefficients $(a_L)^\mu_{ab}$ have dimensions of mass and violate
both Lorentz and CPT symmetry, while the dimensionless coefficients
$(c_L)^{\mu \nu}_{ab}$ violate Lorentz but keep CPT symmetry. The
CPT-odd $(a_L)^\mu_{ab}$ changes sign for antineutrinos and can lead to
differences between neutrino and antineutrino oscillation.

This CPT-violating feature of SME offered an attractive
possible explanation~\cite{Katori:2006mz} for the LSND 
$\nu_{\mu} \to \nu_e$ result~\cite{Aguilar:2001ty}. Moreover, the
vector $(a_L)^\mu_{ab}$ and tensor $(c_L)^{\mu \nu}_{ab}$ coefficients 
introduce directional dependence of neutrino oscillation. If the 
$Z$-axis is chosen as the rotation axis of the Earth, then a sidereal
variation of the neutrino direction in $X$ and $Y$ would occur.
Therefore, a sidereal variation of neutrino oscillation can be caused 
by coefficients
$(a_L)^\mu_{ab}$, $(c_L)^{\mu \nu}_{ab}$, for which at least one of $\mu$ and 
$\nu$ is either $X$ or $Y$. In other words, all coefficients
except $(a_L)^T_{ab}$, $(a_L)^Z_{ab}$, $(c_L)^{TT}_{ab}$, $(c_L)^{TZ}_{ab}$,
and $(c_L)^{ZZ}_{ab}$ can contribute to sidereal variations.

Under SME, the probability for an electron antineutrino $\bar \nu_e$ 
to oscillate to $\bar \nu_x$, where $x$ is $\mu$ or $\tau$, 
is given as~\cite{Kostelecky:2004hg} 
\begin{eqnarray}
  P_{\bar \nu_e \to \bar \nu_x} &\simeq& L^2 [(C)_{\bar e \bar x} + (A_s)_{\bar e \bar x} \sin(\omega_\oplus T_\oplus) \nonumber\\
    &+& (A_c)_{\bar e \bar x} \cos(\omega_\oplus T_\oplus) 
    + (B_s)_{\bar e \bar x} \sin(2 \omega_\oplus T_\oplus) \nonumber \\
    &+& (B_c)_{\bar e \bar x} \cos(2 \omega_\oplus T_\oplus)]^2,
\label{eq:lvosc}
\end{eqnarray}
where $\omega_\oplus$ and $T_\oplus$ are the sidereal frequency and
sidereal time, and $L$ is the baseline. The expressions for the
parameters $A_{s,c}$, $B_{s,c}$, and $C$ consist of the Lorentz-violating
coefficients introduced in Eq. (\ref{eq:lvh}). Expressions analogous
to Eq. (\ref{eq:lvosc}) can be obtained for oscillations involving other
neutrino flavors. For reactor neutrino disappearance experiments, the
probability $P_{\bar \nu_e \to \bar \nu_e}$ is simply 
$P_{\bar \nu_e \to \bar \nu_e} = 1 - P_{\bar \nu_e \to \bar \nu_\mu}
- P_{\bar \nu_e \to \bar \nu_\tau}$.

Searches for Lorentz violations in neutrino oscillation via measurements of
sidereal modulations of oscillation probability have been performed in
accelerator based experiments, including LSND~\cite{Auerbach:2005tq}, 
MINOS~\cite{Adamson:2008aa,Adamson:2010rn,Adamson:2012hp},
and MiniBooNE~\cite{AguilarArevalo:2011yi}, as well as the non-accelerator 
experiment IceCube~\cite{Abbasi:2010kx}. No evidence for 
Lorentz violating sidereal
modulations has been found, setting upper limits on various coefficients 
in Eq. (\ref{eq:lvh}). Combining the analysis of MINOS near-detector (ND) 
data on $\nu_{\mu}$ and $\bar \nu_{\mu}$ disappearance and far-detector (FD) 
data on $\nu_{\mu}$ disappearance, limits on both the real and imaginary parts
of 18 Lorentz-violating coefficients have been obtained~\cite{Adamson:2012hp}.
Effects of the $a_L$-type ($c_L$-type) coefficients are proportional to
$L^2$ and $(E_\nu L)^2$, accounting for the greater sensitivities of the 
FD data~\cite{Adamson:2010rn} for constraining some coefficients, 
despite its lower event rates
compared with the ND data~\cite{Adamson:2008aa}. This consideration 
also favors the IceCube
experiment, which sets a stringent limit for $(c_L)_{\mu \tau}^{TX(TY)}$ at
$3.7 \times 10^{-27}$~\cite{Abbasi:2010kx}.

The only search for Lorentz violation in reactor neutrino experiments
was performed by the Double Chooz 
Collaboration~\cite{Abe:2012gw}. The relatively
low antineutrino energies and short baseline may limit the 
reach of reactor-based neutrino experiments. However, unlike the
long-baseline MINOS and IceCube experiments, the reactor $\bar \nu_e$
disappearance experiments are sensitive to Lorentz-violating 
coefficients in the $e-\tau$ sector. Using 8249 candidate IBD events
collected at the Double Chooz FD, constraints on the upper
limits of various combinations of 14 of the SME coefficients in the
$e-\tau$ sector have been obtained for the first time~\cite{Abe:2012gw}. 
With a much longer
baseline and much larger detector volume, the JUNO reactor-neutrino
experiment~\cite{An:2015jdp} is expected to reach even better 
sensitivities
in the search for Lorentz-violating effects in the $e-\tau$ sector.



\section{Conclusions}\label{sec:conclusion}

In this article, we review the theoretical and experimental physics associated
with man-made reactor neutrinos. Since the discovery of reactor-produced neutrinos
in the 1950s, knowledge of the production of reactor
neutrinos has been significantly improved. The absolute reactor flux
and energy spectrum can now be predicted at the 5\% and 10\% level, respectively.
Inverse beta decay, the primary detection channel of reactor neutrinos,
is the most well-understood reaction, allowing for an accurate determination
of neutrino energy. Benefiting from these important features, reactor neutrinos
have played important roles in establishing the
   current paradigm of three-neutrino flavor mixing.

   At an average baseline of 180 km, the
KamLAND experiment observed neutrino oscillation in the solar sector and
provided an independent constraint in $\theta_{12}$ and an accurate
determination of $\Delta m^2_{21}$. At shorter baselines of 1$\sim$2~km, the Daya Bay, RENO, and
Double Chooz experiments observed neutrino oscillation, establishing
a non-zero value for the last unknown mixing angle, $\theta_{13}$. The
discovery of a non-zero $\theta_{13}$ has opened a gateway to access two of
the remaining unknowns in the neutrino  properties: the CP phase $\delta_{CP}$ that
may provide a new source for CP violation, and the mass hierarchy that may
provide a crucial input to reveal the Dirac or Majorana nature of neutrino. 

The future physics program of reactor neutrinos is quite diversified. On
one hand, the JUNO experiment will precisely measure neutrino oscillation
at a $\sim$55-km baseline with an excellent energy resolution. The simultaneously measured oscillation
caused by ($\theta_{12}$, $\Delta m^2_{21}$) and ($\theta_{13}$, $\Delta m^2_{32}$) will
allow a determination of the neutrino mass hierarchy and a precision
measurement of these mixing parameters. On the other hand, a new generation
of very-short-baseline reactor experiments will search for a light
sterile neutrino. These new measurements together with
those using other neutrino sources are expected to
explore possible new physics beyond the standard
model. As we enter the precision era of neutrino physics, reactor neutrinos
might yet lead to other unexpected major discoveries.

\section{Acknowledgements}
We thank Chao Zhang, Petr Vogel, and Laurence Littenberg for
their helpful comments on the manuscript and helpful discussions. We
thank Wei Tang for his assistance in preparing some figures and Celia Elliott for
her careful reading of the manuscript.
This work was supported in part by the National Science Foundation,
U.S. Department of Energy, Office of Science, Office of High Energy Physics,
under contract number DE-SC0012704.

\bibliographystyle{hunsrt}
\bibliography{reactor_physics}{}

\end{document}